\documentclass[12pt, letter]{article}

\usepackage{geometry}
\usepackage{setspace}
\usepackage{stix}
\usepackage{titling}
\usepackage{authblk}
\usepackage{mathtools}
\usepackage{xcolor, hyperref}
\hypersetup{colorlinks=true, linkcolor=blue, citecolor=blue, urlcolor=cyan}
\usepackage{cite}
\usepackage{algpseudocode} 
\usepackage{algorithmicx}
\usepackage{algorithm}
\usepackage{graphicx}
\usepackage{subcaption}
\usepackage{siunitx}
\usepackage{indentfirst}
\usepackage{pgfgantt}
\usepackage[nottoc,numbib]{tocbibind}
\usepackage{enumitem}
\usepackage{cancel}
\usepackage{cases}
\usepackage{multirow}

\DeclarePairedDelimiter{\ps}{\lparen}{\rparen}
\DeclarePairedDelimiter{\bs}{\lbrack}{\rbrack}

\DeclarePairedDelimiter{\abs}{\lvert}{\rvert} 
 
\DeclarePairedDelimiter{\br}{\{}{\}}

\makeatletter
\newcommand{\pushright}[1]{\ifmeasuring@#1\else\omit$\displaystyle#1$\ignorespaces\fi}
\makeatother
\allowdisplaybreaks

\title{New Plasma Sheath Potential Solutions in Cylindrical and Spherical Coordinates}

\author[1]{Indronil Ghosh\footnote{email: ighosh@g.ucla.edu}}
\author[1]{Timothy S. Fisher}
\affil[1]{University of California, Los Angeles}

\date{ }
\begin{document}
\begin{titlepage}
\maketitle
\begin{abstract}
\noindent Leading edges of hypersonic vehicles can reach temperatures greater than 2000 °C, and radii of curvature smaller than 1 cm, at which thermionic emission (also known as electron transpiration) can play a significant role in cooling the leading edge alongside other heat transfer modes such as convection and radiation. Existing theoretical analyses of thermionic cooling with space-charge effects at a leading edge are limited to one-dimensional (1D), analytical and numerical models that do not capture the influences of geometric curvature of the leading edge or temperature gradients along the leading edge. The key to understanding space-charge effects is development of the plasma sheath potential, and to that end we demonstrate a generalized methodology to calculate the sheath potential space in 1D Cartesian, cylindrical, and spherical coordinate systems. We accomplish this by extending Takamura's approach beyond the Cartesian system, and motivate sheath formation conditions for potential sheathes with and without a virtual cathode similar in nature to how Bohm originally presented his criterion of minimum Mach number for a valid 1D Cartesian sheath. By observing for what parameter inputs we satisfy the sheath formation conditions, we illustrate parameter spaces of minimum Mach number, potential derivative at the wall, and net current, for each coordinate system and for two different input work functions; we also show example potential spaces for each coordinate system. With our numerical approach, generalized to multiple coordinate systems, we enable computationally efficient and higher fidelity analysis of thermionic emission with space-charge effects for more realistic system geometries.

\end{abstract}
\end{titlepage}
\tableofcontents
\newpage
\newgeometry{top=1in,bottom=1in,right=1in,left=1in}

\section{Introduction}
The effect of heat spreading by reflected electrons in thermionic cooling of a leading edge (LE) may be significant for high temperature gradients and small radius LEs (\(<\)1 cm), but this question of heat spreading is largely unexplored. To analyze the significance of heat spreading due to reflected electrons, we must first understand the plasma sheath’s potential space. If the potential space outside a LE presents a virtual cathode, then some fraction of thermionically emitted electrons from the wall will be reflected due to having insufficient kinetic energy to surmount the potential barriers. Current solutions to the problem of the plasma sheath potential outside a leading edge use only a Cartesian coordinate formulation of Poisson’s equation with electron and positive ion source terms originally by Bohm \cite{Bohm1949} and more recently Takamura \cite{Takamura2004} that oversimplifies realistic system geometries. Bohm's chapter on the ``Minimum Ionic Kinetic Energy for a Stable Sheath'' \cite{Bohm1949} considers only constant velocity positive ions interacting with electrons following the Boltzmann relation to develop the eponymous Bohm criterion, that a stable sheath is only possible when the positive ions meet a minimum kinetic energy of half the electron temperature times the Boltzmann constant. While Takamura considers thermionically emitted electrons in addition to the positive ion and electron sources studied by Bohm, many gaps in knowledge remain unaddressed in Takamura’s work for the Cartesian system, e.g., how the shape of the sheath potential depends on system conditions including those of the incident plasma and of the LE, and importantly, what system conditions yield a virtual cathode. We aim to fill these gaps in knowledge, not only for the Cartesian coordinate system but also for the 1D cylindrical and spherical coordinate systems, as hypersonic vehicles typically have cylindrically or spherically shaped leading edges. While Takamura’s solution approach relies on some approximations that can only be applied in Cartesian coordinates, we have developed a unified solution methodology that does not require Taylor approximations and is applicable to Cartesian, cylindrical, and spherical coordinate systems. 

In addition to Takamura and Hagino et al.’s \cite{Takamura2004, Hagino2004} works on space-charge limited current and modified Child-Langmuir formulae, we note the topically relevant work by Oksuz \cite{Oksuz2006} solving for space charge limited current in cylindrical and spherical coordinates for the 1D Poisson's equation, as well as the works by Crespo \cite{Crespo2004,Crespo2006}, Zhu \cite{Zhu2013}, and Torres-Cordoba \cite{Torres-Cordoba2017} on Langmuir-Blodgett solutions in cylindrical and spherical coordinates. More recently, Halpern et. al \cite{Halpern2022} created a coordinate-system-invariant formulation of space-charge limited current with monoenergetic injection velocity, and Chen and Sanchez-Arriaga \cite{Chen2018} modeled cylindrical emissive probes using the Vlasov-Poisson equation and orbital motion theory. In extending Takamura's approach to cylindrical and spherical coordinate systems, part of the novelty in this work compared to the existing literature is the incorporation of Maxwellian distributions of emitted electron velocity, and use of the Runge-Kutta method to solve the second-order differential equation governing the plasma sheath in a manner that is generalized to all three coordinate systems. Additionally, the results provide suitable analytical comparisons for particle-in-cell codes that model plasmas in cylindrical and spherical coordinates. Our methodology not only serves as a benchmark, but also can be easily augmented to handle other highly nonlinear or multiple-term differential equations that underlie plasma sheaths.

\section{Methodology}
\subsection{Wall and plasma source terms}
\noindent The plasma sheath is governed by Poisson’s equation, in which we model the charge density as dependent on the sum of three source terms: the plasma electron density, wall-emitted electron density, and plasma positive ion density. With the Laplacian operator in the Cartesian, cylindrical, and spherical coordinate systems, we construct the following differential equations:
\begin{align}
\frac{d^2 \Phi}{d \xi^2} & = n_{\text{e},\,\text{Car}}^\text{s}\bs*{\Phi\ps*{\xi}} + n_{\text{e},\, \text{Car}}^\text{p}\bs*{\Phi\ps*{\xi}} - n_\text{i}\bs*{\Phi\ps*{\xi}}\qquad \text{(Cartesian)}\label{eq:CartesianDiffEq}\\
\frac{d^2 \Phi}{d\xi^2} + \frac{1}{\xi}\frac{d \Phi}{d \xi} & = n_{\text{e},\,\text{Cyl}}^\text{s}\bs*{\Phi\ps*{\xi}} + n_{\text{e},\, \text{Cyl}}^\text{p}\bs*{\Phi\ps*{\xi}} - n_\text{i}\bs*{\Phi\ps*{\xi}}\qquad \text{(Cylindrical)} \label{eq:CylindricalDiffEq} \\
\frac{d^2 \Phi}{d \xi^2} + \frac{2}{\xi}\frac{d \Phi}{d \xi} & = n_{\text{e},\,\text{Sph}}^\text{s}\bs*{\Phi\ps*{\xi}} + n_{\text{e},\, \text{Sph}}^\text{p}\bs*{\Phi\ps*{\xi}} - n_\text{i}\bs*{\Phi\ps*{\xi}}\qquad \text{(Spherical)} \label{eq:SphericalDiffEq}
\end{align}
\noindent where
\begin{equation}\label{eq:BoltzmannRelation}
n_{\text{e},\,\text{sys}}^\text{p}\ps*{\Phi} = \frac{1}{\gamma_\text{sys}} n^\text{p}_{\text{e}0,\,\text{sys}} \exp \ps*{\Phi}
\end{equation}
is the Boltzmann relation, assuming a collisionless plasma. Next, 
\begin{equation}\label{eq:CurrentRatio}
\gamma_\text{sys} = \abs*{j_{\text{e},\,\text{sys}}^\text{s}/j_{\text{e},\,\text{sys}}^\text{p}}
\end{equation}
is the wall-electron to plasma-electron current ratio, and
\begin{equation}\label{eq:PositiveIonDensity}
n_\text{i}\ps*{\Phi} = \ps*{1 - \frac{2\Phi}{M^2}}^{-1/2},\; M = \frac{V_0}{\sqrt{T_\text{e}/m_\text{i}}}
\end{equation}
derive from positive ion energy conservation.
\(n_\text{e}^\text{p}\) and \(n_\text{e}^\text{s}\) denote the plasma electron density and wall-electron density, respectively, while \(n_\text{i}\) denotes the plasma positive ion density. \(V_0\) is the ion speed at plasma-sheath edge, and \( \sqrt{T_\text{e}/m_\text{i}} \) is the ion acoustic speed. The ``sys'' subscript indicates coordinate-system dependence. In addition to maintaining similar notation to that of Takamura, we normalize the potential variable \(\phi\) using plasma electron energy, \(T_\text{e}\), and the radial position \(r\) with respect to the Debye length \(\lambda_\text{D}\) as:
\begin{equation}\label{eq:Normalization}
\Phi = \frac{e \phi}{T_\text{e}},\; \xi = \frac{r}{\lambda_D} \text{ where }\lambda_D = \sqrt{ \frac{\varepsilon_0 T_\text{e}}{e^2 n_0} }\,,
\end{equation}
where \(e\) is the electron charge, \(\varepsilon_0\) is the vacuum permittivity, and \(n_0\) is the quasi-neutral incident plasma density. While we express the plasma electron density from the Boltzmann relation and the positive ion density through simple ion energy conservation, an additional means to represent the wall-emitted electron density is required. Analogously to Takamura, we assert that the wall electrons emit according to the Maxwell-Boltzmann velocity distribution. Below we derive the velocity distribution \(f\) in each coordinate system, removing all spatial variable dependencies other than the radial. Starting from the three-dimensional, Cartesian Maxwell-Boltzmann distribution:
\begin{equation}\label{eq:3DCartesianMaxwellian}
f\ps*{\vec{v}} = f\ps*{v_x, v_y, v_z} = \ps*{\frac{m}{2\pi T_\text{eW}}}^{3/2} \exp\bs*{\frac{-m\ps*{v_x^2+ v_y^2 + v_z^2}}{2 T_\text{eW}}}\; \text{(3D Cartesian)}\, ,
\end{equation}
where \(T_\text{eW}\) denotes the wall electron emission energy, the one-dimensional distributions are as follows.\\
1D Cartesian:
\begin{align} \label{eq:1DCartesianMaxwellian}
    \begin{split}
    f\ps*{v_x} & = \ps*{\frac{m}{2\pi T_\text{eW} }}^{3/2} \exp\ps*{\frac{-m v_x^2}{2 T_\text{eW} }} \int_{-\infty}^\infty  \int_{-\infty}^\infty \exp\ps*{\frac{-m\bs*{v_x^2+ v_y^2} }{2 T_\text{eW} }} \, d v_y \, d v_z \\
    & = \ps*{\frac{m}{2\pi T_\text{eW} }}^{3/2} \exp\ps*{\frac{-m v_x^2}{2 T_\text{eW} }} \; \frac{2\pi T_\text{eW} }{m} = \ps*{\frac{m}{2\pi T_\text{eW} }}^{1/2} \exp\ps*{\frac{-m v_x^2}{2 T_\text{eW} }} 
    \end{split}
\end{align}
Correcting PDF normalization of Cartesian $f\ps*{v_x}$ to account for $v_x \in [0, \infty) $:
\begin{align} \label{eq:1DCartesianMaxwellianNormalized}
    \begin{split}
    & 1 = \int_0^\infty \ps*{\text{const}} f\ps*{v_x} \, d v_x = \int_0^\infty \ps*{\text{const}} \exp\ps*{\frac{-m v_x^2}{2 T_\text{eW} }} \, d v_x = \ps*{\text{const}} \sqrt{\frac{\pi}{2}} \sqrt{\frac{m}{T_\text{eW} }} \\
    & \Rightarrow \text{const} = \sqrt{\frac{2}{\pi}} \sqrt{\frac{T_\text{eW} }{m}}, \text{ correct coefficient to } \ps*{\frac{2m}{\pi T_\text{eW} }}^{1/2} \Rightarrow f\ps*{v_x} = \ps*{\frac{2m}{\pi T_\text{eW} }}^{1/2} \exp\ps*{\frac{-m v_x^2}{2 T_\text{eW} }}
    \end{split}
\end{align}
1D Cylindrical:
\begin{align} \label{eq:1DCylindricalMaxwellian}
    \begin{split}
    & f\ps*{\vec{v}} = f\ps*{v_r, v_z} = \ps*{\frac{m}{2\pi T_\text{eW} }}^{3/2} \exp\bs*{\frac{-m\ps*{v_r^2 + v_z^2}}{2 T_\text{eW} }}\\
    & d v_x \, d v_y \, d v_z = v_r \, d v_r \, d \theta \, d v_z \\
    & f\ps*{v_r} = \ps*{\frac{m}{2\pi T_\text{eW} }}^{3/2} \exp\ps*{\frac{-mv_r^2}{2 T_\text{eW} }} v_r \int_0^{2\pi} d \theta \int_{-\infty}^\infty \exp\ps*{\frac{-mv_z^2}{2 T_\text{eW} }} \, d v_z \\
    & = \ps*{\frac{m}{2\pi T_\text{eW} }}^{3/2} \exp\ps*{\frac{-mv_r^2}{2 T_\text{eW} }} v_r 2\pi \sqrt{\frac{2\pi T_\text{eW} }{m}} = \frac{m}{T_\text{eW} } v_r \exp \ps*{\frac{-m v_r^2}{2 T_\text{eW} }}
    \end{split}
\end{align}
1D Spherical:
\begin{align} \label{eq:1DSphericalMaxwellian}
    \begin{split}
    & f\ps*{\vec{v}} = f\ps*{v_r} = \ps*{\frac{m}{2\pi T_\text{eW} }}^{3/2} \exp\ps*{\frac{-m v_r^2}{2 T_\text{eW} }}\\
    & d v_x \, d v_y \, d v_z = v_r^2 \sin \theta \, d v_r \, d \theta \, d \varphi \\
    & f\ps*{v_r} = f\ps*{\vec{v}} v_r^2 \int_0^{\pi} \sin \theta \, d \theta \int_0^{2\pi} d \varphi \\
    & = f\ps*{\vec{v}}v_r^2 \; 4\pi = \frac{4}{\sqrt{\pi}}\ps*{\frac{m}{2 T_\text{eW} }}^{3/2} v_r^2 \exp \ps*{\frac{-m v_r^2}{2 T_\text{eW} }}
    \end{split}
\end{align}
By imposing current continuity for the wall-emitted electrons from the material surface to arbitrary radial position \(r\), we can express the differential of the emitted electron density in terms of the initial velocity distribution:
\begin{equation} \label{eq:CurrentAndEnergyConservation}
d j_\text{e}^\text{s} = e v\ps*{r, v_0} \, d n_\text{e}^\text{s}\ps*{r} = e v_0 N_0 f\ps*{v_0} \, d v_0 \Rightarrow d n_\text{e}^\text{s}\ps*{r} = N_0 \frac{v_0 f\ps*{v_0}}{v\ps*{r, v_0}} \, d v_0\,.
 \end{equation}
where \(N_0\) is a function of the quasi-neutral density \(n_0\), both with units of \(\text{m}^{-3}\) as discussed later along with the \(n_\text{e0}^\text{p}\) term. Also, \(v_0\) is the wall electron emission velocity, while \(v\) represents the wall electron velocity at arbitrary \(r\). Note that the velocity distribution \(f\) has units of s/m, yielding appropriate units of \(\text{m}^{-3}\) for \(d n_\text{e}^\text{s}\) above. Then, by requiring energy conservation for the emitted electrons:
\begin{equation} \label{eq:EmittedElectrongEnergyConservation}
\frac{1}{2} m_\text{e} v^2\ps*{r, v_0} - e\phi\ps*{r} =  \frac{1}{2} m_\text{e} v_0^2 - e\phi_\text{W} \,.
\end{equation}
The denominator of \(dn_\text{e}^\text{s}\) can now be written in terms of \(v_0\), and we may proceed to integrate the differential emitted electron density from the minimum electron velocity necessary to overcome a virtual cathode (VC) to the upper bound of infinity.
\begin{equation}\label{eq:IntegratedDifferentialEmittedElectronDensity}
n_\text{e}^\text{s}\ps*{r} = N_0 \int^\infty_{v^\text{s}_\text{VC}} \frac{v_0 f\ps*{v_0}}{v\ps*{r, v_0}}\, d v_0 = N_0 \int^\infty_{v^\text{s}_\text{VC}} \frac{v_0 f\ps*{v_0}}{\sqrt{v_0^2 + \frac{2e\ps*{\phi - \phi_\text{W}}}{m_\text{e}}}} \, d v_0 
\end{equation}
The velocity distribution creates coordinate system dependence for the plasma sheath problem, in addition to the coordinate system dependence from the Laplacian. Bringing the specific velocity distribution into the integrand above, we must solve the following integral for each coordinate system:
\begin{equation}\label{eq:CurrentDensityIntegral}
\int^\infty_{v^\text{s}_\text{VC}} \frac{v_0^g \exp\ps*{\frac{-m v^2}{2 T_\text{eW}}} }{\sqrt{v_0^2 + \frac{2e\ps*{\phi - \phi_\text{W}}}{m_\text{e}}}}\,d v_0 \text{, where } \frac{1}{2} m_\text{e} \ps*{v^\text{s}_\text{VC}}^2 = e\ps*{\phi_\text{W} - \phi_\text{VC}} \Rightarrow v^\text{s}_\text{VC} = \sqrt{\frac{2e\ps*{\phi_\text{W} - \phi_\text{VC}}}{m_\text{e}}} 
\end{equation}
where \(g = 1\) in Cartesian coordinates, \(g = 2\) in cylindrical coordinates, and \(g = 3\) in spherical coordinates. The basic forms of the integrals appear below, temporarily ignoring the velocity distributions’ front constants for simplicity, and using the shorthands \(a = m_\text{e}/\ps*{2T_\text{eW}}\), \(b = \bs*{2e\ps*{\phi - \phi_\text{W}}}/m_\text{e}\).\\
Cartesian (indefinite):
\begin{equation}\label{eq:CartesianBasicIntegral}
    \int \frac{x \exp\ps*{-a x^2} }{\sqrt{x^2 + b}} \, d x = - \frac{1}{2}\sqrt{\frac{\pi}{a}}\exp\ps*{ab}\,\text{erfc}\sqrt{a \ps*{x^2 + b}}
\end{equation}
Cylindrical (definite, to illustrate \textit{u} substitution effect on bounds): 
\begin{equation}\label{eq:CylindricalBasicIntegral}
\int_{x'}^{\infty}\frac{x^2 \exp\ps*{-a x^2} }{\sqrt{x^2 + b}} \, d x = \frac{1}{2a} \int_{a{x'}^2}^{\infty}\frac{\sqrt{u} \exp\ps*{-u} }{\sqrt{u + ab}} \, d u \text{ where } a{x'}^2 = \frac{e\ps*{\phi_\text{W} - \phi_\text{VC}}}{T_\text{eW}} \text{ and } ab = \frac{e\ps*{\phi - \phi_\text{W}}}{T_\text{eW}}
\end{equation}
(numerically evaluated with Gaussian quadrature due to lack of a closed form)\\
Spherical (indefinite):
\begin{equation}\label{eq:SphericalBasicIntegral}
\int \frac{x^3 \exp\ps*{-a x^2} }{\sqrt{x^2 + b}} \, d x = \frac{\exp\ps*{ab}}{4 a^{3/2}} \bs*{\sqrt{\pi} \ps*{1 - 2 ab}\, \text{erf}\sqrt{a \ps*{x^2 + b}} - 2\sqrt{a \ps*{x^2 + b}} \exp\bs*{-a\ps*{x^2 + b}} }
\end{equation}
Expanding the \(a\) and \(b\) shorthands and evaluating the definite integrals in detail with \(v^\text{s}_\text{VC} = \sqrt{\frac{2e\ps*{\phi_\text{W} - \phi_\text{VC}}}{m_\text{e}}}\), results in the following.\\
Cartesian: 
\begin{align}\label{eq:CartesianDetailedIntegral}
    \begin{split}
    & n_{\text{e},\,\text{Car}}^\text{s}\ps*{\phi} = N_{0,\,\text{Car}} \ps*{\frac{2 m_\text{e}}{\pi T_\text{eW}}}^{1/2} \int_{v^\text{s}_\text{VC}}^\infty \frac{x \exp\ps*{-a x^2} }{\sqrt{x^2 + b}}\, d x\\
    & = N_{0,\,\text{Car}} \ps*{\frac{2 m_\text{e}}{\pi T_\text{eW}}}^{1/2} \frac{1}{2}\sqrt{\frac{2\pi T_\text{eW}}{m_\text{e}}} \exp\bs*{\frac{m_\text{e}}{2T_\text{eW}} \frac{2e\ps*{\phi - \phi_\text{W}}}{m_\text{e}}}\,\text{erfc}\sqrt{ \frac{m_\text{e}}{2T_\text{eW}} \ps*{\frac{2e\ps*{\phi_\text{W} - \phi_\text{VC}}}{m_\text{e}} + \frac{2e\ps*{\phi - \phi_\text{W}}}{m_\text{e}} } } \\
    & = N_{0,\,\text{Car}} \exp \bs*{\frac{e\ps*{\phi - \phi_\text{W}}}{T_\text{eW}}}\,\text{erfc}\sqrt{ \frac{e \ps*{\phi - \phi_\text{VC}} }{T_\text{eW}}}
    \end{split}
\end{align}
Cylindrical:
\begin{align}\label{eq:CylindricalDetailedIntegral}
    \begin{split}
    n_{\text{e},\,\text{Cyl}}^\text{s}\ps*{\phi} & = N_{0,\,\text{Cyl}} \frac{m_\text{e}}{T_\text{eW}} \frac{1}{2a} \int_{\frac{e\ps*{\phi_\text{W} - \phi_\text{VC}}}{T_\text{eW}}}^{\infty}\frac{\sqrt{u} \exp\ps*{-u} }{\sqrt{u + ab}}\, d u = N_{0,\,\text{Cyl}}  \int_{\frac{e\ps*{\phi_\text{W} - \phi_\text{VC}}}{T_\text{eW}}}^{\infty}\frac{\sqrt{u} \exp\ps*{-u} }{\sqrt{u + ab}} \, d u\\
    & = N_{0,\,\text{Cyl}} I\bs*{u' = \frac{e\ps*{\phi_\text{W} - \phi_\text{VC}}}{T_\text{eW}}, \, ab = \frac{e\ps*{\phi - \phi_\text{W}}}{T_\text{eW}}},\\
    & \text{ where } I\ps*{u', ab} = \int_{u'}^{\infty}\frac{\sqrt{u} \exp\ps*{-u} }{\sqrt{u + ab}}\, d u\;\; \text{[using scipy.integrate.quad()]}
    \end{split}
\end{align}
Spherical:
\begin{align}\label{eq:SphericalDetailedIntegral}
    \begin{split}
    n_{\text{e},\,\text{Sph}}^\text{s}\ps*{\phi}& = N_{0,\,\text{Sph}} \frac{4}{\sqrt{\pi}}\ps*{\frac{m_\text{e}}{2 T_\text{eW}}}^{3/2} \int_{v^\text{s}_\text{VC}}^{\infty} \frac{x^3 \exp\ps*{-a x^2} }{\sqrt{x^2 + b}} \, d x\\
    & = N_{0,\,\text{Sph}} \frac{4}{\sqrt{\pi}}\ps*{\frac{m_\text{e}}{2 T_\text{eW}}}^{3/2} \frac{1}{4} \ps*{\frac{2 T_\text{eW}}{m_\text{e}}}^{3/2} \exp\bs*{ \frac{e \ps*{\phi - \phi_\text{W}} }{T_\text{eW}} } \\
    & \times \Bigg \{ \sqrt{\pi} \bs*{1 - 2 \frac{e \ps*{\phi - \phi_\text{W}} }{T_\text{eW}} }\,\text{erfc}\sqrt{\frac{m_\text{e}}{2 T_\text{eW}} \ps*{x^2 + \frac{2e \ps*{\phi - \phi_\text{W}} }{m_\text{e}} } } \\
    & + 2\sqrt{ \frac{m_\text{e}}{2 T_\text{eW}} \ps*{x^2 + \frac{2e \ps*{\phi - \phi_\text{W}} }{m_\text{e}} } } \exp \bs*{- \frac{m_\text{e}}{2 T_\text{eW}} \ps*{x^2 + \frac{2e \ps*{\phi - \phi_\text{W}} }{m_\text{e}} } } \Bigg \}\\
    & = N_{0,\,\text{Sph}} \frac{1}{\sqrt{\pi}} \exp\bs*{ \frac{e \ps*{\phi - \phi_\text{W}} }{T_\text{eW}} } \\
    & \times \Bigg \{\sqrt{\pi} \bs*{1 - 2 \frac{e \ps*{\phi - \phi_\text{W}} }{T_\text{eW}} }\,\text{erfc}\sqrt{\frac{e\ps*{\phi - \phi_\text{VC}}}{T_\text{eW}} } + 2\sqrt{\frac{e\ps*{\phi - \phi_\text{VC}}}{T_\text{eW}} } \exp \bs*{- \frac{e\ps*{\phi - \phi_\text{VC}}}{T_\text{eW}}} \Bigg \}
    \end{split}
\end{align}
Next, we relate the \(N_0\) and \(n_{\text{e},0}^\text{p}\) terms through the definition of the current ratio \(\gamma\). Prescribing quasi-neutrality at the sheath edge allows us to write the \(n_{\text{e},0}^\text{p}\) term explicitly in terms of \(n_0\). For convenience, we modularize the mathematical expressions as much as possible, beginning with the \(G\) term defined below, as the following derivation becomes increasingly complicated for the cylindrical and spherical coordinate systems. Along with the \(G\) term, \(N_0\) and \(n_{\text{e},0}^\text{p}\) for each coordinate system, can be derived as follows. \\
\noindent Cartesian:
\begin{align}\label{eq:CartesianWallCurrent}
    \begin{split}
    j_{\text{e},\,\text{Car}}^\text{s} & = \int_{v^\text{s}_{\text{VC}}}^\infty e v_0 N_{0,\,\text{Car}} \ps*{\frac{2 m_\text{e}}{\pi T_\text{eW}}}^{1/2} \exp\ps*{-\frac{m_\text{e} v_0^2}{2T_\text{eW}}} \, d v_0,\;v^\text{s}_\text{VC} = \sqrt{\frac{2e\ps*{\phi_\text{W} - \phi_\text{VC}}}{m_\text{e}}}\\
    & = e N_{0,\,\text{Car}} \ps*{\frac{2 m_\text{e}}{\pi T_\text{eW}}}^{1/2}\cdot \frac{2T_\text{eW}}{2m_\text{e}} \exp\bs*{-\frac{m_\text{e}}{2T_\text{eW}}\frac{2e\ps*{\phi_\text{W} - \phi_\text{VC}}}{m_\text{e}} }\\
    & = e N_{0,\,\text{Car}} \sqrt{\frac{2 T_\text{eW}}{\pi m_\text{e}}} \exp\bs*{\frac{e\ps*{\phi_\text{VC} - \phi_\text{W}}}{T_\text{eW}} }\\
    \end{split}
\end{align}
\begin{align}\label{eq:CartesianPlasmaCurrent}
    \begin{split}
    j_{\text{e},\,\text{Car}}^\text{p} & = \int_{v^\text{p}_{\text{VC}}}^\infty e v_p n^\text{p}_{\text{e}0,\,\text{Car}} \ps*{\frac{2 m_\text{e}}{\pi T_\text{e}}}^{1/2} \exp\ps*{-\frac{m_\text{e} v_p^2}{2T_\text{e}}} \, d v_p,\;v^\text{p}_\text{VC} = \sqrt{\frac{-2e \phi_\text{VC}}{m_\text{e}}}\\
    & = e n^\text{p}_{\text{e}0,\,\text{Car}} \ps*{\frac{2 m_\text{e}}{\pi T_\text{e}}}^{1/2}\cdot \frac{2T_\text{e}}{2m_\text{e}} \exp\ps*{-\frac{m_\text{e}}{2T_\text{eW}}\frac{-2e\phi_\text{VC}}{m_\text{e}}}\\
    & = e n^\text{p}_{\text{e}0,\,\text{Car}}\sqrt{\frac{2 T_\text{e}}{\pi m_\text{e}}} \exp\ps*{\frac{e\phi_\text{VC}}{T_\text{e}}}
    \end{split}
\end{align}
Setting \(\gamma_\text{Car} = \abs*{j_{\text{e},\,\text{Car}}^\text{s}/j_{\text{e},\,\text{Car}}^\text{p}}\),
\begin{equation}\label{eq:CartesianN0}
N_{0,\,\text{Car}} = \gamma n^\text{p}_{\text{e}0,\,\text{Car}} \sqrt{\frac{T_\text{e}}{T_\text{eW}}} \exp\bs*{\frac{e\ps*{\phi_\text{W} - \phi_\text{VC}}}{T_\text{eW}}} \exp\ps*{\frac{e\phi_\text{VC}}{T_\text{e}}}
\end{equation}
Taking quasi-neutrality at the plasma-sheath edge, \(n_0 = n^\text{p}_{\text{e}0,\,\text{Car}} + n^\text{s}_{\text{e},\,\text{Car}}\ps*{\phi=0}\). Substituting the above \(N_{0,\,\text{Car}}\) into this yields,
\begin{align}\label{eq:CartesianG}
    \begin{split}
    n_0 & = n^\text{p}_{\text{e}0,\,\text{Car}}\br*{ 1 + \gamma_\text{Car} \sqrt{\frac{T_\text{e}}{T_\text{eW}}} \exp\bs*{\frac{e\ps*{\phi_\text{W} - \phi_\text{VC}}}{T_\text{eW}}} \exp\ps*{\frac{e\phi_\text{VC}}{T_\text{e}}} \exp \bs*{\frac{e\ps*{\phi - \phi_\text{W}}}{T_\text{eW}}}\,\text{erfc}\bs*{\sqrt{ \frac{e \ps*{\phi - \phi_\text{VC}} }{T_\text{eW}}}} }\\
    & = n^\text{p}_{\text{e}0,\,\text{Car}}\bs*{ 1 + \gamma_\text{Car} \sqrt{\frac{T_\text{e}}{T_\text{eW}}}\exp\ps*{\frac{-e \phi_\text{VC}}{T_\text{eW}}} \exp\ps*{\frac{e\phi_\text{VC}}{T_\text{e}}}\,\text{erfc}\ps*{\sqrt{ \frac{- e\phi_\text{VC} }{T_\text{eW}}}}} \\
    & \Rightarrow n^\text{p}_{\text{e}0,\,\text{Car}} = \frac{n_0}{1/\gamma_\text{Car} + G_\text{Car}\ps*{\phi_\text{VC}}},\\
    & \text{where}\; G_\text{Car}\ps*{\phi_\text{VC}} = \sqrt{\frac{T_\text{e}}{T_\text{eW}}}\exp\ps*{\frac{-e \phi_\text{VC}}{T_\text{eW}}} \exp\ps*{\frac{e\phi_\text{VC}}{T_\text{e}}}\,\text{erfc}\ps*{\sqrt{ \frac{- e\phi_\text{VC} }{T_\text{eW}}}}
    \end{split}
\end{align}

\noindent Cylindrical:
\begin{align}\label{eq:CylindricalWallCurrent}
    \begin{split}
    j_\text{e}^\text{s} = & \int_{v^\text{s}_{\text{VC}}}^\infty e v_0 N_0 \frac{m_\text{e}}{T_\text{eW}} v_0 \exp\ps*{-\frac{m_\text{e} v_0^2}{2T_\text{eW}}} \, d v_0 \\
    = &\; e N_0 \frac{m_\text{e}}{T_\text{eW}} \Bigg\{\frac{\sqrt{\pi}}{4} \ps*{\frac{2T_\text{eW}}{m_\text{e}}}^{3/2}\,\text{erfc}\sqrt{\frac{m_\text{e}}{2T_\text{eW}} \sqrt{\frac{2e\ps*{\phi_\text{W} - \phi_\text{VC}}}{m_\text{e}}}}\\
    & + \sqrt{\frac{2e\ps*{\phi_\text{W} - \phi_\text{VC}}}{m_\text{e}}} \frac{2T_\text{eW}}{2m_\text{e}} \exp\bs*{ -\frac{m_\text{e}}{2T_\text{eW}} \frac{2e\ps*{\phi_\text{W} - \phi_\text{VC}}}{m_\text{e}}}\Bigg\}  \\
    = &\; e N_0 \frac{m_\text{e}}{T_\text{eW}} \br*{ \frac{\sqrt{\pi}}{4} \ps*{\frac{2T_\text{eW}}{m_\text{e}}}^{3/2}\,\text{erfc}\sqrt{\frac{e\ps*{\phi_\text{W} - \phi_\text{VC}}}{T_\text{eW}}} + \sqrt{\frac{2e\ps*{\phi_\text{W} - \phi_\text{VC}}}{m_\text{e}}} \frac{T_\text{eW}}{m_\text{e}} \exp\bs*{ \frac{ e\ps*{\phi_\text{VC} - \phi_\text{W}}}{T_\text{eW}} }} 
    \end{split}
\end{align}
\begin{align}\label{eq:CylindricalPlasmaCurrent}
    \begin{split}
    j_\text{e}^\text{p} = & \int_{v^\text{p}_{\text{VC}}}^\infty e v_p n^\text{p}_{\text{e}0} \frac{m_\text{e}}{T_\text{e}} v_p \exp\ps*{-\frac{m_\text{e} v_p^2}{2T_\text{e}}} \, d v_p \\
    = & \; e n^\text{p}_{\text{e}0} \frac{m_\text{e}}{T_\text{e}} \bs*{ \frac{\sqrt{\pi}}{4} \ps*{\frac{2T_\text{e}}{m_\text{e}}}^{3/2}\,\text{erfc}\sqrt{\frac{-e \phi_\text{VC}}{T_\text{e}}} + \sqrt{\frac{-2e\phi_\text{VC}}{m_\text{e}}} \frac{T_\text{e}}{m_\text{e}} \exp\ps*{ \frac{ e\phi_\text{VC}}{T_\text{e}} }} 
    \end{split}
\end{align}
Setting \(\gamma_\text{Cyl} = \abs*{j_\text{e}^\text{s}/j_\text{e}^\text{p}}\),
\begin{equation}
N_{0,\,\text{Cyl}} = \gamma_\text{Cyl} n^\text{p}_{\text{e}0}\frac{T_\text{eW}}{T_\text{e}} \frac{\frac{\sqrt{\pi}}{4} \ps*{\frac{2T_\text{e}}{m_\text{e}}}^{3/2}\,\text{erfc}\sqrt{\frac{-e \phi_\text{VC}}{T_\text{e}}} + \sqrt{\frac{-2e\phi_\text{VC}}{m_\text{e}}} \frac{T_\text{e}}{m_\text{e}} \exp\ps*{ \frac{ e\phi_\text{VC}}{T_\text{e}} }}{\frac{\sqrt{\pi}}{4} \ps*{\frac{2T_\text{eW}}{m_\text{e}}}^{3/2}\,\text{erfc}\sqrt{\frac{e\ps*{\phi_\text{W} - \phi_\text{VC}}}{T_\text{eW}}} + \sqrt{\frac{2e\ps*{\phi_\text{W} - \phi_\text{VC}}}{m_\text{e}}} \frac{T_\text{eW}}{m_\text{e}} \exp\bs*{ \frac{ e\ps*{\phi_\text{VC} - \phi_\text{W}}}{T_\text{eW}} }}  
\end{equation}\label{eq:CylindricalN0}
Taking quasi-neutrality at the plasma-sheath edge, \(n_0 = n^\text{p}_{\text{e}0,\,\text{Cyl}} + n^\text{s}_{\text{e},\,\text{Cyl}}\ps*{\phi=0}\). Substituting the above \(N_{0,\,\text{Cyl}}\) into this yields,
\begin{align}\label{eq:CylindricalG}
    \begin{split}
    n_0 & = n^\text{p}_{\text{e}0} \Bigg\{ 1 + \gamma_\text{Cyl}\frac{T_\text{eW}}{T_\text{e}} \frac{\frac{\sqrt{\pi}}{4} \ps*{\frac{2T_\text{e}}{m_\text{e}}}^{3/2}\,\text{erfc}\sqrt{\frac{-e \phi_\text{VC}}{T_\text{e}}} + \sqrt{\frac{-2e\phi_\text{VC}}{m_\text{e}}} \frac{T_\text{e}}{m_\text{e}} \exp\ps*{ \frac{ e\phi_\text{VC}}{T_\text{e}} }}{\frac{\sqrt{\pi}}{4} \ps*{\frac{2T_\text{eW}}{m_\text{e}}}^{3/2}\,\text{erfc}\sqrt{\frac{e\ps*{\phi_\text{W} - \phi_\text{VC}}}{T_\text{eW}}} + \sqrt{\frac{2e\ps*{\phi_\text{W} - \phi_\text{VC}}}{m_\text{e}}} \frac{T_\text{eW}}{m_\text{e}} \exp\bs*{ \frac{ e\ps*{\phi_\text{VC} - \phi_\text{W}}}{T_\text{eW}} }}\\
    & \cdot I\bs*{u' = \frac{e\ps*{\phi_\text{W} - \phi_\text{VC}}}{T_\text{eW}}, \, ab = \frac{-e \phi_\text{W}}{T_\text{eW}}} \Bigg\}\\
    & \Rightarrow n^\text{p}_{\text{e}0} = \frac{n_0}{1/\gamma_\text{Cyl} + G_\text{Cyl}\ps*{\phi_\text{VC}}},\\
    & \text{where}\; G_\text{Cyl}\ps*{\phi_\text{VC}} = \frac{T_\text{eW}}{T_\text{e}} \frac{\frac{\sqrt{\pi}}{4} \ps*{\frac{2T_\text{e}}{m_\text{e}}}^{3/2}\,\text{erfc}\sqrt{\frac{-e \phi_\text{VC}}{T_\text{e}}} + \sqrt{\frac{-2e\phi_\text{VC}}{m_\text{e}}} \frac{T_\text{e}}{m_\text{e}} \exp\ps*{ \frac{ e\phi_\text{VC}}{T_\text{e}} }}{\frac{\sqrt{\pi}}{4} \ps*{\frac{2T_\text{eW}}{m_\text{e}}}^{3/2}\,\text{erfc}\sqrt{\frac{e\ps*{\phi_\text{W} - \phi_\text{VC}}}{T_\text{eW}}} + \sqrt{\frac{2e\ps*{\phi_\text{W} - \phi_\text{VC}}}{m_\text{e}}} \frac{T_\text{eW}}{m_\text{e}} \exp\bs*{ \frac{ e\ps*{\phi_\text{VC} - \phi_\text{W}}}{T_\text{eW}} }}\\
    & \cdot I\bs*{u' = \frac{e\ps*{\phi_\text{W} - \phi_\text{VC}}}{T_\text{eW}}, \, ab = \frac{-e \phi_\text{W}}{T_\text{eW}}} 
    \end{split}
\end{align}
\noindent Spherical:
\begin{align}\label{eq:SphericalWallCurrent}
    \begin{split}
    j_\text{e}^\text{s} = & \int_{v^\text{s}_{\text{VC}}}^\infty e v_0 N_0 \frac{4}{\sqrt{\pi}}\ps*{\frac{m_\text{e}}{2 T_\text{eW}}}^{3/2} v^2_0 \exp\ps*{-\frac{m_\text{e} v_0^2}{2T_\text{eW}}} \, d v_0 \\
    = &\; e N_0 \frac{4}{\sqrt{\pi}}\ps*{\frac{m_\text{e}}{2 T_\text{eW}}}^{3/2} \frac{1}{2} \ps*{\frac{2 T_\text{eW}}{m_\text{e}}}^2 \exp\bs*{- \frac{m_\text{e}}{2T_\text{eW}} \frac{2e\ps*{\phi_\text{W} - \phi_\text{VC}}}{m_\text{e}} } \bs*{\frac{m_\text{e}}{2T_\text{eW}} \frac{2e\ps*{\phi_\text{W} - \phi_\text{VC}}}{m_\text{e}} + 1} \\
    = &\; e N_0 \frac{2\sqrt{2}}{\sqrt{\pi}} \sqrt{\frac{T_\text{eW}}{m_\text{e}}} \exp\bs*{-\frac{e\ps*{\phi_\text{W} - \phi_\text{VC}}}{T_\text{eW}} } \bs*{\frac{e\ps*{\phi_\text{W} - \phi_\text{VC}}}{T_\text{eW}} + 1}
    \end{split}
\end{align}
\begin{align}\label{eq:SphericalPlasmaCurrent}
    \begin{split}
    j_\text{e}^\text{p} = & \int_{v^\text{p}_{\text{VC}}}^\infty e v_p n^\text{p}_{\text{e}0} \frac{4}{\sqrt{\pi}}\ps*{\frac{m_\text{e}}{2 T_\text{e}}}^{3/2} v^2_p \exp\ps*{-\frac{m_\text{e} v_p^2}{2T_\text{e}}} \, d v_p \\
    = &\; e n^\text{p}_{\text{e}0} \frac{4}{\sqrt{\pi}}\ps*{\frac{m_\text{e}}{2 T_\text{e}}}^{3/2} \frac{1}{2} \ps*{\frac{2 T_\text{e}}{m_\text{e}}}^2 \exp\bs*{- \frac{m_\text{e}}{2T_\text{e}} \frac{-2e\phi_\text{VC}}{m_\text{e}} } \bs*{\frac{m_\text{e}}{2T_\text{e}} \frac{-2e\phi_\text{VC}}{m_\text{e}} + 1} \\
    = &\; e n^\text{p}_{\text{e}0} \frac{2\sqrt{2}}{\sqrt{\pi}} \sqrt{\frac{T_\text{e}}{m_\text{e}}} \exp\ps*{\frac{\phi_\text{VC}}{T_\text{e}} } \ps*{1 - \frac{e\phi_\text{VC}}{T_\text{e}}}
    \end{split}
\end{align}
Setting \(\gamma_\text{Sph} = \abs*{j_\text{e}^\text{s}/j_\text{e}^\text{p}}\),
\begin{equation}\label{eq:SphericalN0}
N_{0,\,\text{Sph}} = \gamma_\text{Sph} n^\text{p}_{\text{e}0}\sqrt{\frac{T_\text{e}}{T_\text{eW}}} \frac{\exp\ps*{\frac{\phi_\text{VC}}{T_\text{e}} } \ps*{1 - \frac{e\phi_\text{VC}}{T_\text{e}}}}{\exp\bs*{-\frac{e\ps*{\phi_\text{W} - \phi_\text{VC}}}{T_\text{eW}} } \bs*{\frac{e\ps*{\phi_\text{W} - \phi_\text{VC}}}{T_\text{eW}} + 1}}
\end{equation}
Taking quasi-neutrality at the plasma-sheath edge, \(n_0 = n^\text{p}_{\text{e}0,\,\text{Sph}} + n^\text{s}_{\text{e},\,\text{Sph}}\ps*{\phi=0}\).\\ Substituting the above \(N_{0,\,\text{Sph}}\) into this yields,
\begin{align}\label{eq:SphericalG}
    \begin{split}
    n_0 & = n^\text{p}_{\text{e}0,\,\text{Sph}} \Bigg\{ 1 + \gamma_\text{Sph} \sqrt{\frac{T_\text{e}}{T_\text{eW}}} \frac{\exp\ps*{\frac{\phi_\text{VC}}{T_\text{e}} } \ps*{1 - \frac{e\phi_\text{VC}}{T_\text{e}}}}{\exp\bs*{-\frac{e\ps*{\phi_\text{W} - \phi_\text{VC}}}{T_\text{eW}} } \bs*{\frac{e\ps*{\phi_\text{W} - \phi_\text{VC}}}{T_\text{eW}} + 1}} \cdot \frac{1}{\sqrt{\pi}} \exp\ps*{ \frac{-e  \phi_\text{W} }{T_\text{eW}} } \\
    & \cdot \bs*{\sqrt{\pi} \ps*{1 + \frac{2 e \phi_\text{W} }{T_\text{eW}} }\,\text{erfc}\sqrt{\frac{-e \phi_\text{VC}}{T_\text{eW}} } + 2\sqrt{\frac{- e \phi_\text{VC}}{T_\text{eW}} } \exp \ps*{\frac{e \phi_\text{VC}}{T_\text{eW}}} } \Bigg \} \\
    & \Rightarrow n^\text{p}_{\text{e}0,\,\text{Sph}} = \frac{n_0}{1/\gamma_\text{Sph} + G_\text{Sph}\ps*{\phi_\text{VC}}},\\
    & \text{where}\; G_\text{Sph}\ps*{\phi_\text{VC}} = \sqrt{\frac{T_\text{e}}{T_\text{eW}}} \frac{\exp\ps*{\frac{\phi_\text{VC}}{T_\text{e}} } \ps*{1 - \frac{e\phi_\text{VC}}{T_\text{e}}}}{\exp\bs*{-\frac{e\ps*{\phi_\text{W} - \phi_\text{VC}}}{T_\text{eW}} } \bs*{\frac{e\ps*{\phi_\text{W} - \phi_\text{VC}}}{T_\text{eW}} + 1}} \cdot \frac{1}{\sqrt{\pi}} \exp\ps*{ \frac{-e  \phi_\text{W} }{T_\text{eW}} } \\
    & \cdot \bs*{\sqrt{\pi} \ps*{1 + \frac{2 e \phi_\text{W} }{T_\text{eW}} }\,\text{erfc}\sqrt{\frac{-e \phi_\text{VC}}{T_\text{eW}} } + 2\sqrt{\frac{- e \phi_\text{VC}}{T_\text{eW}} } \exp \ps*{\frac{e \phi_\text{VC}}{T_\text{eW}}} }
    \end{split}
\end{align}
Upon deriving the \(N_0\) and \(n_{\text{e}0}^\text{p}\) terms for each coordinate system, all of the density \(n_{\text{e},\,\text{sys}}\) and current density \(j_\text{e,\,\text{sys}}\) functions can be expressed compactly, strategically composing the \(G\) and \(G_\phi\) terms to incorporate them in multiple expressions and avoid redundancy. Note that the potential \(\phi\) and position \(r\) variables normalize to \(\Phi = \frac{e \phi}{T_\text{e}},\; \xi = \frac{r}{\lambda_D}\), where \(\lambda_D = \sqrt{ \frac{\varepsilon_0 T_\text{e}}{e^2 n_0} }\) and \(C = \frac{T_\text{e}}{T_\text{eW}}\).\\
\noindent Cartesian:
\begin{align}\label{eq:CartesianModularization}
    \begin{split}
    G_{\phi, \text{Car}} & = \sqrt{C}\exp\bs*{\ps*{1 - C} \Phi_\text{VC}} \\
    G_\text{Car} & = G_{\phi, \text{Car}} \,\text{erfc}\sqrt{-C \Phi_\text{VC}}\\
    \gamma_\text{Car} & = \abs*{j_{\text{e},\,\text{Car}}^\text{s}/j_{\text{e},\,\text{Car}}^\text{p}} \\
    n^\text{p}_{\text{e}0,\,\text{Car}} & = \ps*{1/\gamma_\text{Car} + G_\text{Car} }^{-1} \\
    N_{0,\,\text{Car}} & = n^\text{p}_{\text{e}0,\,\text{Car}} G_{\phi, \text{Car}} \\
    \end{split}
    \begin{split}
    n_{\text{e},\,\text{Car}}^\text{s}\ps*{\Phi} & = N_{0,\,\text{Car}} \exp \ps*{C\Phi}\,\text{erfc}\sqrt{C\ps*{\Phi - \Phi_\text{VC}}  } \\ 
    n_{\text{e},\,\text{Car}}^\text{p}\ps*{\Phi} & = \frac{1}{\gamma} n^\text{p}_{\text{e}0,\,\text{Car}} \exp \ps*{\Phi}\\
    j_{\text{e},\,\text{Car, over } n_0}^\text{s} & = e N_{0,\,\text{Car}} \sqrt{\frac{2T_\text{eW}}{\pi m_\text{e}}} \exp\ps*{C \Phi_\text{VC}} \\
    j_{\text{e},\,\text{Car, over } n_0}^\text{p} & = \frac{1}{\gamma_\text{Car}}e n^\text{p}_{\text{e}0,\,\text{Car}}\sqrt{\frac{2T_\text{e}}{\pi m_\text{e}}} \exp\ps*{\Phi_\text{VC}}
    \end{split}
\end{align}
Cylindrical:
\begin{align}\label{eq:CylindricalModularization}
    \begin{split}
    G_{\phi, \text{Cyl}} & = \frac{T_\text{e}^{3/2}}{C} \frac{ \sqrt{\frac{\pi}{2}} \,\text{erfc}\sqrt{- \Phi_\text{VC}} + \sqrt{-2\Phi_\text{VC}}  \exp\ps*{\Phi_\text{VC} }}{  \sqrt{\frac{\pi}{2}}T_\text{eW}^{3/2}\,\text{erfc}\sqrt{C\ps*{\Phi_\text{W} - \Phi_\text{VC}}} + \sqrt{T_\text{e}} T_\text{eW} \sqrt{2\ps*{\Phi_\text{W} - \Phi_\text{VC}}} \ \exp\bs*{C \ps*{\Phi_\text{VC} - \Phi_\text{W}} }} \\
    G_\text{Cyl} & = G_{\phi, \text{Cyl}} I\bs*{u' = C\ps*{\Phi_\text{W} - \Phi_\text{VC}}, \, ab = -C\Phi_\text{W} } \\
    \gamma_\text{Cyl} & = \abs*{j_{\text{e},\,\text{Cyl}}^\text{s}/j_{\text{e},\,\text{Cyl}}^\text{p}} \\
    n^\text{p}_{\text{e}0,\,\text{Cyl}} & = \ps*{1/\gamma_\text{Cyl} + G_\text{Cyl} }^{-1} \\
    N_{0,\,\text{Cyl}} & = n^\text{p}_{\text{e}0,\,\text{Cyl}} G_{\phi, \text{Cyl}} \\
    n_{\text{e},\,\text{Cyl}}^\text{s}\ps*{\Phi} & = N_{0,\,\text{Cyl}} I\bs*{u' = C\ps*{\Phi_\text{W} - \Phi_\text{VC}}, \, ab = C\ps*{\Phi - \Phi_\text{W}} }\\
    n_{\text{e},\,\text{Cyl}}^\text{p}\ps*{\Phi} & = \frac{1}{\gamma_\text{Cyl}} n^\text{p}_{\text{e}0,\,\text{Cyl}} \exp \ps*{\Phi_\text{VC} }\\
    j_{\text{e},\,\text{Cyl, over } n_0}^\text{s} & = e N_{0,\,\text{Cyl}}\frac{m_e}{T_\text{eW}}\\ 
    & \cdot \br*{\sqrt{\frac{\pi}{2}} \ps*{\frac{T_\text{eW}}{m_\text{e}}}^{3/2}   \,\text{erfc}\sqrt{ C\ps*{\Phi_\text{W} - \Phi_\text{VC}} } + \ps*{\frac{\sqrt{T_e} T_\text{eW}}{m_\text{e}^{3/2}}} \sqrt{2\ps*{\Phi_\text{W} - \Phi_\text{VC}}}  \exp\bs*{C\ps*{\Phi_\text{VC} - \Phi_\text{W}} } } \\
    j_{\text{e},\,\text{Cyl, over } n_0}^\text{p} & = \frac{1}{\gamma_\text{Cyl}}e n^\text{p}_{\text{e}0,\,\text{Cyl}} \sqrt{\frac{T_\text{e}}{m_\text{e}}} \bs*{ \sqrt{\frac{\pi}{2}} \,\text{erfc}\sqrt{-\Phi_\text{VC}} + \sqrt{-2\Phi_\text{VC}}  \exp\ps*{\Phi_\text{VC} }}
    \end{split}
\end{align}
Spherical:
\begin{align}\label{eq:SphericalModularization}
    \begin{split}
    G_{\phi, \text{Sph}} & =\sqrt{C} \frac{\exp\ps*{\Phi_\text{VC}} \ps*{1-\Phi_\text{VC}}}{\exp\bs*{-C \ps*{\Phi_W -\Phi_\text{VC}} } \bs*{C \ps*{\Phi_W -\Phi_\text{VC}} + 1}} \\
    G_\text{Sph} & = G_{\phi, \text{Sph}} \frac{1}{\sqrt{\pi}} \exp\ps*{-C\Phi_\text{W}} \bs*{\sqrt{\pi} \ps*{1 + 2 C\Phi_\text{W}}\,\text{erfc}\sqrt{ -C \Phi_\text{VC} } + 2\sqrt{-C \Phi_\text{VC} } \exp \ps*{C\Phi_\text{VC}} } \\
    \gamma_\text{Sph} & = \abs*{j_{\text{e},\,\text{Sph}}^\text{s}/j_{\text{e},\,\text{Sph}}^\text{p}} \\
    n^\text{p}_{\text{e}0,\,\text{Sph}} & = \ps*{1/\gamma_\text{Sph} + G_\text{Sph} }^{-1} \\
    N_{0,\,\text{Sph}} & = n^\text{p}_{\text{e}0,\,\text{Sph}}  G_{\phi, \text{Sph}} \\
    n_{\text{e},\,\text{Sph}}^\text{s}\ps*{\Phi} & = N_{0,\,\text{Sph}} \frac{1}{\sqrt{\pi}} \exp\bs*{ C\ps*{\Phi - \Phi_\text{W}}}\\
    & \cdot \Bigg \{\sqrt{\pi} \bs*{1 - 2C \ps*{\Phi - \Phi_\text{W}} }\,\text{erfc}\sqrt{C\ps*{\Phi - \Phi_\text{VC}}} + 2\sqrt{ C\ps*{\Phi - \Phi_\text{VC}}}  \exp \bs*{- C\ps*{\Phi - \Phi_\text{VC}} } \Bigg\} \\
    n_{\text{e},\,\text{Sph}}^\text{p}\ps*{\Phi} & = \frac{1}{\gamma_\text{Sph}} n^\text{p}_{\text{e}0,\,\text{Sph}} \exp \ps*{\Phi}\\
    j_{\text{e},\,\text{Sph, over } n_0}^\text{s} & = e N_{0,\,\text{Sph}} \frac{2\sqrt{2}}{\sqrt{\pi}} \sqrt{\frac{T_\text{eW}}{m_\text{e}}} \exp\bs*{-C\ps*{\Phi_\text{W} - \Phi_\text{VC}}} \bs*{C\ps*{\Phi_\text{W} - \Phi_\text{VC}} + 1} \\
    j_{\text{e},\,\text{Sph, over } n_0}^\text{p} & = \frac{1}{\gamma_\text{Sph}}e n^\text{p}_{\text{e}0,\,\text{Sph}} \frac{2\sqrt{2}}{\sqrt{\pi}} \sqrt{\frac{T_\text{e}}{m_\text{e}}} \exp\ps*{\Phi_\text{VC}} \ps*{1 - \Phi_\text{VC}}
    \end{split}
\end{align}
As in Takamura’s approach \cite{Takamura2004}, the product of Richardson's current density and the exponential current dropout because of the virtual cathode is equal to the wall current density \(j_{\text{e},\,\text{sys}}^\text{s}\), which depends on the coordinate system. Also, note that the wall and virtual cathode potentials assume negative values (hence, the lack of negative sign in the exponential term of Richardson's current density). The current equivalence below can constrain the quasi-neutral plasma density \(n_0\) to be a function of all other inputs, notably the current ratio \(\gamma\) and again the coordinate system in question. Denoting \(A_R\) as Richardson's constant and \(\phi_\text{work}\) as the material work function: 
\begin{align}\label{eq:CurrentEquivalence}
    \begin{split}
    & A_R \ps*{\frac{T_\text{eW}}{k_B}}^2 \exp\bs*{\frac{e \ps*{\phi_\text{work} + \phi_\text{VC} - \phi_\text{W}}}{T_\text{eW}}} = n_{0,\,\text{sys}} j_{\text{e},\,\text{sys, over } n_0}^\text{s} \\
    & \Rightarrow n_{0,\,\text{sys}}  = A_R \ps*{\frac{T_\text{eW}}{k_B}}^2 \exp\bs*{\frac{e \ps*{\phi_\text{work} + \phi_\text{VC} - \phi_\text{W}}}{T_\text{eW}}}/j_{\text{e},\,\text{sys, over } n_0}^\text{s}\,.
    \end{split}
\end{align}
However, in the cylindrical and spherical coordinate systems, the leading edge mathematically starts at some radial offset \(r_\text{LE}\) (which gives rise to normalized offset \(\xi_0\)), so instead of imposing Richardson's current the density \(n_0\) can be alternatively constrained by the leading edge radius \(r_\text{LE}\) through
\begin{equation}\label{eq:LERadiusConstraintOnDensity}
\xi_0 = r_\text{LE}\sqrt{\frac{e^2 n_0}{\varepsilon_0 T_e}}
\end{equation}
which follows from Eq. \eqref{eq:Normalization}. The application of this constraint with leading radius \(r_\text{LE}\) to compute the density \(n_0\) would in turn fix the wall current density \(j_\text{e}^\text{s}\). With the source terms developed in Eqs. \eqref{eq:CartesianModularization}-\eqref{eq:SphericalModularization} for the differential equation in each coordinate system, we can investigate the conditions that yield a valid sheath potential.
\subsection{New sheath formation criteria}
\subsubsection{Cartesian coordinates}
Let us consider the differential equation that describes the sheath potential space in Cartesian coordinates, with the specific source terms developed in the previous section.
\begin{align}\label{eq:CartesianDiffEqWithSources}
    \begin{split}
    \frac{d^2 \Phi}{d \xi^2} & = n_{\text{e},\, \text{Car}}^\text{s}\bs*{\Phi\ps*{\xi}} + n_{\text{e},\,\text{Car}}^\text{p}\bs*{\Phi\ps*{\xi}} - n_\text{i}\bs*{\Phi\ps*{\xi}}\\
    \frac{d^2 \Phi}{d \xi^2} & = N_{0,\,\text{Car}} \exp \ps*{C\Phi}\,\text{erfc}\sqrt{C\ps*{\Phi - \Phi_\text{VC}} } + \frac{n^\text{p}_{\text{e}0,\,\text{Car}}}{\gamma} \exp \ps*{\Phi}  - \ps*{1 - \frac{2\Phi}{M^2}}^{-1/2}
    \end{split}
\end{align}
Similarly to Bohm's original approach \cite{Bohm1949} which was also taken by Takamura  \cite{Takamura2004}, this ODE can be integrated once by multiplying \(d \Phi/d \xi\) to both sides. We integrate the left side first as follows.
\begin{equation}\label{eq:CartesianLHSIntegration}
    \int \frac{d \Phi}{d \xi}\frac{d^2 \Phi}{d \xi^2} \, d \xi = \int \frac{d^2 \Phi}{d \xi^2}\, d \Phi = \int d \ps*{\frac{d \Phi}{d \xi}} \frac{d \Phi}{d \xi} = \int \Phi'\, d\ps*{\Phi'} = \frac{1}{2} \ps*{\Phi'}^2
\end{equation}
Now considering the right side to be a function of the normalized potential \(f_\text{Car}\ps*{\Phi}\) as a shorthand, the indefinite integral of the right side times \(d \Phi/d \xi\) becomes: 
\begin{align}\label{eq:CartesianRHSIntegration}
    \begin{split}
    \int \frac{d \Phi}{d \xi} f_\text{Car}\ps*{\Phi}\,d \xi & = \int f_\text{Car}\ps*{\Phi}\,d \Phi  \\
    & = \frac{N_{0,\,\text{Car}}}{C} \bs*{\exp\ps*{C\Phi} \,\text{erfc}\sqrt{C\ps*{\Phi - \Phi_\text{VC}}} + \frac{2}{\sqrt{\pi}}\exp\ps*{C\Phi_\text{VC}}\sqrt{C\ps*{\Phi - \Phi_\text{VC} } }}\\
    & + \frac{n^\text{p}_{\text{e}0,\,\text{Car}}}{\gamma} \exp\ps*{\Phi} + M^2 \sqrt{1 - \frac{2\Phi}{M^2}}\,.
    \end{split}
\end{align}
Asserting that \(d \Phi/d \xi \rightarrow 0\) as \(\Phi \rightarrow 0\), i.e., that the plasma sheath yields zero electric field as position \(\xi \rightarrow \infty\), we obtain \(\frac{1}{2} \ps*{\Phi'}^2\) with the plasma sheath edge boundary condition:
\begin{align}\label{eq:CartesianIntegratedWithBC}
    \begin{split}
    \frac{1}{2} \ps*{\Phi'}^2 & = \frac{N_{0,\,\text{Car}}}{C} \bigg\{ \exp\ps*{C\Phi} \,\text{erfc}\sqrt{C\ps*{\Phi - \Phi_\text{VC}}} - \text{erfc}\sqrt{-C\Phi_\text{VC}}\\
    & + \frac{2}{\sqrt{\pi}}\exp\ps*{C\Phi_\text{VC}}\bs*{\sqrt{C\ps*{\Phi - \Phi_\text{VC} } } - \sqrt{-C\Phi_\text{VC}}} \bigg\} + \frac{n^\text{p}_{\text{e}0,\,\text{Car}}}{\gamma} \bs*{\exp\ps*{\Phi} - 1} + M^2 \ps*{\sqrt{1 - \frac{2\Phi}{M^2}} - 1}\,.
    \end{split}
\end{align}

\begin{figure}[H]
    \centering
    \includegraphics[width=0.9\linewidth]{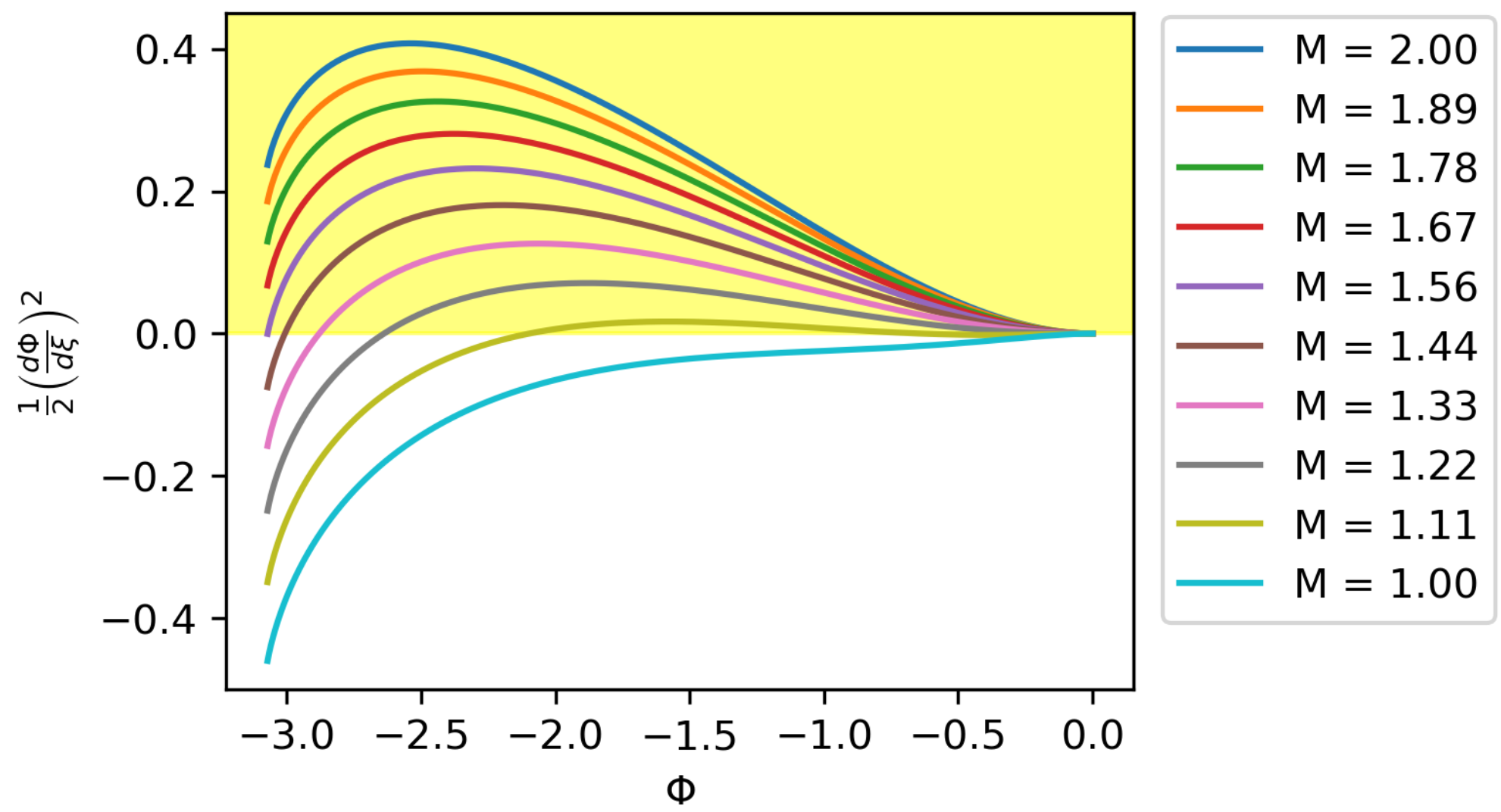}
    \caption{Right side of Eq. \eqref{eq:CartesianIntegratedWithBC}, with \(\Phi_\text{W} = -2.57\), \(\Phi_\text{VC} = \Phi_\text{W} - 0.5\), \(T_\text{e} = \SI{1.75}{eV}\), \(T_\text{eW} = \SI{0.175}{eV}\); only sheathes that satisfy the highlighted region where \(\ps*{d\Phi/d\xi}^2/2 > 0\) are valid.}
    \label{fig:CarRHS}
\end{figure}
\noindent Given that the left side is squared, the right side must be greater than or equal to zero for all \(\Phi\) to achieve valid sheath formation. By graphing the right side for varying ion Mach number \(M\), we observe that \(M\) must be above a certain minimum to maintain a non-negative right side for all \(\Phi\). We find that this minimum \(M\) is achieved when both \(\frac{1}{2} \ps*{\Phi_\text{VC}'}^2 \geq 0\) and \(\frac{1}{2} \ps*{\Phi'}^2 > 0\) as \(\Phi \rightarrow 0\) are satisfied. The first condition is as follows.\\
\underline{New condition 1 (Cartesian)}
\begin{align}\label{eq:CartesianBohm1}
    \begin{split}
    & \frac{1}{2} \ps*{\Phi_\text{VC}'}^2 \geq 0 \Rightarrow\\
    & 0 \leq \frac{N_{0,\,\text{Car}}}{C} \br*{ \exp\ps*{C\Phi_\text{VC}} - \text{erfc}\sqrt{-C\Phi_\text{VC}} - \frac{2}{\sqrt{\pi}}\exp\ps*{C\Phi_\text{VC}} \sqrt{-C\Phi_\text{VC}} }\\ 
    & + \frac{n^\text{p}_{\text{e}0,\,\text{Car}}}{\gamma} \bs*{\exp\ps*{\Phi_\text{VC}} - 1} + M^2 \ps*{\sqrt{1 - \frac{2\Phi_\text{VC}}{M^2}} - 1}
    \end{split}
\end{align}
For the second condition, by numerical inspection we observe that it is equivalent to requiring \(\bs*{\frac{1}{2} \ps*{\Phi'}^2}' = \Phi' \Phi'' < 0\) as \(\Phi \rightarrow 0\). To fulfill this, note that because \(\Phi' > 0\) as \(\Phi \rightarrow 0\), \(\Phi''\) must be negative, i.e., the potential curve must be concave down as \(\Phi \rightarrow 0\). As \(\Phi'' = 0\) at \(\Phi = 0\), because the quasi-neutral plasma has zero net charge density for zero potential, \(\Phi'' < 0\) as \(\Phi \rightarrow 0\) implies that the third derivative \(\Phi''' > 0\) at \(\Phi = 0\). We can impose this by Taylor expanding \(\Phi''\) to first-order, and requiring that the first-order term is greater than zero. Remember, that the constant term of the Taylor expansion of \(\Phi''\) vanishes upon differentiation.\\
\underline{New condition 2 (Cartesian)}
\begin{align}\label{eq:CartesianBohm2}
    \begin{split}
    & \frac{1}{2} \bs*{\Phi'\ps*{\Phi \rightarrow 0}}^2 > 0 \Rightarrow \\
    & \Phi''\ps*{\Phi \rightarrow 0} \approx N_{0,\,\text{Car}} \text{erfc}\sqrt{-C\Phi_\text{VC}} + \frac{n^\text{p}_{\text{e}0,\,\text{Car}}}{\gamma}  - 1\\
    & + \Phi\bs*{C N_{0,\,\text{Car}}\text{erfc}\sqrt{-C\Phi_\text{VC}} + \frac{N_{0,\,\text{Car}} \exp \ps*{C\Phi_\text{VC}}\sqrt{-C\Phi_\text{VC}}}{\sqrt{\pi} \Phi_\text{VC}} + \frac{n^\text{p}_{\text{e}0,\,\text{Car}}}{\gamma} - \frac{1}{M^2}}\\
    & \Rightarrow M > \bs*{C N_{0,\,\text{Car}}\text{erfc}\sqrt{-C\Phi_\text{VC}} + \frac{N_{0,\,\text{Car}} \exp \ps*{C\Phi_\text{VC}}\sqrt{-C\Phi_\text{VC}}}{\sqrt{\pi} \Phi_\text{VC}} + \frac{n^\text{p}_{\text{e}0,\,\text{Car}}}{\gamma}}^{-1/2}
    \end{split}
\end{align}
Now that we have an expression for the minimum Mach number, let us consider what conditions yield a virtual cathode and a lack thereof. A potential sheath without VC requires \(\Phi_\text{VC} = \Phi_\text{W}\) and a non-negative derivative at the wall, i.e., \(\Phi_\text{W}' \geq 0\).\\
\underline{Condition for NO virtual cathode formation (Cartesian case)}
\begin{equation}\label{eq:CartesianNoVC}
    \Phi_\text{VC} = \Phi_\text{W} \Rightarrow \Phi_\text{W}' > 0
\end{equation}
Next, consider how to construct a virtual cathode in the potential sheath. If a VC is present, there is some VC potential \(\Phi_\text{VC} < \Phi_\text{W}\), which can only occur if there is a negative derivative at the wall \(\ps*{\Phi_\text{W}' < 0}\) and if the virtual cathode is in fact a potential minimum \(\ps*{\Phi_\text{VC}' = 0}\).\\
\underline{Condition for virtual cathode formation (Cartesian case)}
\begin{align}\label{eq:CartesianVC}
    \begin{split}
    & \Phi_\text{VC} < \Phi_\text{W} \Rightarrow \Phi_\text{W}' < 0, \text{ and}\\
    & \Phi_\text{VC}' = 0 \Rightarrow \\
    & 0 = \frac{N_{0,\,\text{Car}}}{C} \br*{\exp\ps*{C\Phi_\text{VC}} - \text{erfc}\sqrt{-C\Phi_\text{VC}} - \frac{2}{\sqrt{\pi}}\exp\ps*{C\Phi_\text{VC}}  \sqrt{-C\Phi_\text{VC}} }\\
    & + \frac{n^\text{p}_{\text{e}0,\,\text{Car}}}{\gamma} \bs*{\exp\ps*{\Phi_\text{VC}} - 1} + M^2 \ps*{\sqrt{1 - \frac{2\Phi_\text{VC}}{M^2}} - 1}
    \end{split}
\end{align}
As a benchmark of our Cartesian methodology and Runge-Kutta based potential sheath solution technique described in detail in subsection \ref{Cylindrical}, we deduce input parameters that generate potential sheathes similar to those reported by Takamura while demonstrating the generalized application of Runge-Kutta in subsection \ref{Cylindrical} to create potential sheathes for any of the three coordinate systems under consideration. We can find similar potential space phenomena to those shown by Takamura, while only confidently assuming that Takamaura set \(\gamma = 0\) for the case without electron emission, \(\gamma > 0\) for the case with electron emission, and normalized wall potential \(\Phi_\text{W} = -0.99\).
\begin{figure}[H]
\centering
\begin{subfigure}{.52\textwidth}
    \centering
    \includegraphics[width=1\linewidth]{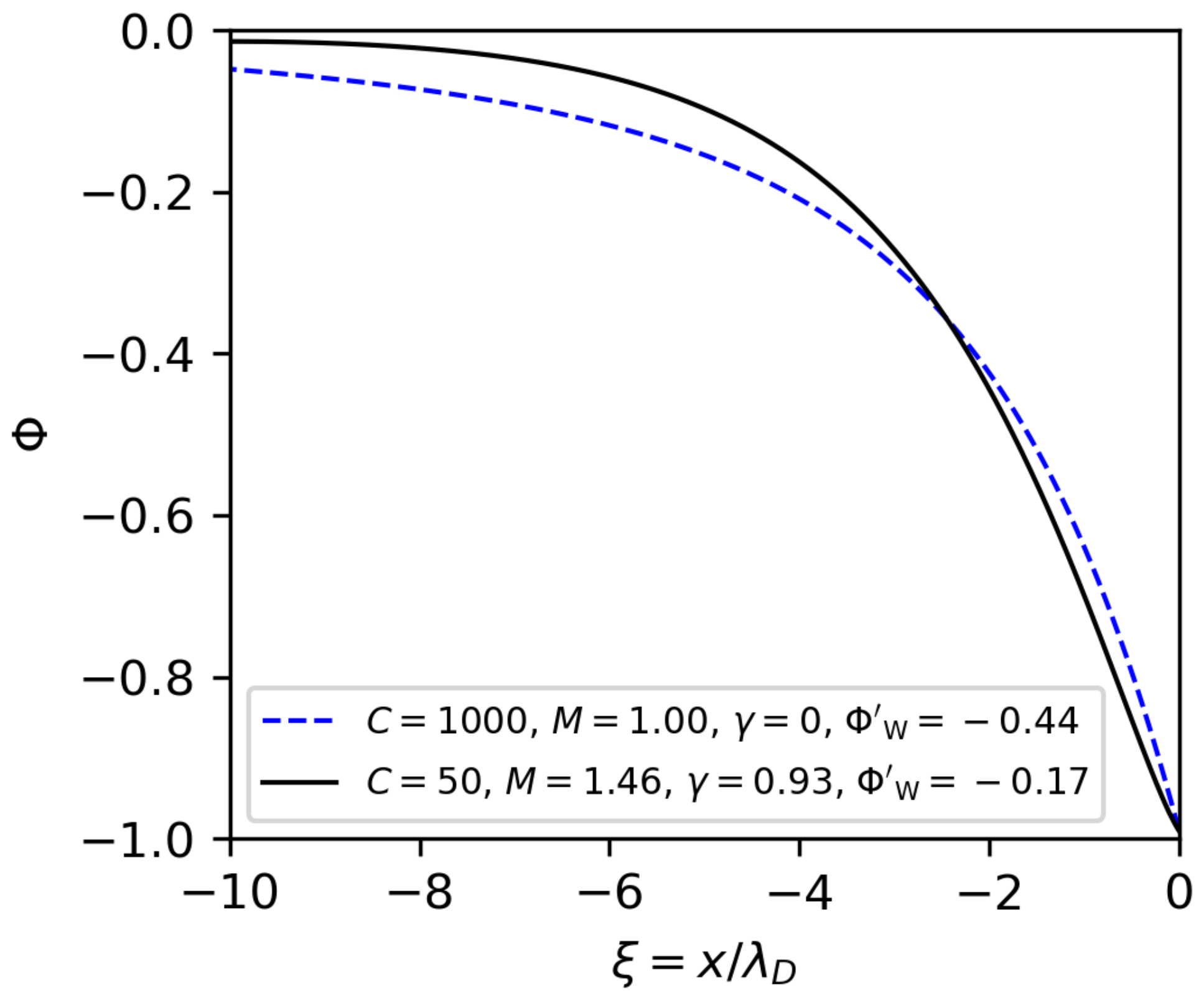}
    \caption{present work\vspace{13px}}
    \label{fig:PresentSheath}
\end{subfigure}
\hfill
\begin{subfigure}{.47\textwidth}
    \centering
    \includegraphics[width=1\linewidth]{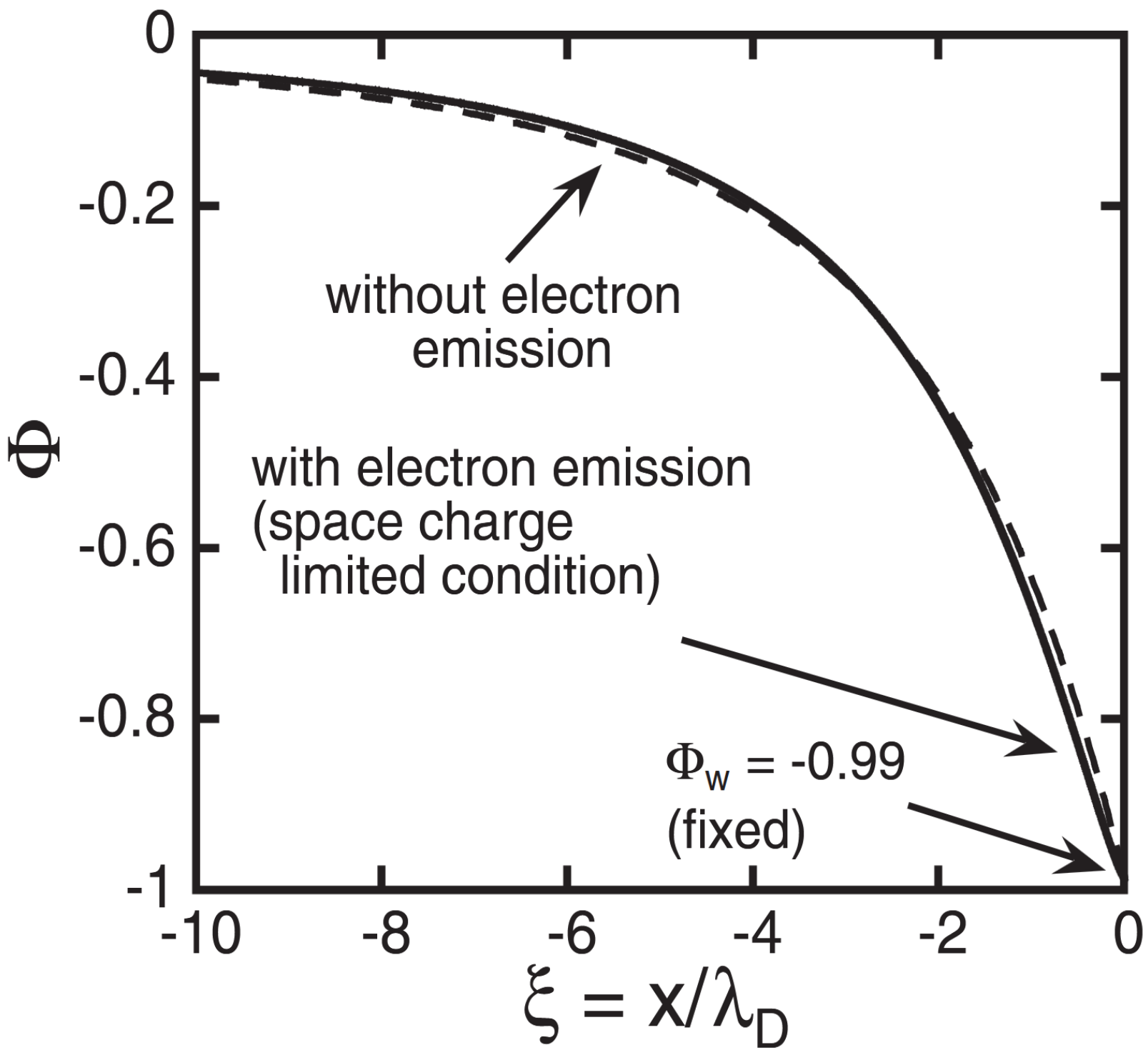}
    \caption{Takamura's work \cite{Takamura2004}, reproduced with permission}
    \label{fig:TakamuraSheath}
\end{subfigure}
\caption{Example potential sheathes similar qualitatively to those shown by Takamura, without virtual cathode and for Cartesian coordinates}
\label{fig:SheathReproduction}
\end{figure}

\begin{figure}[H]
\centering
\begin{subfigure}{.51\textwidth}
    \centering
    \includegraphics[width=1\linewidth]{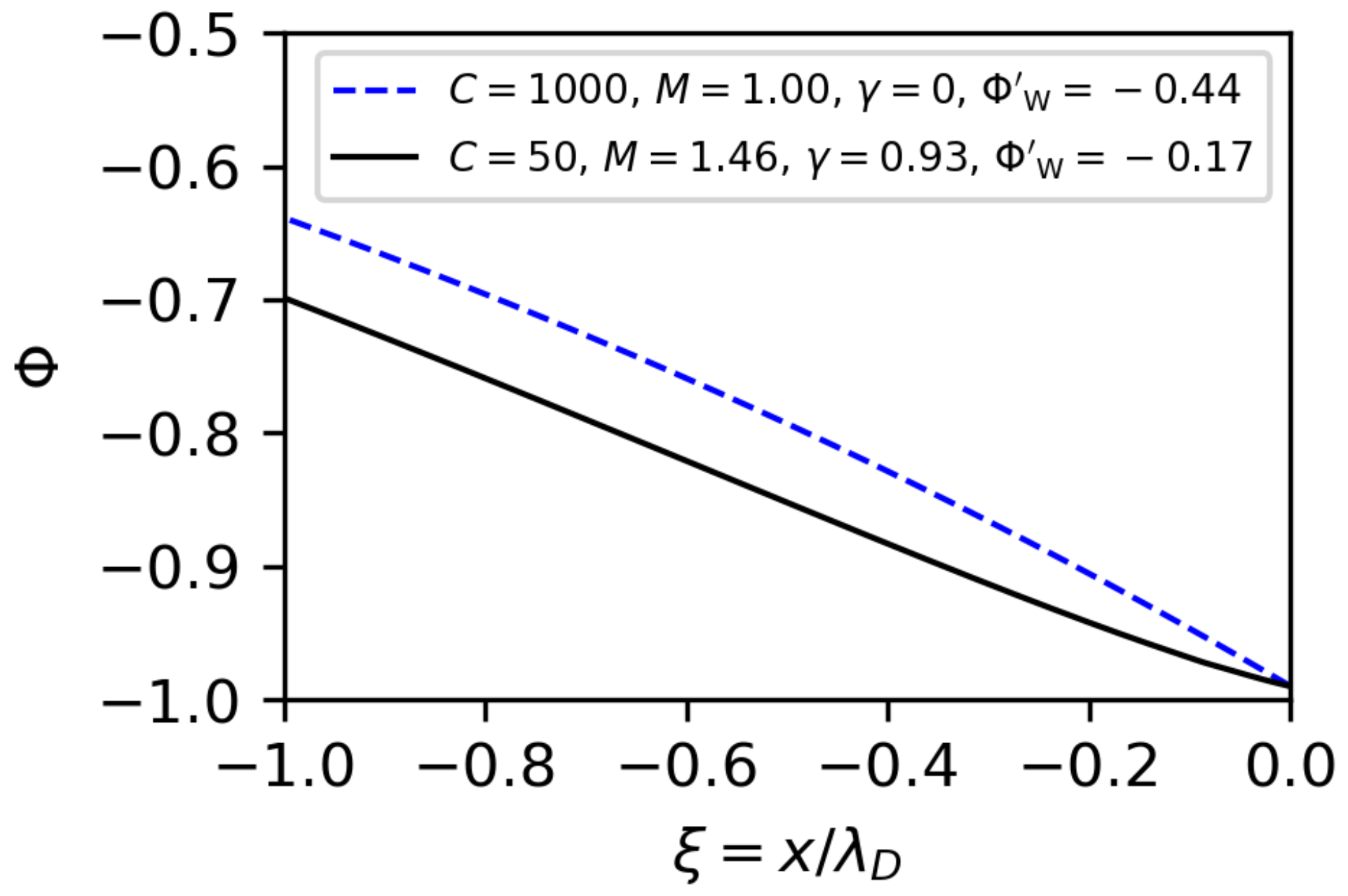}
    \caption{present work}
    \label{fig:PresentSheathZoom}
\end{subfigure}
\hfill
\begin{subfigure}{.48\textwidth}
    \centering
    \includegraphics[width=1\linewidth]{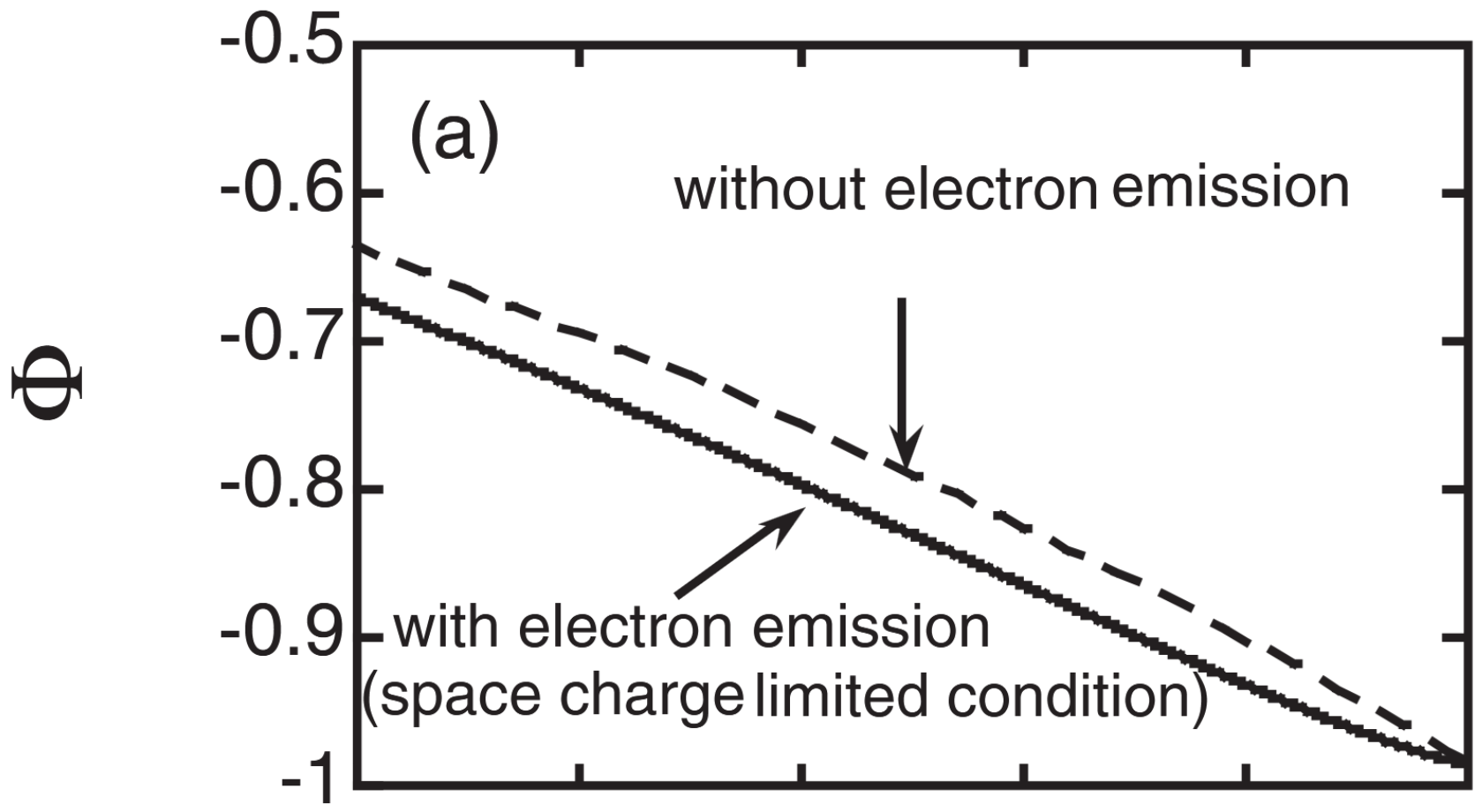}
    \caption{Takamura's work \cite{Takamura2004}, reproduced with permission\vspace{16px}}
    \label{fig:TakamuraSheathZoom}
\end{subfigure}
\caption{Sheath potentials focused on the region within one Debye length of the wall in Fig. \ref{fig:SheathReproduction}}
\label{fig:ZoomedSheathReproduction}
\end{figure}

\subsubsection{Cylindrical coordinates} \label{Cylindrical}
\begin{align}\label{eq:CylindricalDiffEqWithSources}
    \begin{split}
    \frac{d^2 \Phi}{d \xi^2} + \frac{1}{\xi}\frac{d \Phi}{d \xi}  & = n_{\text{e},\, \text{Cyl}}^\text{s}\bs*{\Phi\ps*{\xi}} + n_{\text{e},\,\text{Cyl}}^\text{p}\bs*{\Phi\ps*{\xi}} - n_\text{i}\bs*{\Phi\ps*{\xi}}\\
    \frac{d^2 \Phi}{d \xi^2} + \frac{1}{\xi}\frac{d \Phi}{d \xi}  & = N_{0,\,\text{Cyl}} I\bs*{u' = C\ps*{\Phi_\text{W} - \Phi_\text{VC}}, \, ab = C\ps*{\Phi - \Phi_\text{W}} } + \frac{n^\text{p}_{\text{e}0,\,\text{Cyl}}}{\gamma} \exp \ps*{\Phi}  - \ps*{1 - \frac{2\Phi}{M^2}}^{-1/2}
    \end{split}
\end{align}
As we did in the Cartesian case, we start by integrating the left side. Let us replace \(1/\xi\) with \(n/\xi\), so that we can generalize to both the cylindrical coordinates case where \(n = 1\) and the spherical coordinates case where \(n = 2\).
\begin{equation}\label{eq:CylindricalLHSIntegration}
    \int \frac{d \Phi}{d \xi}\frac{d^2 \Phi}{d \xi^2} \, d \xi + \int \frac{n}{\xi} \frac{d \Phi}{d \xi}\frac{d \Phi}{d \xi} \, d \xi = \frac{1}{2} \ps*{\frac{d \Phi}{d \xi}}^2 + \int \frac{n}{\xi} \ps*{\frac{d \Phi}{d \xi}}^2 \, d \xi 
\end{equation}
For the right side, we obtain the following.
\begin{align}\label{eq:CylindricalRHSIntegration}
    \begin{split}
    \int \frac{d \Phi}{d \xi} f_\text{Cyl}\ps*{\Phi} \, d \xi & = \int f_\text{Cyl}\ps*{\Phi} \, d \Phi  \\
    & = N_{0,\,\text{Cyl}} I_\text{int}\bs*{u' = C\ps*{\Phi_\text{W} - \Phi_\text{VC}}, \, \Phi } + \frac{n^\text{p}_{\text{e}0,\,\text{Cyl}}}{\gamma} \exp\ps*{\Phi} + M^2 \sqrt{1 - \frac{2\Phi}{M^2}}, \\ 
    & \text{where } I_\text{int}\ps*{u',\,\Phi} = \int_{u'}^\infty \frac{2}{C}\exp\ps*{-u}\sqrt{u}\sqrt{u + C\ps*{\Phi - \Phi_\text{W}}} \, d u
    \end{split}
\end{align}
As we did in the Cartesian case, we apply the boundary condition of \(d \Phi/d \xi \rightarrow 0\) as \(\Phi \rightarrow 0\) (i.e., \(\xi \rightarrow \infty\)). But unlike the Cartesian case, we must explicitly express the left side integral's bounds.
\begin{align}\label{eq:CylindricalIntegratedWithBC}
    \begin{split}
    & \frac{1}{2} \ps*{\frac{d \Phi}{d \xi}}^2  - \cancelto{0}{\bs*{\frac{1}{2} \ps*{\frac{d \Phi}{d \xi}}^2}_{\Phi=0}} + \int_{\infty}^{\xi'} \frac{n}{\xi} \ps*{\frac{d \Phi}{d \xi}}^2 \, d \xi \\
    & = N_{0,\,\text{Cyl}}\br*{ I_\text{int}\bs*{u' = C\ps*{\Phi_\text{W} - \Phi_\text{VC}}, \, \Phi } - I_\text{int}\bs*{u' = C\ps*{\Phi_\text{W} - \Phi_\text{VC}}, \, 0 }} \\
    & + \frac{n^\text{p}_{\text{e}0,\,\text{Cyl}}}{\gamma} \bs*{\exp\ps*{\Phi} - 1} + M^2 \ps*{\sqrt{1 - \frac{2\Phi}{M^2}} - 1}\\
    \Rightarrow \; & \frac{1}{2} \ps*{\frac{d \Phi}{d \xi}}^2 - \int_{\xi'}^{\infty} \frac{n}{\xi} \ps*{\frac{d \Phi}{d \xi}}^2 \, d \xi \\
    & = N_{0,\,\text{Cyl}}\br*{ I_\text{int}\bs*{u' = C\ps*{\Phi_\text{W} - \Phi_\text{VC}}, \, \Phi } - I_\text{int}\bs*{u' = C\ps*{\Phi_\text{W} - \Phi_\text{VC}}, \, 0 }} \\
    & + \frac{n^\text{p}_{\text{e}0,\,\text{Cyl}}}{\gamma} \bs*{\exp\ps*{\Phi} - 1} + M^2 \ps*{\sqrt{1 - \frac{2\Phi}{M^2}} - 1}
    \end{split}
\end{align}

\begin{figure}[H]
    \centering
    \includegraphics[width=0.9\linewidth]{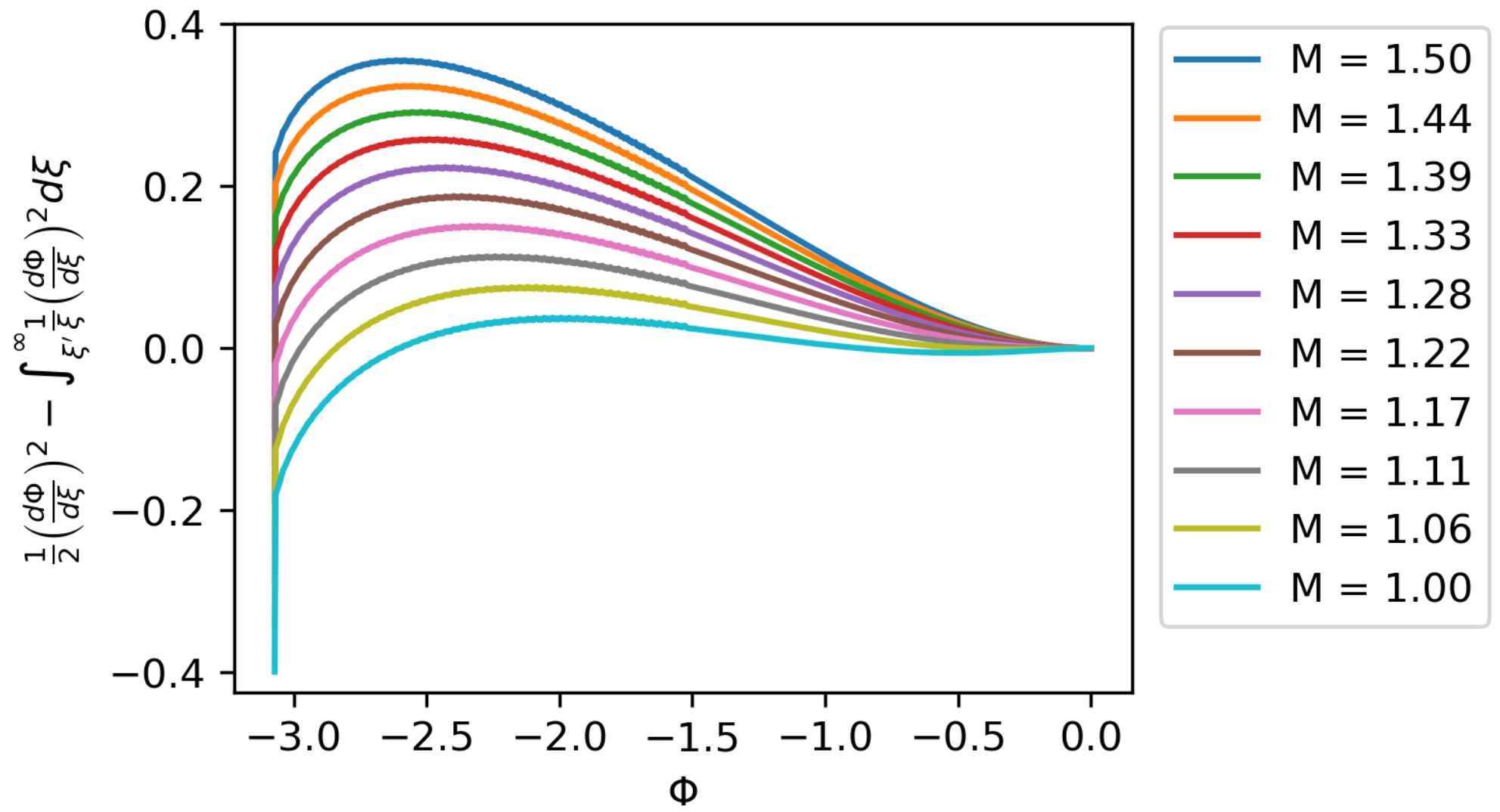}
    \caption{Right side of Eq. \eqref{eq:CylindricalIntegratedWithBC}, with \(\Phi_\text{W} = -2.57\), \(\Phi_\text{VC} = \Phi_\text{W} - 0.5\), \(T_\text{e} = \SI{1.75}{eV}\), \(T_\text{eW} = \SI{0.175}{eV}\); a sheath may be valid even for \(\ps*{d\Phi/d\xi}^2/2 \leq 0\).}
    \label{fig:CylRHS}
\end{figure}
\noindent Now, because the left side integrated term has a positive integrand and integration taken in the negative \(\xi\) direction, the integral term is necessarily negative, resulting in the left side being not necessarily positive. Therefore, the minimum Mach number \(M\) is not dependent on the positivity or negativity of the left and right side expressions, but instead on whether the input parameters of normalized virtual cathode potential \(\Phi_\text{VC}\), current ratio \(\gamma\), and wall derivative \(\Phi'_\text{W}\) yield a valid sheath potential in which \(\Phi \rightarrow 0\) and \(\Phi' \rightarrow 0\)  as \(\xi \rightarrow \infty\). \\
Let us evaluate the integral term on the left side. To do this accurately, we start by computing a numerical solution to the potential curve \(\Phi_\text{num}\ps*{x;\,\Phi_\text{W},\Phi'_\text{W}}\) by constructing an initial value problem that begins at the wall. To solve the initial value problem, each second order equation is decomposed into a system of two first order differential equations, which are numerically solved by an explicit Runge-Kutta method of order 8 (DOP853, \verb|scipy.integrate.solve_ivp| \cite{Virtanen2020}). The source terms developed for each coordinate system are encapsulated in the \(f\) functions per each coordinate system below.
\begin{equation}\label{eq:CartesianDiffEqDecomp}
    \text{Cartesian:}\; u'[2] = f_\text{Car}\ps*{\Phi},\; u'[1] = u[2]
\end{equation}
\begin{equation}\label{eq:CylindricalDiffEqDecomp}
    \text{Cylindrical:}\; u'[2] = f_\text{Cyl}\ps*{\Phi} - \frac{1}{\xi},\; u'[1] = u[2]
\end{equation}
\begin{equation}\label{eq:SphericalDiffEqDecomp}
    \text{Spherical:}\; u'[2] = f_\text{Sph}\ps*{\Phi} - \frac{2}{\xi},\; u'[1] = u[2]
\end{equation}
With a numerical solution in hand, we can express a definite integral from the virtual cathode to infinity as follows. 
\begin{equation}\label{eq:CylindricalNumIntWithVC}
\int_{\xi_\text{VC}}^{\infty} \frac{n}{\xi} \ps*{\frac{d \Phi}{d \xi}}^2 \, d \xi = \int_{\xi_\text{VC}}^{\infty}\frac{n}{\xi} \bs*{\frac{d \Phi_\text{num}\ps*{\xi ;\,\Phi_\text{W},\Phi'_\text{W}} }{d \xi}}^2 \, d \xi
\end{equation}
On the other hand, if there is no virtual cathode present, then \(\xi_\text{VC} = \xi_0\) and we can express a definite integral from the wall to the sheath edge.
\begin{equation}\label{eq:CylindricalNumIntWithoutVC}
\int_{\xi_\text{W}}^{\infty} \frac{n}{\xi} \ps*{\frac{d \Phi}{d \xi}}^2 \, d \xi = \int_{\xi_\text{W}}^{\infty}\frac{n}{\xi} \bs*{\frac{d \Phi_\text{num}\ps*{\xi ;\,\Phi_\text{W},\Phi'_\text{W}} }{d \xi}}^2 \, d \xi
\end{equation}
Whether a virtual cathode is present or not, by using the fundamental theorem of calculus, we can assemble the left side with the right side source term function (\(f_\text{Cyl}\) or \(f_\text{Sph}\)) and find a constraint between the the virtual cathode potential \(\Phi_\text{VC} \leq \Phi_\text{W}\), wall derivative \(\Phi'_\text{W}\), and leading edge radial offset \(\xi_\text{0}\). Therefore, in cylindrical and spherical coordinates, for \textit{no} virtual cathode formation (i.e., \(\Phi_\text{W}' > 0\)), we include the following constraint based on Equation \eqref{eq:CylindricalIntegratedWithBC}.\\
\underline{Condition for NO virtual cathode formation (cylindrical case, \(n=1, \; \Phi_\text{VC} = \Phi_\text{W} \Rightarrow \Phi_\text{W}' > 0\))}
\begin{align}\label{eq:CylindricalNoVC}
    \begin{split}
    & \frac{1}{2} \ps*{\Phi'_\text{W}}^2 -   \int_{\xi_\text{W}}^{\infty}\frac{n}{\xi} \bs*{\frac{d \Phi_\text{num}\ps*{\xi ;\,\Phi_\text{W},\Phi'_\text{W}} }{d \xi}}^2 \, d \xi\\
    & = N_{0,\,\text{Cyl}}\br*{ I_\text{int}\bs*{u' = 0, \, \Phi_\text{W}} - I_\text{int}\bs*{u' = 0, \, 0 }} \\
    & + \frac{n^\text{p}_{\text{e}0,\,\text{Cyl}}}{\gamma} \bs*{\exp\ps*{\Phi_\text{W}} - 1} + M^2 \ps*{\sqrt{1 - \frac{2\Phi_\text{W}}{M^2}} - 1}
    \end{split}
\end{align}
Similarly, if there is virtual cathode formation (i.e., \(\Phi_\text{W}' < 0\)), then we must satisfy Equation \eqref{eq:CylindricalIntegratedWithBC} both between the wall and sheath edge, as well as between the virtual cathode and sheath edge.\\
\underline{Conditions for virtual cathode formation (cylindrical case, \(n=1,\;\Phi_\text{VC} < \Phi_\text{W} \Rightarrow \Phi_\text{W}' < 0\))}
\begin{align}\label{eq:CylindricalVC}
    \begin{split}
    & \text{For the space between the wall and sheath edge:}\\
    & \frac{1}{2} \ps*{\Phi'_\text{W}}^2 -   \int_{\xi_\text{W}}^{\infty}\frac{n}{\xi} \bs*{\frac{d \Phi_\text{num}\ps*{\xi ;\,\Phi_\text{W},\Phi'_\text{W}} }{d \xi}}^2 \, d \xi\\
    & = N_{0,\,\text{Cyl}}\br*{ I_\text{int}\bs*{u' = C\ps*{\Phi_\text{W} - \Phi_\text{VC}}, \, \Phi_\text{W}} - I_\text{int}\bs*{u' = C\ps*{\Phi_\text{W} - \Phi_\text{VC}}, \, 0 }} \\
    & + \frac{n^\text{p}_{\text{e}0,\,\text{Cyl}}}{\gamma} \bs*{\exp\ps*{\Phi_\text{W}} - 1} + M^2 \ps*{\sqrt{1 - \frac{2\Phi_\text{W}}{M^2}} - 1}\\
    & \text{For the space between the VC and sheath edge:}\\
    & - \int_{\xi_\text{VC}}^{\infty}\frac{n}{\xi} \bs*{\frac{d \Phi_\text{num}\ps*{\xi ;\,\Phi_\text{W},\Phi'_\text{W}} }{d \xi}}^2 \, d \xi\\
    & = N_{0,\,\text{Cyl}}\br*{ I_\text{int}\bs*{u' = C\ps*{\Phi_\text{W} - \Phi_\text{VC}}, \, \Phi_\text{VC}} - I_\text{int}\bs*{u' = C\ps*{\Phi_\text{W} - \Phi_\text{VC}}, \, 0 }} \\
    & + \frac{n^\text{p}_{\text{e}0,\,\text{Cyl}}}{\gamma} \bs*{\exp\ps*{\Phi_\text{VC}} - 1} + M^2 \ps*{\sqrt{1 - \frac{2\Phi_\text{VC}}{M^2}} - 1}
    \end{split}
\end{align}
These equations describing virtual cathode constraints enable us to construct the condition for virtual cathode formation and lack thereof in both cylindrical and spherical coordinate systems. The only difference for the spherical coordinate system besides \(n = 2\), is its integrated right side which we derive next.
\subsubsection{Spherical coordinates} \label{Spherical}
\begin{align}\label{eq:SphericalDiffEqWithSources}
    \begin{split}
    \frac{d^2 \Phi}{d \xi^2} + \frac{2}{\xi}\frac{d \Phi}{d \xi} & = n_{\text{e},\, \text{Sph}}^\text{s}\bs*{\Phi\ps*{\xi}} + n_{\text{e},\,\text{Sph}}^\text{p}\bs*{\Phi\ps*{\xi}} - n_\text{i}\bs*{\Phi\ps*{\xi}}\\
    \frac{d^2 \Phi}{d \xi^2} + \frac{2}{\xi}\frac{d \Phi}{d \xi} & = N_{0,\,\text{Sph}} \frac{1}{\sqrt{\pi}} \exp\bs*{ C\ps*{\Phi - \Phi_\text{W}}}\\
    & \cdot \Bigg \{\sqrt{\pi} \bs*{1 - 2C \ps*{\Phi - \Phi_\text{W}} }\,\text{erfc}\sqrt{C\ps*{\Phi - \Phi_\text{VC}}} + 2\sqrt{ C\ps*{\Phi - \Phi_\text{VC}}}  \exp \bs*{- C\ps*{\Phi - \Phi_\text{VC}} } \Bigg\}\\
    & + \frac{n^\text{p}_{\text{e}0,\,\text{Sph}}}{\gamma} \exp \ps*{\Phi} - \ps*{1 - \frac{2\Phi}{M^2}}^{-1/2}
    \end{split}
\end{align}
Since we have already obtained the left side integral expression generalized for both cylindrical \(n = 1\) and spherical \(n = 2\) coordinate systems, we move on to the integral of the right side.
\begin{align}\label{eq:SphericalRHSIntegration}
    \begin{split}
    & \int \frac{d \Phi}{d \xi} f_\text{Sph}\ps*{\Phi} \, d \xi = \int f_\text{Sph}\ps*{\Phi} \, d \Phi  \\
    & = \frac{1}{C\sqrt{\pi}}\exp\ps*{-C \Phi_\text{W}} \bigg\{\sqrt{\pi}\exp\ps*{C\Phi} \bs*{2C\ps*{\Phi_\text{W} - \Phi} + 3} \text{erfc}\sqrt{C\ps*{\Phi - \Phi_\text{VC}}} \\
    & - 2\exp\ps*{C\Phi_\text{VC}}\bs*{2 C \ps*{\Phi_\text{VC} - \Phi_\text{W}} - 3}\sqrt{C \ps*{\Phi - \Phi_\text{VC}}}\bigg\}\\
    & + \frac{n^\text{p}_{\text{e}0,\,\text{Sph}}}{\gamma} \exp\ps*{\Phi} + M^2 \sqrt{1 - \frac{2\Phi}{M^2}}
    \end{split}
\end{align}
As we did for the cylindrical coordinate system, we apply the boundary condition of \(d \Phi/d \xi \rightarrow 0\) as \(\Phi \rightarrow 0\).
\begin{align}\label{eq:SphericalIntegratedWithBC}
    \begin{split}
    & \frac{1}{2} \ps*{\frac{d \Phi}{d \xi}}^2 - \int_{\xi'}^\infty\frac{n}{\xi} \ps*{\frac{d \Phi}{d \xi}}^2  \, d \xi \\
    & = \frac{1}{C\sqrt{\pi}}\exp\ps*{-C \Phi_\text{W}} \bigg\{\sqrt{\pi}\exp\ps*{C\Phi} \bs*{2C\ps*{\Phi_\text{W} - \Phi} + 3} \text{erfc}\sqrt{C\ps*{\Phi - \Phi_\text{VC}}} \\
    & - \sqrt{\pi} \ps*{2 C \Phi_\text{W} + 3} \text{erfc}\sqrt{-C\Phi_\text{VC}} - 2\exp\ps*{C\Phi_\text{VC}}\bs*{2 C \ps*{\Phi_\text{VC} - \Phi_\text{W}} - 3}\bs*{\sqrt{C \ps*{\Phi - \Phi_\text{VC}}} -\sqrt{-C \Phi_\text{VC}}}\bigg\}\\
    & + \frac{n^\text{p}_{\text{e}0,\,\text{Sph}}}{\gamma} \bs*{\exp\ps*{\Phi} - 1} + M^2 \ps*{\sqrt{1 - \frac{2\Phi}{M^2}} - 1}
    \end{split}
\end{align}
\begin{figure}[H]
    \centering
    \includegraphics[width=0.9\linewidth]{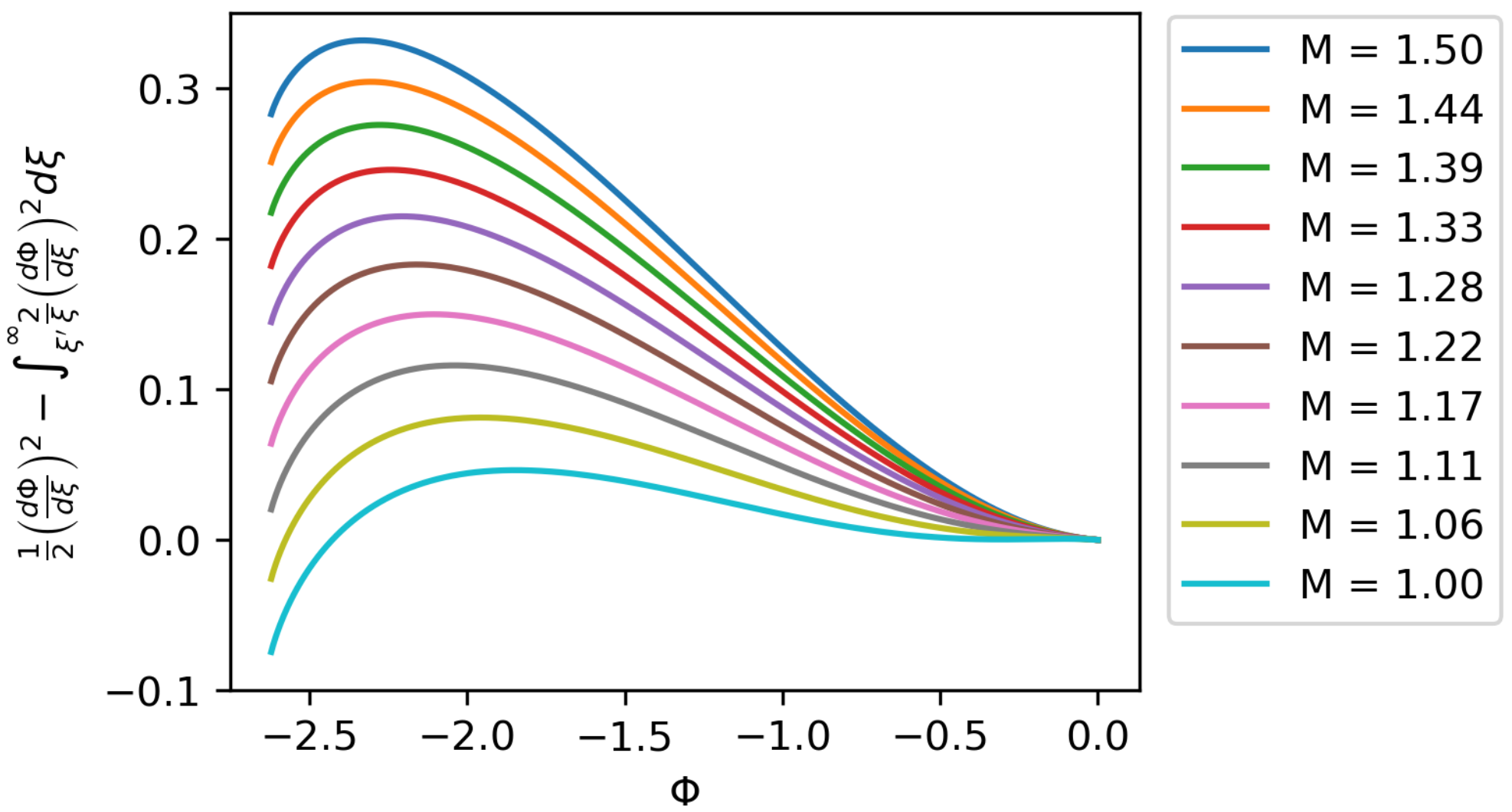}
    \caption{Right-hand side of Eq. \eqref{eq:SphericalIntegratedWithBC}, with \(\Phi_\text{W} = -2.57\), \(\Phi_\text{VC} = \Phi_\text{W} - 0.05\), \(T_\text{e} = \SI{1.75}{eV}\), \(T_\text{eW} = \SI{0.175}{eV}\); a sheath may be valid even for \(\ps*{d\Phi/d\xi}^2/2 \leq 0\).}
    \label{fig:SphRHS}
\end{figure}
\noindent Because the integrated left side is generalized to both cylindrical and spherical coordinates, the arguments leading to the conditions for virtual cathode formation and lack thereof for cylindrical coordinates are reused now in spherical coordinates.\\
\underline{Condition for NO virtual cathode formation (spherical case, \(n=2,\;\Phi_\text{VC} = \Phi_\text{W} \Rightarrow \Phi_\text{W}' > 0\))}
\begin{align}\label{eq:SphericalNoVC}
    \begin{split}
    & \frac{1}{2} \ps*{\Phi'_\text{W}}^2 -  \int_{\xi_\text{W}}^{\infty}\frac{n}{\xi} \bs*{\frac{d \Phi_\text{num}\ps*{\xi ;\,\Phi_\text{W},\Phi'_\text{W}} }{d \xi}}^2 \, d \xi\\
    & = \frac{1}{C\sqrt{\pi}}\exp\ps*{-C \Phi_\text{W}} \bigg\{3\sqrt{\pi}\exp\ps*{C\Phi_\text{W}} \\
    & - \sqrt{\pi} \ps*{2 C \Phi_\text{W} + 3} \text{erfc}\sqrt{-C\Phi_\text{W}} - 6 \exp\ps*{C\Phi_\text{W}} \sqrt{-C \Phi_\text{W}}\bigg\}\\
    & + \frac{n^\text{p}_{\text{e}0,\,\text{Sph}}}{\gamma} \bs*{\exp\ps*{\Phi_\text{W}} - 1} + M^2 \ps*{\sqrt{1 - \frac{2\Phi_\text{W}}{M^2}} - 1}
    \end{split}
\end{align}
\underline{Conditions for virtual cathode formation (spherical case, \(n=2,\;\Phi_\text{VC} < \Phi_\text{W} \Rightarrow \Phi_\text{W}' < 0\))}
\begin{align}\label{eq:SphericalVC}
    \begin{split}
    & \text{For the space between the wall and sheath edge:}\\
    & \frac{1}{2} \ps*{\Phi'_\text{W}}^2 -   \int_{\xi_\text{W}}^{\infty}\frac{n}{\xi} \bs*{\frac{d \Phi_\text{num}\ps*{\xi ;\,\Phi_\text{W},\Phi'_\text{W}} }{d \xi}}^2 \, d \xi\\
    & =  \frac{1}{C\sqrt{\pi}}\exp\ps*{-C \Phi_\text{W}} \bigg\{ 3\sqrt{\pi}\exp\ps*{C\Phi_\text{W}}  \text{erfc}\sqrt{C\ps*{\Phi_\text{W} - \Phi_\text{VC}}} \\
    & - \sqrt{\pi} \ps*{2 C \Phi_\text{W} + 3} \text{erfc}\sqrt{-C\Phi_\text{VC}} - 2\exp\ps*{C\Phi_\text{VC}}\bs*{2 C \ps*{\Phi_\text{VC} - \Phi_\text{W}} - 3}\bs*{\sqrt{C \ps*{\Phi_\text{W} - \Phi_\text{VC}}} -\sqrt{-C \Phi_\text{VC}}}\bigg\}\\
    & + \frac{n^\text{p}_{\text{e}0,\,\text{Sph}}}{\gamma} \bs*{\exp\ps*{\Phi_\text{W}} - 1} + M^2 \ps*{\sqrt{1 - \frac{2\Phi_\text{W}}{M^2}} - 1}\\
    & \text{For the space between the VC and sheath edge:}\\
    & - \int_{\xi_\text{VC}}^{\infty}\frac{n}{\xi} \bs*{\frac{d \Phi_\text{num}\ps*{\xi ;\,\Phi_\text{W},\Phi'_\text{W}} }{d \xi}}^2 \, d \xi\\
    & = \frac{1}{C\sqrt{\pi}}\exp\ps*{-C \Phi_\text{W}} \bigg\{\sqrt{\pi}\exp\ps*{C\Phi_\text{VC}} \bs*{2C\ps*{\Phi_\text{W} - \Phi_\text{VC}} + 3} \\
    & - \sqrt{\pi} \ps*{2 C \Phi_\text{W} + 3} \text{erfc}\sqrt{-C\Phi_\text{VC}} + 2\exp\ps*{C\Phi_\text{VC}}\bs*{2 C \ps*{\Phi_\text{VC} - \Phi_\text{W}} - 3}\sqrt{-C \Phi_\text{VC}}\bigg\}\\
    & + \frac{n^\text{p}_{\text{e}0,\,\text{Sph}}}{\gamma} \bs*{\exp\ps*{\Phi_\text{VC}} - 1} + M^2 \ps*{\sqrt{1 - \frac{2\Phi_\text{VC}}{M^2}} - 1}
    \end{split}
\end{align}

\section{Results \& Discussion}

\begin{table}[H]
\centering
\begin{tabular}{|| c | c ||} 
 \hline\hline
 \(\phi_\text{w} = \SI{3.0}{eV}\) (low), \(\phi_\text{w} = \SI{4.5}{eV} \) (high) & work function energy \\ \hline
 \(T_\text{e} = \SI{1.75}{eV}\) & plasma electron energy \\ \hline
 \(T_\text{eW} = \SI{0.175}{eV}\) &  wall electron energy \\ \hline
 \(\xi_0 = 5000\lambda_\text{D}\) & LE radius \\ \hline
 \(m_\text{i} = 4\times\text{proton mass} = 4\times 1.67\times 10^{-27}\) kg & positive ion mass \\ \hline
 \(\text{threshold}_\text{NO VC} = 1\times 10^{-10}\) (Cartesian) & \multirow{2}{0.5\linewidth}{Cartesian model's residual threshold for a sheath \textit{without} virtual cathode} \\ & \\ \hline
 \(\text{threshold}_\text{VC} = 1\times 10^{-4}\) (Cartesian) & \multirow{2}{0.5\linewidth}{Cartesian model's residual threshold for a sheath \textit{with} virtual cathode} \\ & \\ \hline
 \(\text{threshold}_\text{NO VC} = 0.15\) (cyl. \& sph.) & \multirow{2}{0.5\linewidth}{cylindrical and spherical models' residual threshold for a sheath \textit{without} virtual cathode} \\ & \\ \hline
 \(\text{threshold}_\text{VC} = 0.30\) (cyl. \& sph.) & \multirow{2}{0.5\linewidth}{cylindrical and spherical models' residual threshold for a sheath \textit{with} virtual cathode} \\ & \\[0.5ex]
 \hline\hline
\end{tabular}
\caption{Input parameters}
\label{table:ParametersSheathCriteria}
\end{table}

In order to discover what combinations of current ratio \(\gamma\), virtual cathode potential \(\Phi_\text{VC}\), and minimum Mach number \(M_\text{min}\) yield a valid sheath, we take a brute force approach of trying many \(\gamma, \, \Phi_\text{VC},\, M\) combinations and observing the parameter combinations that satisfy the sheath formation conditions for each coordinate system. For the Cartesian coordinate system, both of the new conditions Eqs. \eqref{eq:CartesianBohm1} and \eqref{eq:CartesianBohm2} must be met; and, if there is a virtual cathode present, the virtual cathode condition Eq. \eqref{eq:CartesianVC} right-hand side side is not allowed to exceed a prescribed threshold. In the cases of the cylindrical and spherical coordinate systems, we constrain the potential value \(\Phi\) and potential derivative \(\Phi'\) as position goes to infinity, in addition to the conditions outlined in sections \ref{Cylindrical} and \ref{Spherical} because there are no simple inequalities that only maintain sheath validity as in the Cartesian case. Thus for cylindrical and spherical coordinates, a parameter combination that meets the sheath formation conditions yields one of the two following residuals depending on if a virtual cathode can form, and these residuals are not allowed to exceed prescribed thresholds. LHS and RHS denote the left side and right side, respectively, of the conditions for virtual cathode formation and lack thereof in sections \ref{Cylindrical} and \ref{Spherical}. The LHS and RHS terms are subscripted with ``W'' and ``VC'' to imply that the LHS and RHS expressions are evaluated at the wall or virtual cathode, respectively.
\begin{equation}\label{eq:Cyl&SphSheathResidualExpressionNoVC}
\abs*{\text{LHS}_\text{W} - \text{RHS}_\text{W}} + \abs*{\lim_{\xi \to \infty} \Phi} + \abs*{\lim_{\xi \to \infty} \Phi'} < \text{threshold}_\text{NO VC}
\end{equation}
\begin{equation}\label{eq:Cyl&SphSheathResidualExpressionVC}
\abs*{\text{LHS}_\text{VC} - \text{RHS}_\text{VC}} + \abs*{\text{LHS}_\text{W} - \text{RHS}_\text{W}} + \abs*{\lim_{\xi \to \infty} \Phi} + \abs*{\lim_{\xi \to \infty} \Phi'} < \text{threshold}_\text{VC}
\end{equation}
For two different work functions and constant wall and plasma energies, we generated datasets for minimum Mach number \(M_\text{min}\), the corresponding derivative of potential at the wall \(\Phi'_\text{W}\), and the corresponding net current \(j_\text{net}/n_0\) in each coordinate system. We apply the net current expression defined by Takamura [1], but negated to represent electron current emitted in the direction pointing away from the material surface:
\begin{equation}\label{eq:NetCurrentExpression}
\frac{j_\text{net, sys}}{n_0} = \frac{j_\text{e, sys}^\text{s} - j_\text{e, sys}^\text{p} + j_\text{i}^+}{n_0},\;\text{where } \frac{j_\text{i}^+}{n_0} = M e
\sqrt{\frac{T_\text{e}}{m_\text{i}}}
\end{equation}
is the (positive) ion saturation current divided by the quasi-neutral plasma density and multiplied by the Mach number.

\subsection{Cartesian model spaces} \label{Cartesian Model Spaces}

In this subsection and the ones that follow, we present plots of the minimum Mach number \(M_\text{min}\), the corresponding derivative of potential at the wall \(\Phi'_\text{W}\), and the corresponding net current \(j_\text{net}/n_0\) for each coordinate system and for two separate work functions of \(\phi_\text{w} = \SI{4.5}{eV}\) and \(\phi_\text{w} = \SI{3.0}{eV}\). First, we note that we limit the ranges of current ratio \(\gamma_\text{sys} = \abs*{j_{\text{e},\,\text{sys}}^\text{s}/j_{\text{e},\,\text{sys}}^\text{p}}\) to focus on input parameters that yield potentials with a virtual cathode, as we notice that most current ratios queried yield potentials without a virtual cathode.

\subsubsection{Cartesian model spaces for high work function \(\phi_\text{w} = \SI{4.5}{eV}\)}
\begin{figure}[H]
    \centering
    \includegraphics[width=0.675\linewidth]{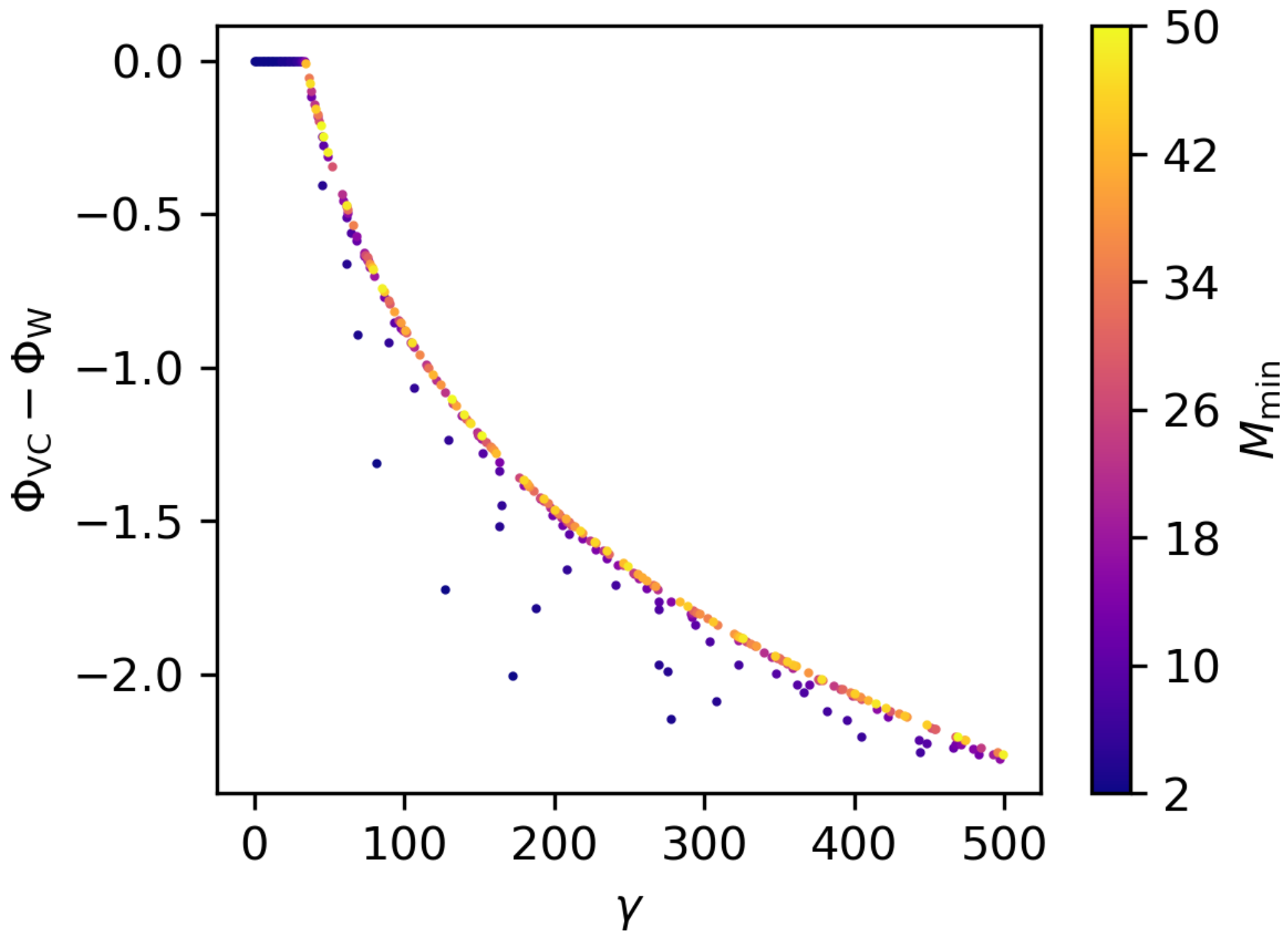}
    \caption{\(M_\text{min}\) as a function of current ratio \(\gamma \in \bs*{0,\,500}\) and virtual cathode potential with respect to the wall \(\Phi_\text{VC} - \Phi_\text{W}\in \bs*{0,-2.5}\), for material work function \(\phi_\text{w} = \SI{4.5}{eV}\).}
    \label{fig:CarHighM}
\end{figure}

\begin{figure}[H]
    \centering
    \includegraphics[width=0.675\linewidth]{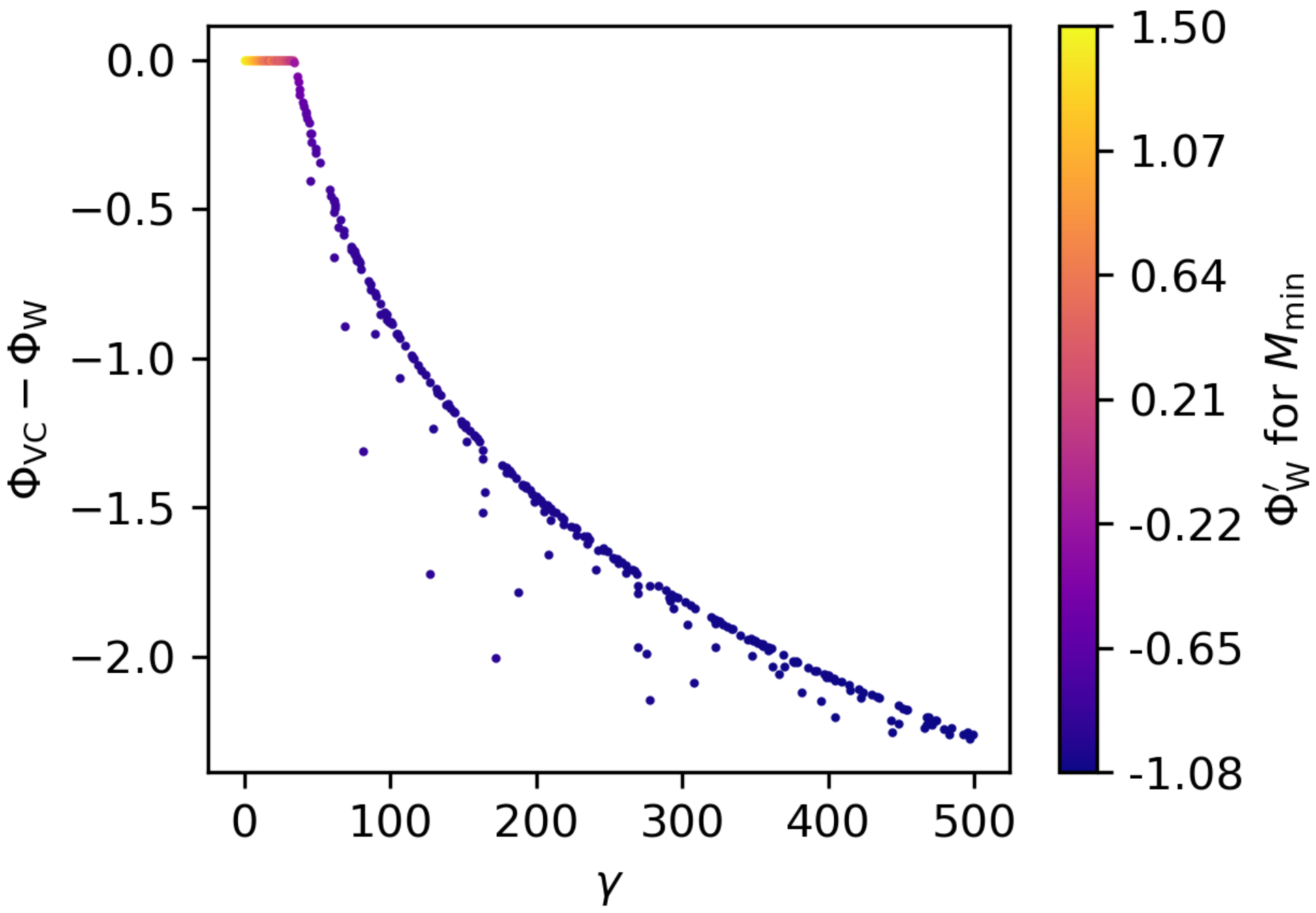}
    \caption{\(\Phi'_\text{W}\) for \(M_\text{min}\) in Fig. \ref{fig:CarHighM} as a function of current ratio \(\gamma \in \bs*{0,\,500}\) and virtual cathode potential with respect to the wall \(\Phi_\text{VC} - \Phi_\text{W}\in \bs*{0,-2.5}\), for material work function \(\phi_\text{w} = \SI{4.5}{eV}\).}
    \label{fig:CarHighPhiWPrime}
\end{figure}

\begin{figure}[H]
    \centering
    \includegraphics[width=0.7\linewidth]{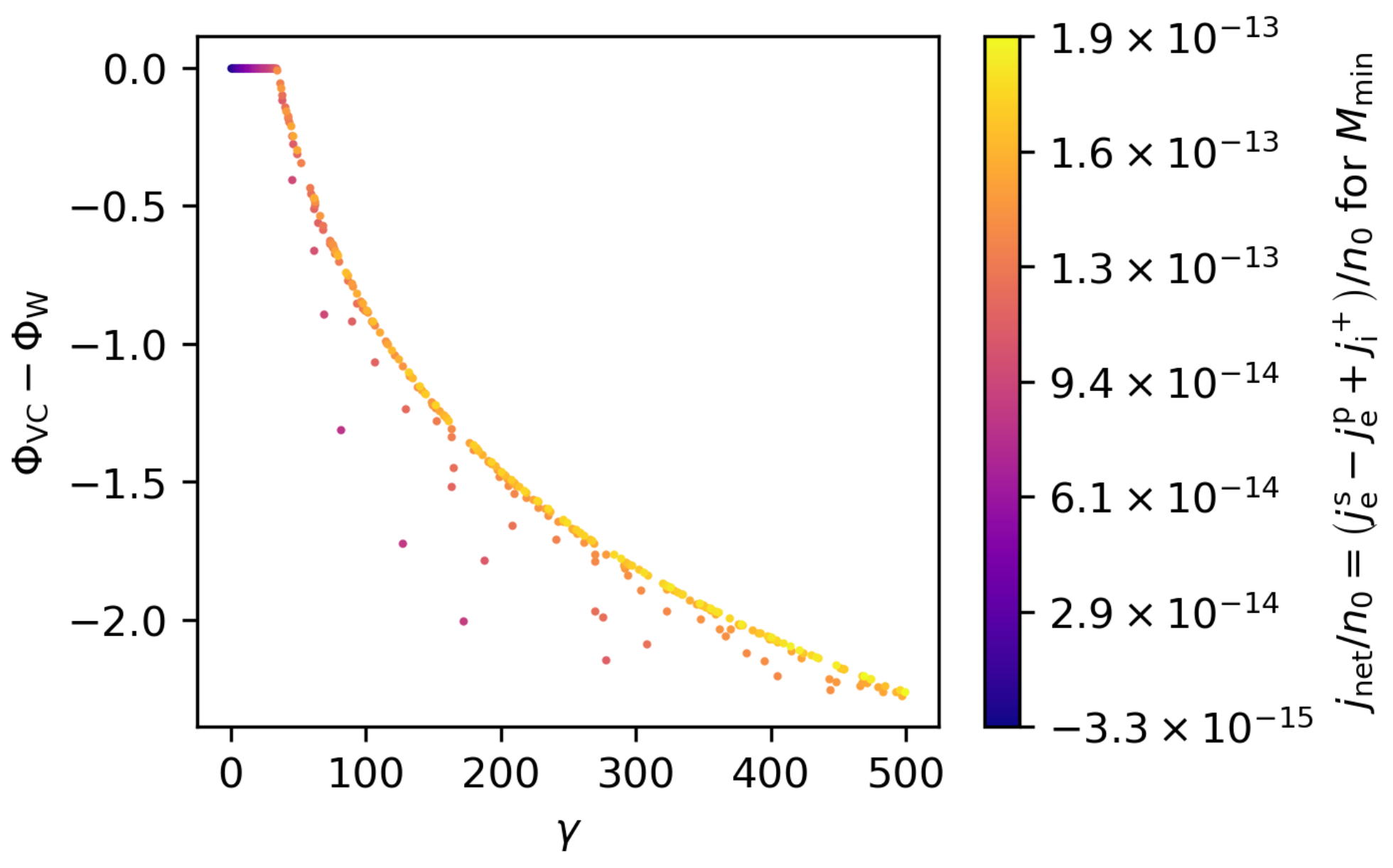}
    \caption{\(j_\text{net}/n_0\) for \(M_\text{min}\) in Fig. \ref{fig:CarHighM} as a function of current ratio \(\gamma \in \bs*{0,\,500}\) and virtual cathode potential with respect to the wall \(\Phi_\text{VC} - \Phi_\text{W}\in \bs*{0,-2.5}\), for material work function \(\phi_\text{w} = \SI{4.5}{eV}\).}
    \label{fig:CarHighJnet}
\end{figure}

\begin{figure}[H]
    \centering
    \begin{subfigure}{0.4\textwidth}
         \includegraphics[width=\textwidth]{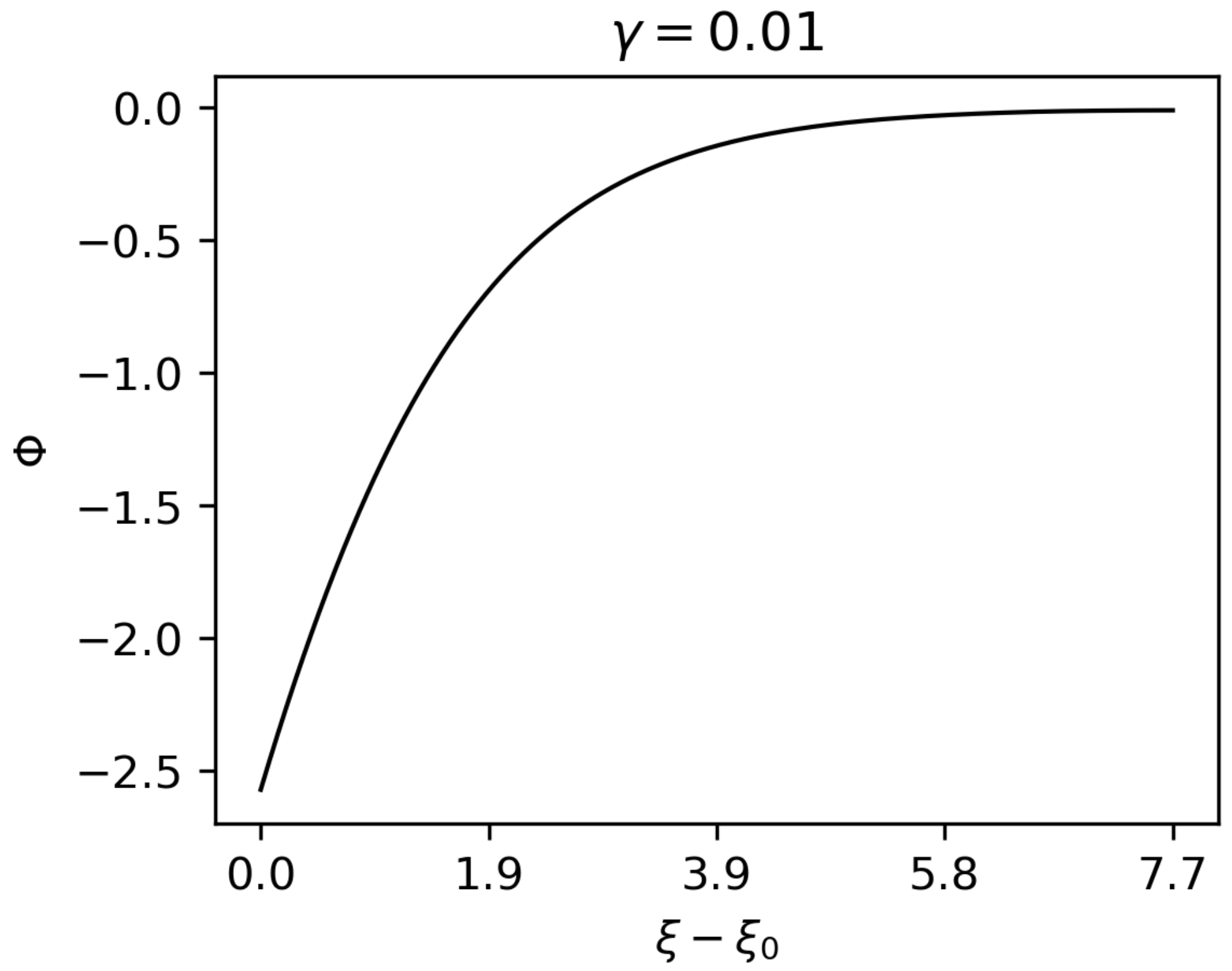}
    \end{subfigure}
    \begin{subfigure}{0.4\textwidth}
         \includegraphics[width=\textwidth]{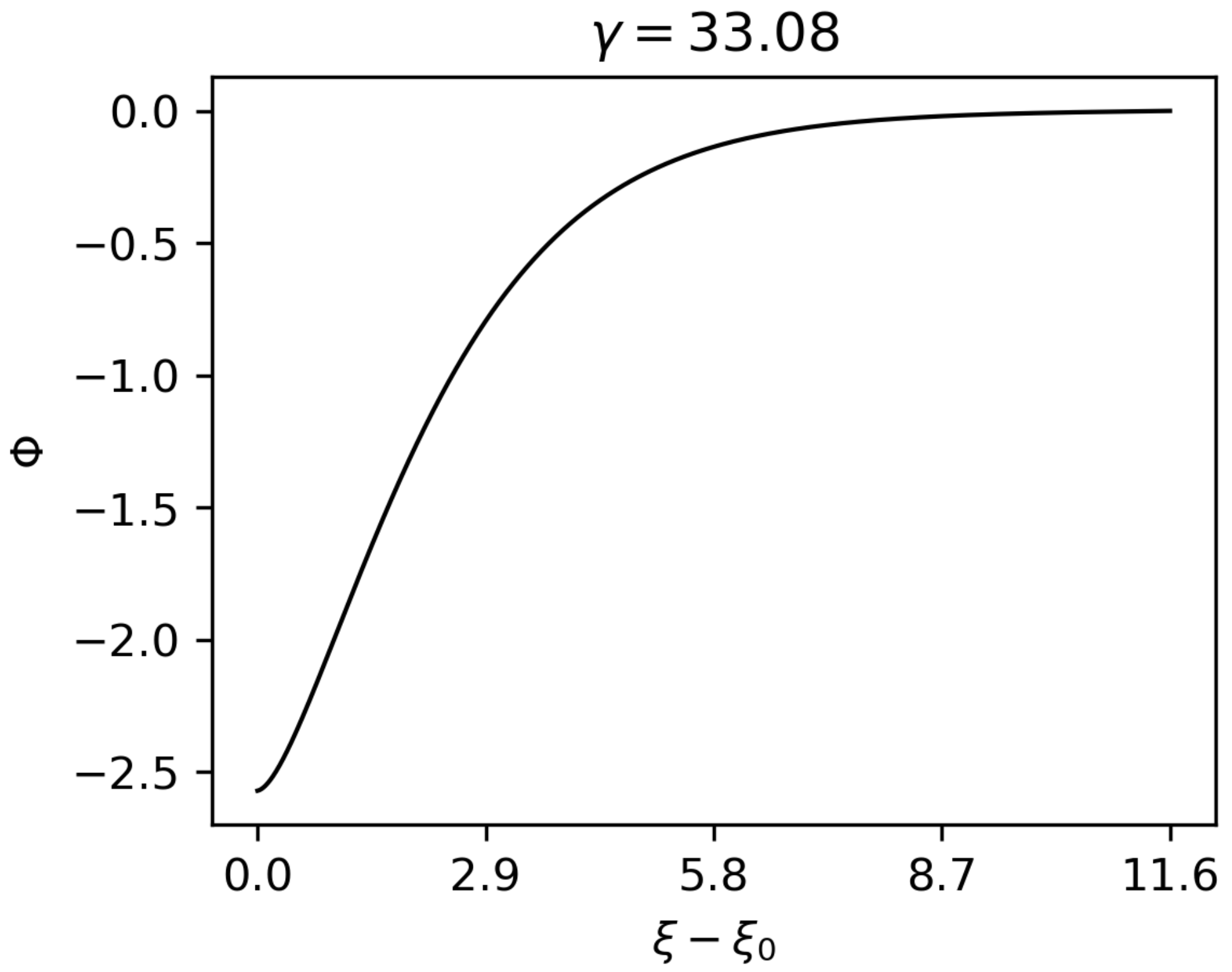}
    \end{subfigure}
    \hfill
    \begin{subfigure}{0.4\textwidth}
         \includegraphics[width=\textwidth]{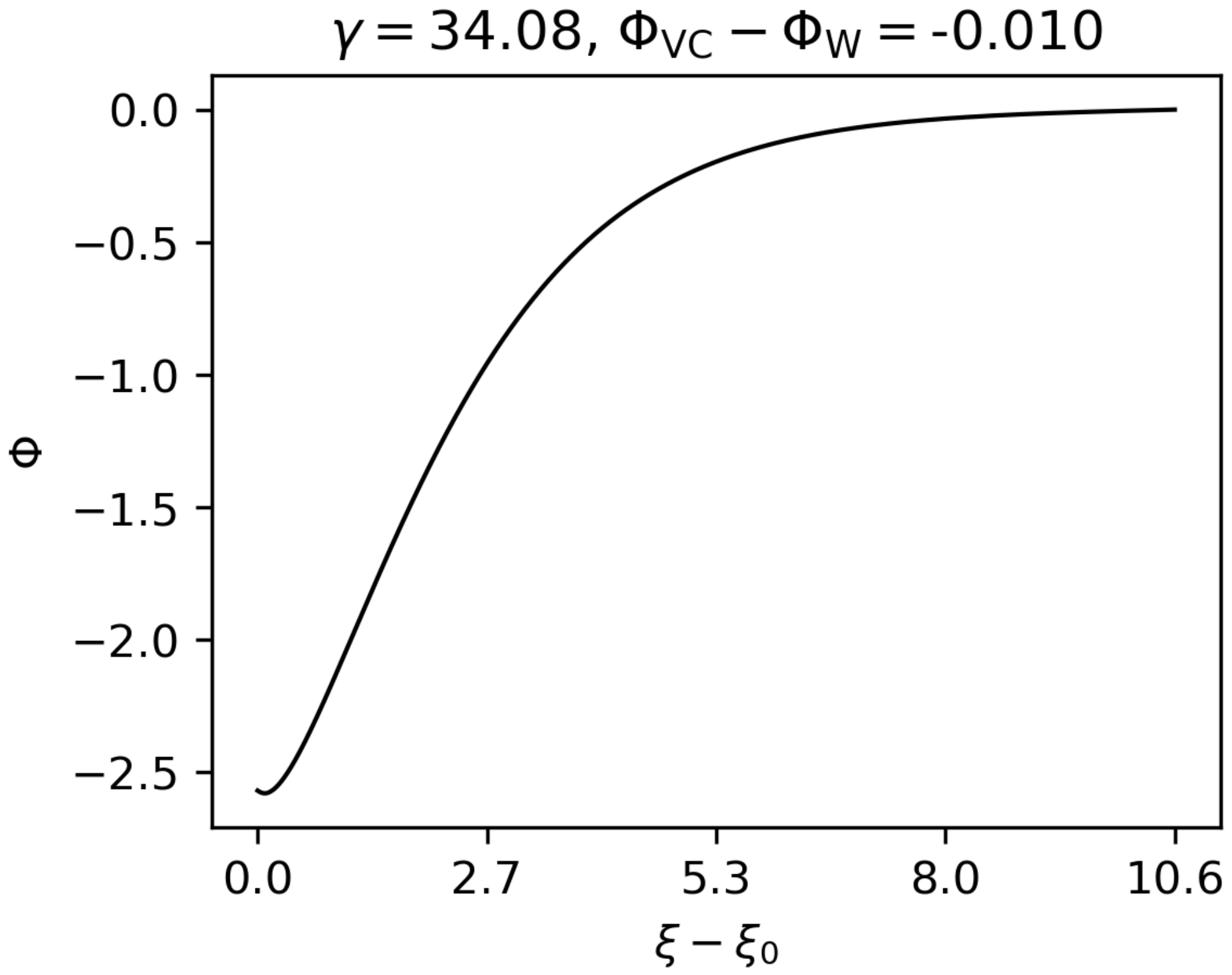}
    \end{subfigure}
    \begin{subfigure}{0.4\textwidth}
         \includegraphics[width=\textwidth]{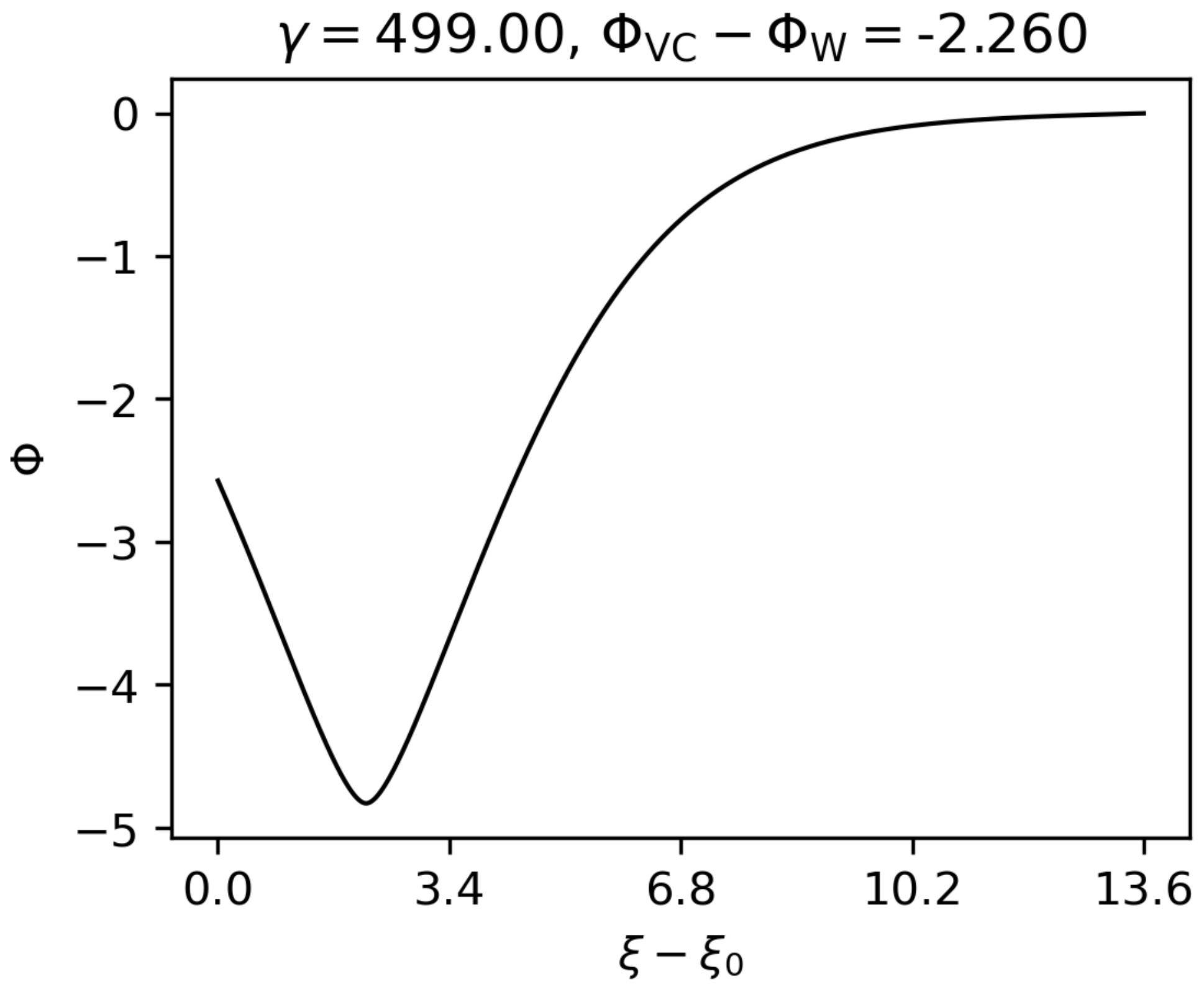}
    \end{subfigure}
\caption{Example potential sheaths at extremes of the \(\ps*{\gamma,\,\Phi_\text{VC} - \Phi_\text{W}}\) parameter space, for material work function \(\phi_\text{w} = \SI{4.5}{eV}\); (top) potential spaces without virtual cathode, (bottom) potential spaces with virtual cathode.}
\label{fig:CarHighExamplePotentials}
\end{figure}

\subsubsection{Cartesian model spaces for low work function \(\phi_\text{w} = \SI{3.0}{eV}\)}

\begin{figure}[H]
    \centering
    \includegraphics[width=0.7\linewidth]{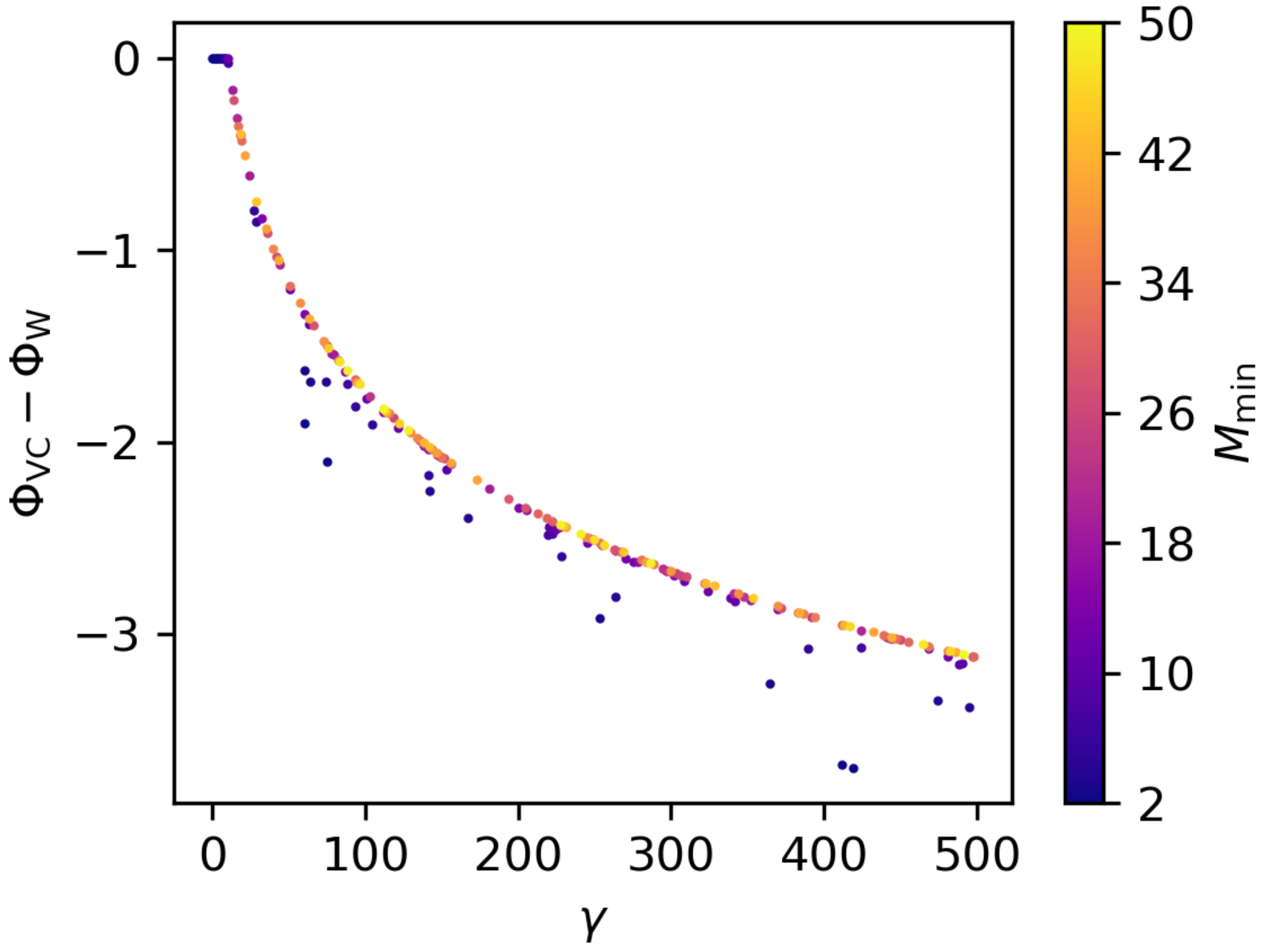}
    \caption{\(M_\text{min}\) as a function of current ratio \(\gamma \in \bs*{0,\,500}\) and virtual cathode potential with respect to the wall \(\Phi_\text{VC} - \Phi_\text{W}\in \bs*{0,-4}\), for material work function \(\phi_\text{w} = \SI{3.0}{eV}\).}
    \label{fig:CarLowM}
\end{figure}

\begin{figure}[H]
    \centering
    \includegraphics[width=0.7\linewidth]{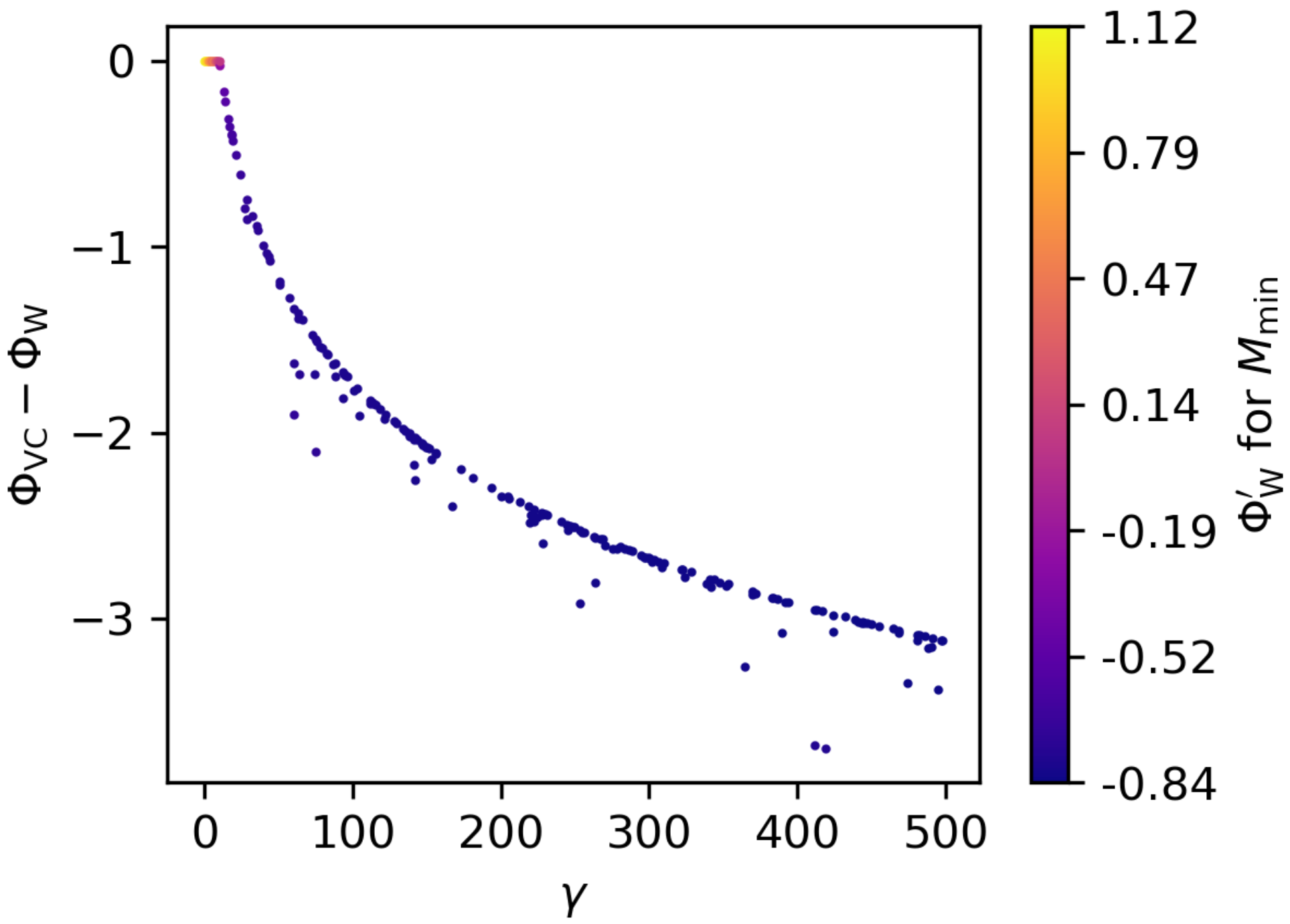}
    \caption{\(\Phi'_\text{W}\) for \(M_\text{min}\) in Fig. \ref{fig:CarLowM} as a function of current ratio \(\gamma \in \bs*{0,\,500}\) and virtual cathode potential with respect to the wall \(\Phi_\text{VC} - \Phi_\text{W}\in \bs*{0,-4}\), for material work function \(\phi_\text{w} = \SI{3.0}{eV}\).}
    \label{fig:CarLowPhiWPrime}
\end{figure}

\begin{figure}[H]
    \centering
    \includegraphics[width=0.7\linewidth]{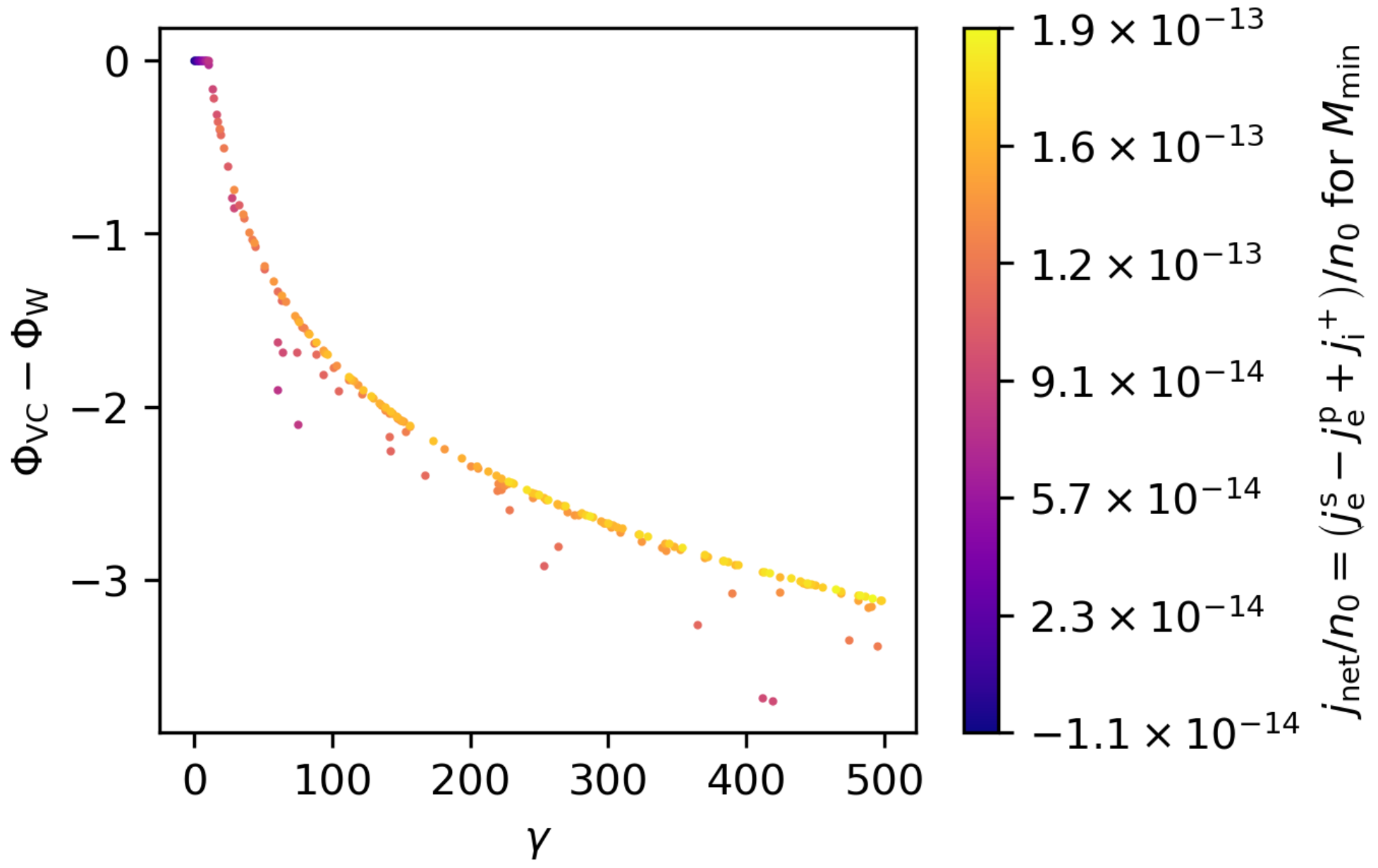}
    \caption{\(j_\text{net}/n_0\) for \(M_\text{min}\) in Fig. \ref{fig:CarLowM} as a function of current ratio \(\gamma \in \bs*{0,\,500}\) and virtual cathode potential with respect to the wall \(\Phi_\text{VC} - \Phi_\text{W}\in \bs*{0,-4}\), for material work function \(\phi_\text{w} = \SI{3.0}{eV}\).}
    \label{fig:CarLowJnet}
\end{figure}

\begin{figure}[H]
    \centering
    \begin{subfigure}{0.4\textwidth}
         \includegraphics[width=\textwidth]{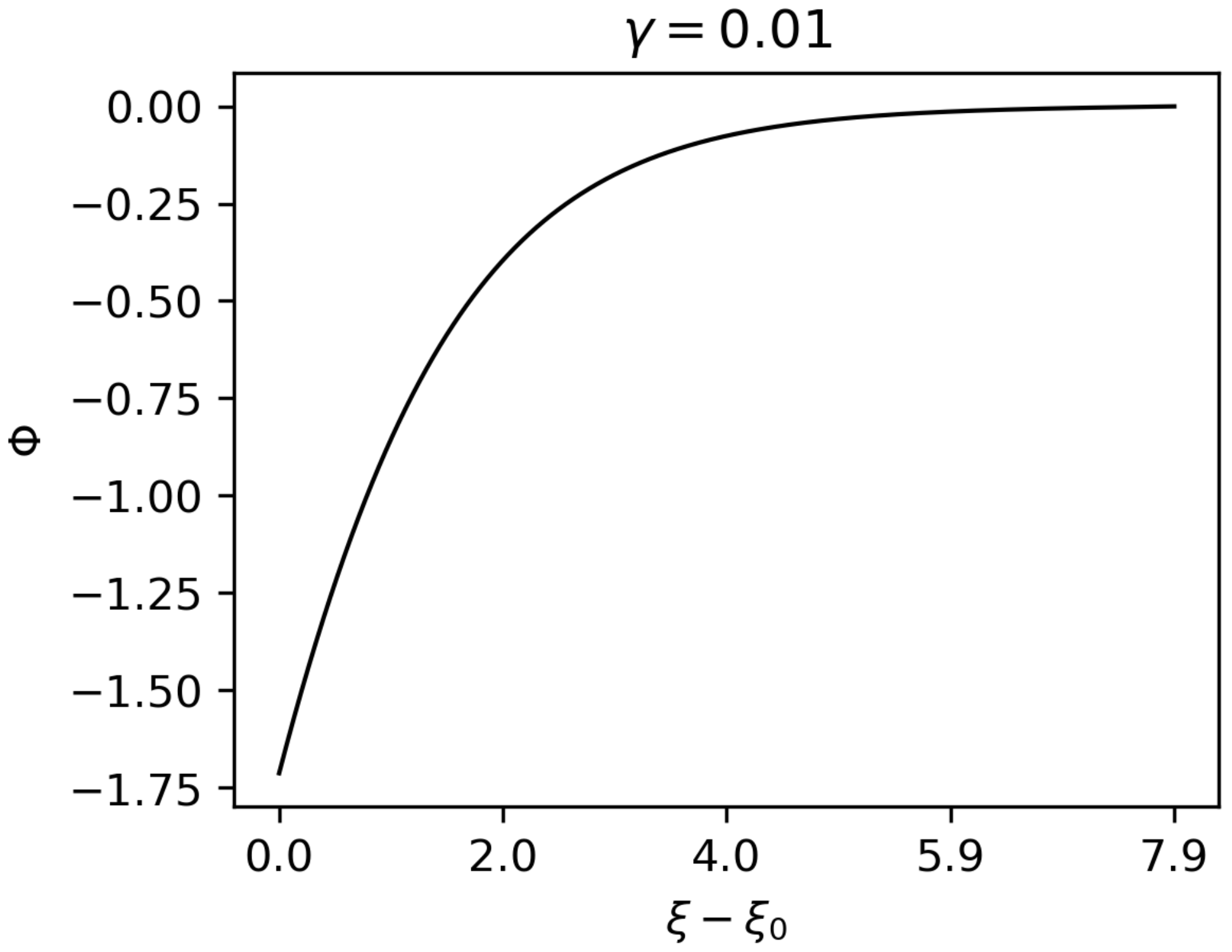}
    \end{subfigure}
    \begin{subfigure}{0.4\textwidth}
         \includegraphics[width=\textwidth]{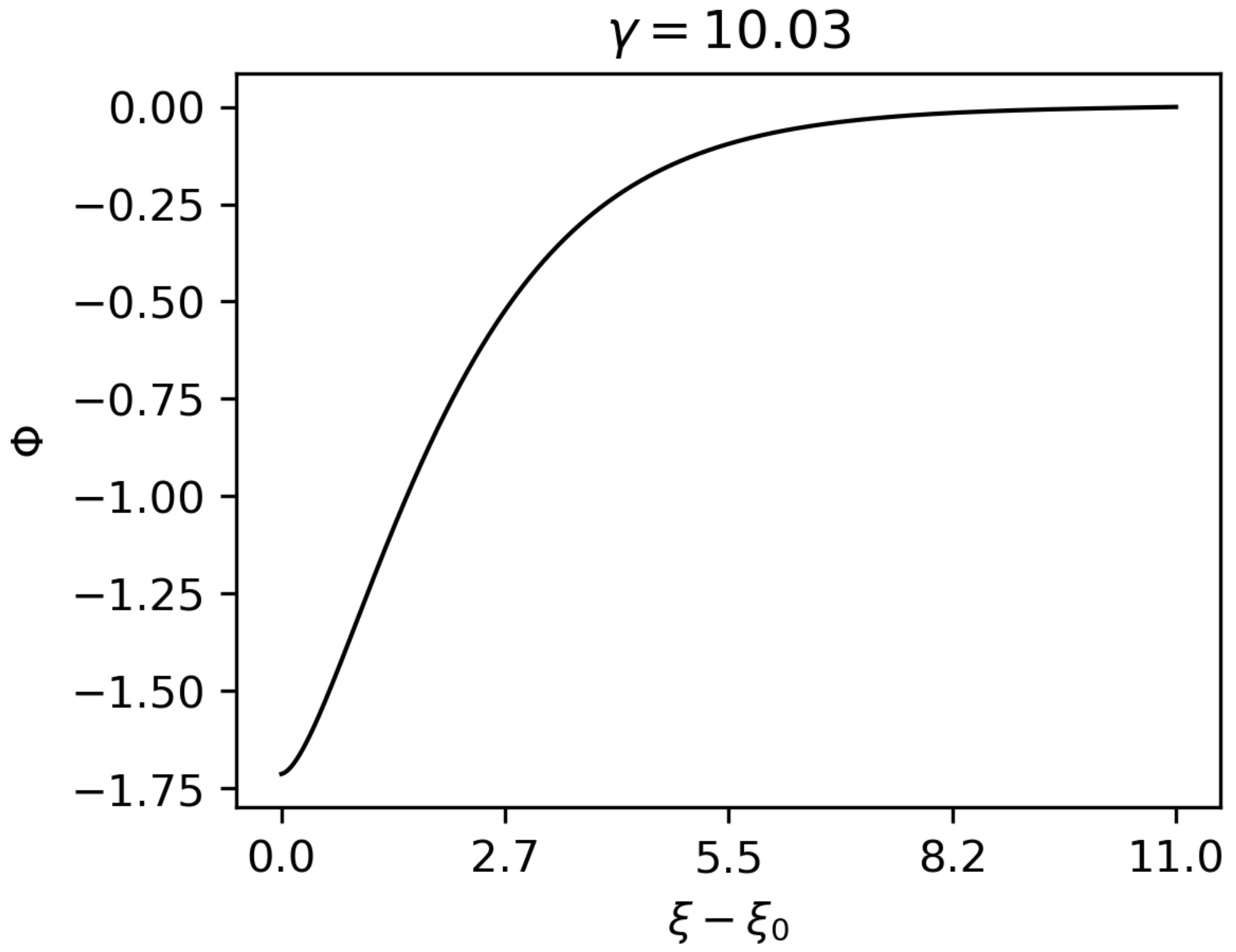}
    \end{subfigure}
    \hfill
    \begin{subfigure}{0.4\textwidth}
         \includegraphics[width=\textwidth]{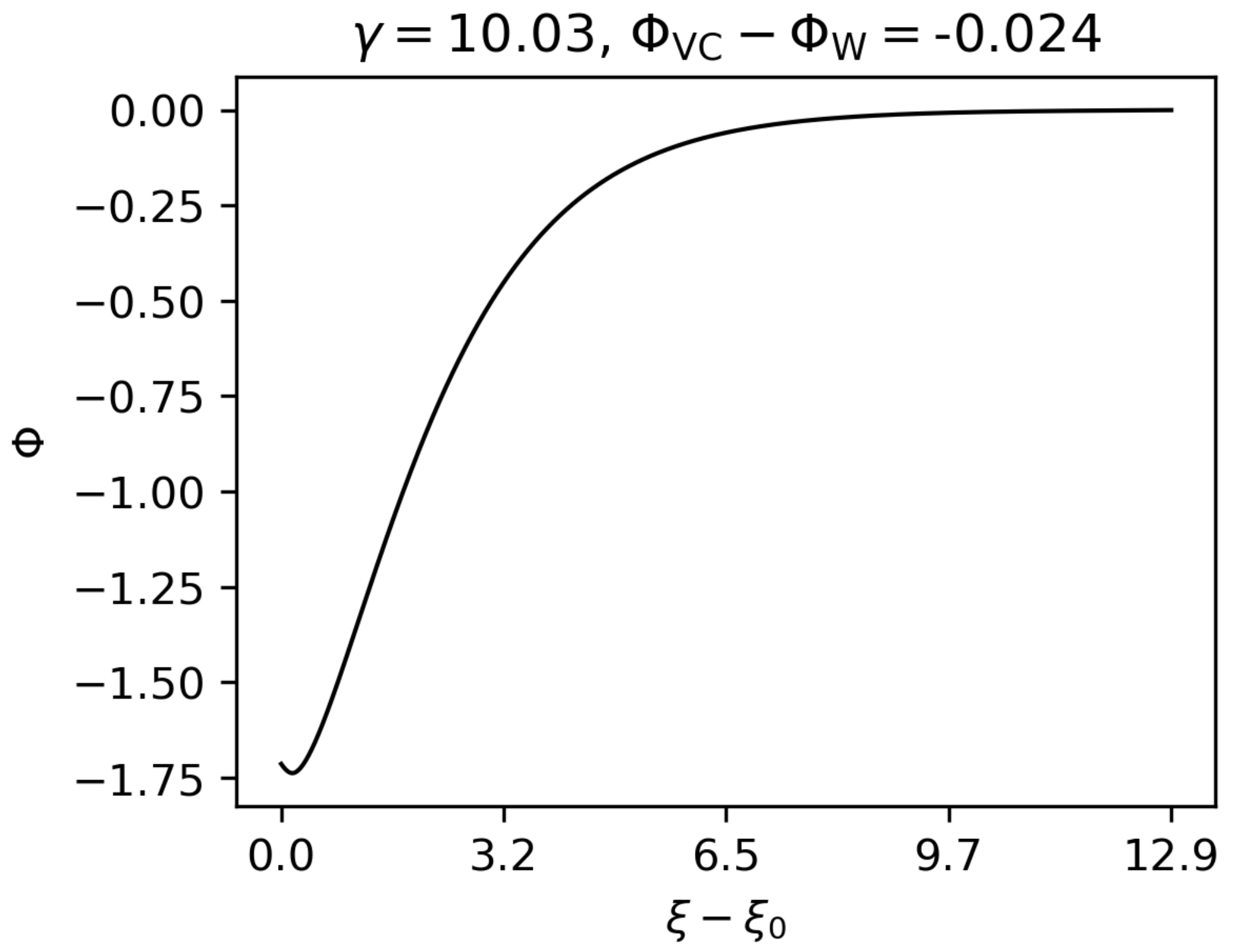}
    \end{subfigure}
    \begin{subfigure}{0.4\textwidth}
         \includegraphics[width=\textwidth]{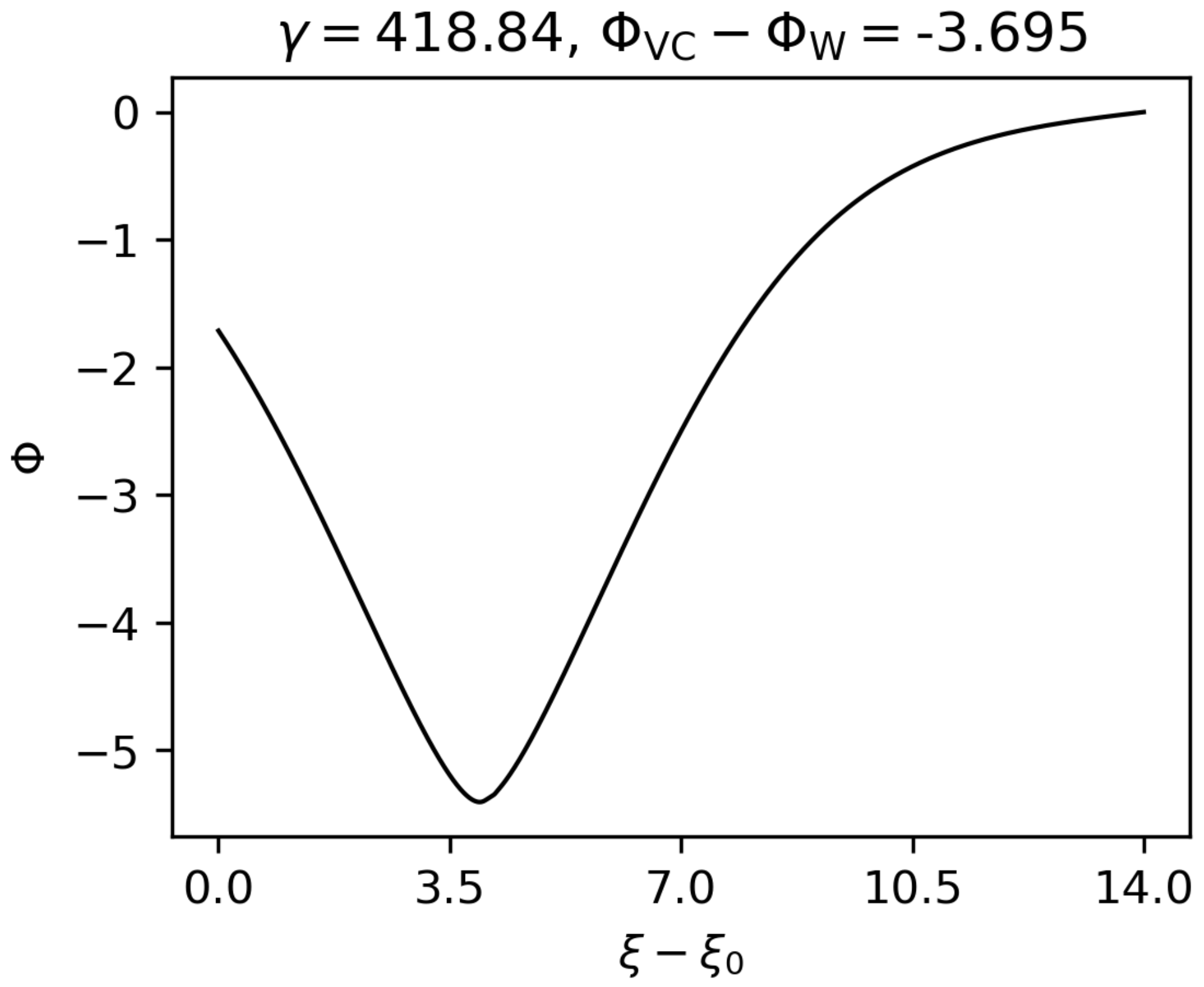}
    \end{subfigure}
\caption{Example potential sheaths at extremes of the \(\ps*{\gamma,\,\Phi_\text{VC} - \Phi_\text{W}}\) parameter space, for material work function \(\phi_\text{w} = \SI{3.0}{eV}\); (top) potential spaces without virtual cathode, (bottom) potential spaces with virtual cathode.}
\label{fig:CarLowExamplePotentials}
\end{figure}

For the Cartesian parameter spaces above, we search for valid sheathes for current ratio \(\gamma\) between 0 and 500, and find that valid sheathes seem to mostly lie along a nonlinear curve. However, some sheathes also exist under that curve near particular \(\gamma\) that support multiple virtual cathode magnitudes in which lower \(\Phi_\text{VC} - \Phi_\text{W}\) yields lower \(M_\text{min}\). The Cartesian parameter spaces are notably different from the cylindrical and spherical ones that follow, in that the largest \(\gamma\) that yields a sheath without a virtual cathode does not exceed \(\gamma = 33.08\) in Fig. \ref{fig:CarHighExamplePotentials} and \(\gamma = 10.03\) in Fig. \ref{fig:CarLowExamplePotentials}, beyond which sheathes only exist with a virtual cathode. Whereas, in the cylindrical and spherical parameter spaces that follow, typically a \(\gamma\) that gives a sheath with a virtual cathode will also give a sheath without one. Regarding the minimum Mach number \(M_\text{min}\) for the Cartesian parameter space, \(M_\text{min}\) can reach or potentially exceed 50 for large enough \(\gamma\). Also, \(M_\text{min}\) converges to a minimum or very near it for sheathes without a virtual cathode; on the other hand, if a virtual cathode is present, \(M_\text{min}\) varies quite significantly. 

Considering net electron current, we observe that it increases with increase in current ratio \(\gamma\), which is intuitive because higher wall current should be directly proportional to the net current, as the net current takes a contribution of wall current that is directly proportional to \(\gamma\). However, there are cases of \(\gamma\) that allow multiple virtual cathodes, in which decreased \(\Phi_\text{VC} - \Phi_\text{W}\) leads to lower net electron current, which is reasonable because there is a greater potential barrier at the same \(\gamma\). On the effect of low versus high work function, we see that low work function leads to lower possible \(\Phi_\text{VC} - \Phi_\text{W}\) compared to the case of high work function.

\subsection{Cylindrical model spaces} \label{Cylindrical Model Spaces}

\subsubsection{Cylindrical model spaces for high work function \(\phi_\text{w} = \SI{4.5}{eV}\)}

\begin{figure}[H]
    \centering
    \includegraphics[width=0.675\linewidth]{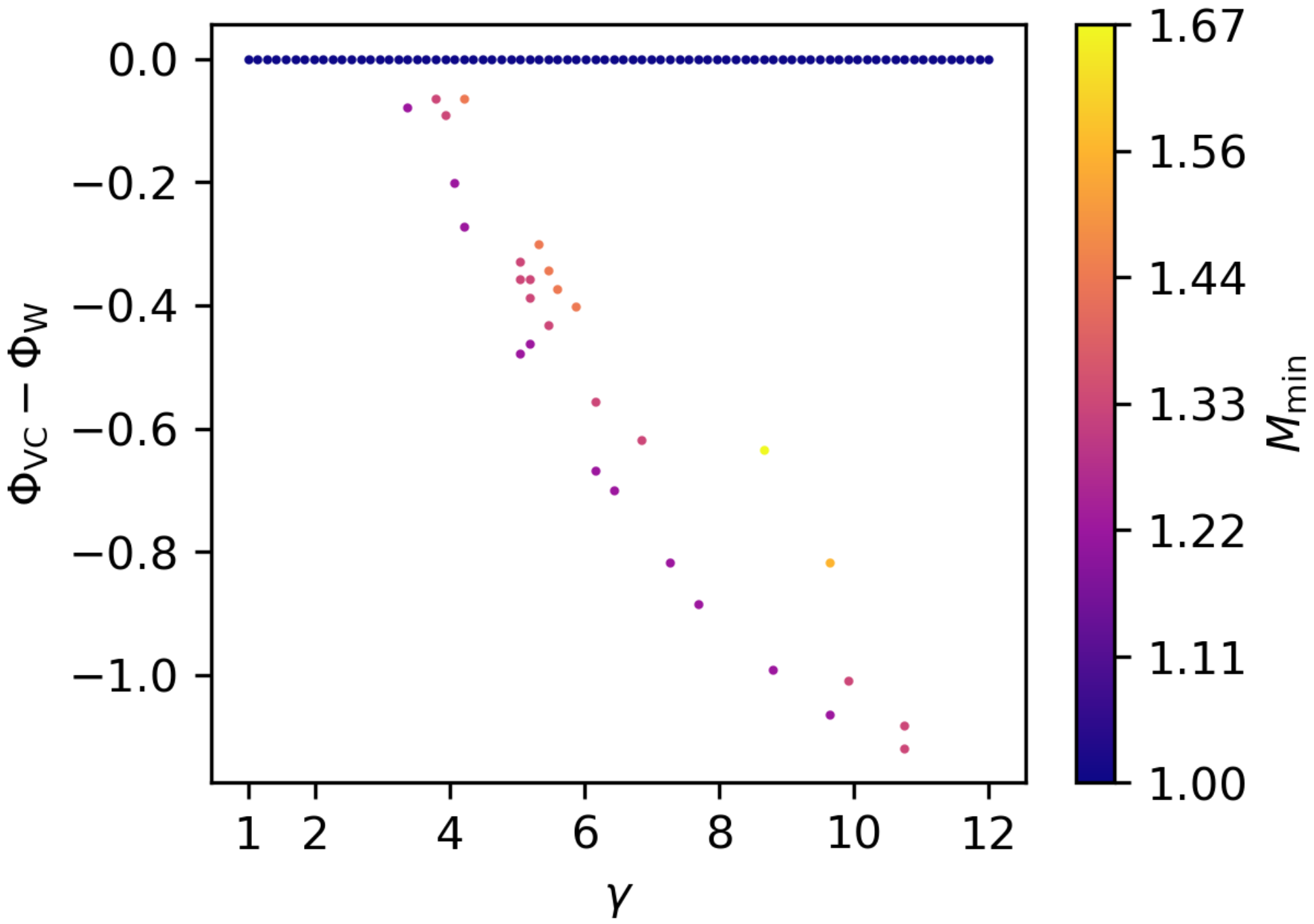}
    \caption{\(M_\text{min}\) as a function of current ratio \(\gamma \in \bs*{1,\,12}\) and virtual cathode potential with respect to the wall \(\Phi_\text{VC} - \Phi_\text{W}\in \bs*{0,-1.25}\), for material work function \(\phi_\text{w} = \SI{4.5}{eV}\).}
    \label{fig:CylHighM}
\end{figure}

\begin{figure}[H]
    \centering
    \includegraphics[width=0.675\linewidth]{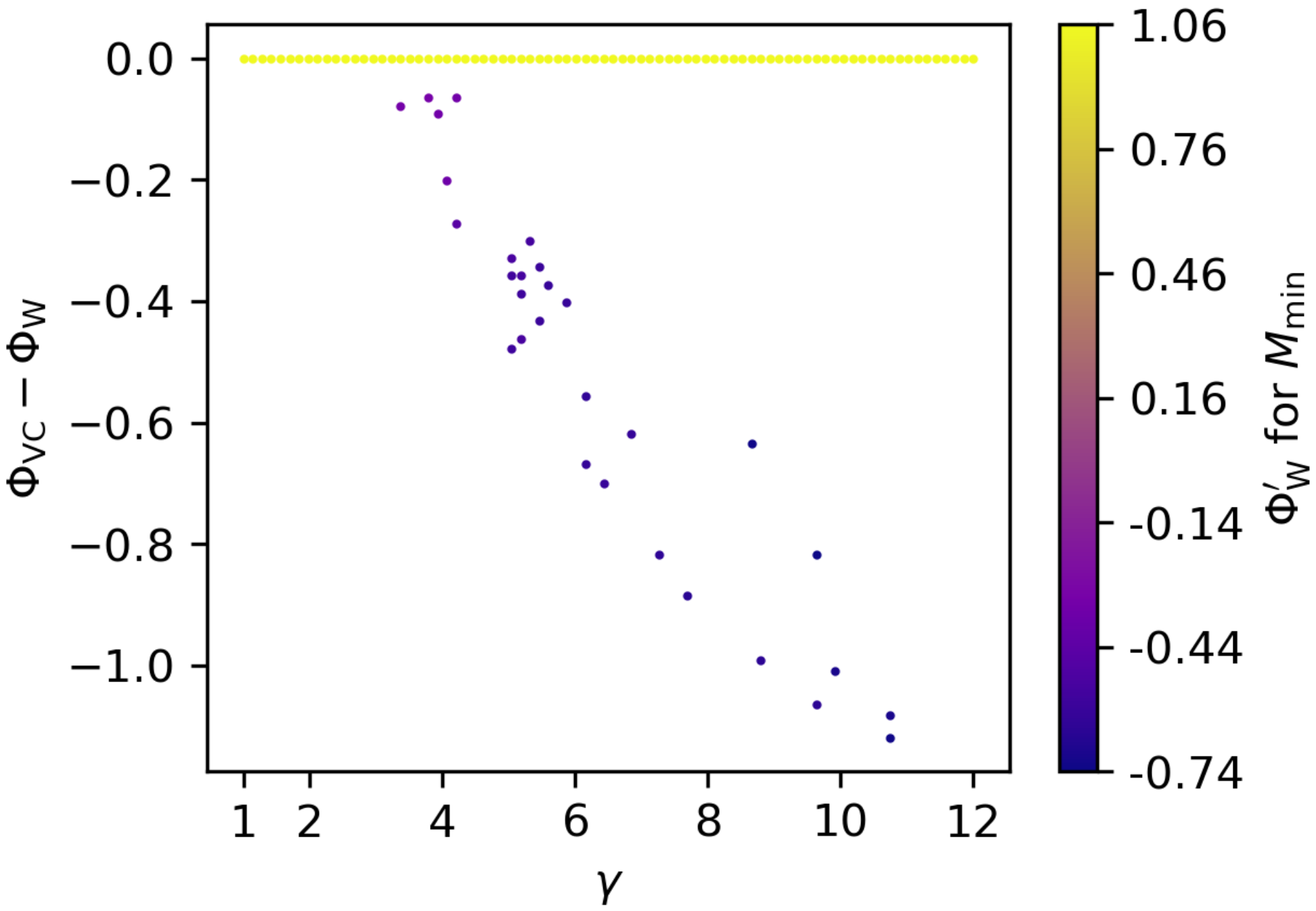}
    \caption{\(\Phi'_\text{W}\) for \(M_\text{min}\) in Fig. \ref{fig:CylHighM} as a function of current ratio \(\gamma \in \bs*{1,\,12}\) and virtual cathode potential with respect to the wall \(\Phi_\text{VC} - \Phi_\text{W}\in \bs*{0,-1.25}\), for material work function \(\phi_\text{w} = \SI{4.5}{eV}\).}
    \label{fig:CylHighPhiWPrime}
\end{figure}

\begin{figure}[H]
    \centering
    \includegraphics[width=0.7\linewidth]{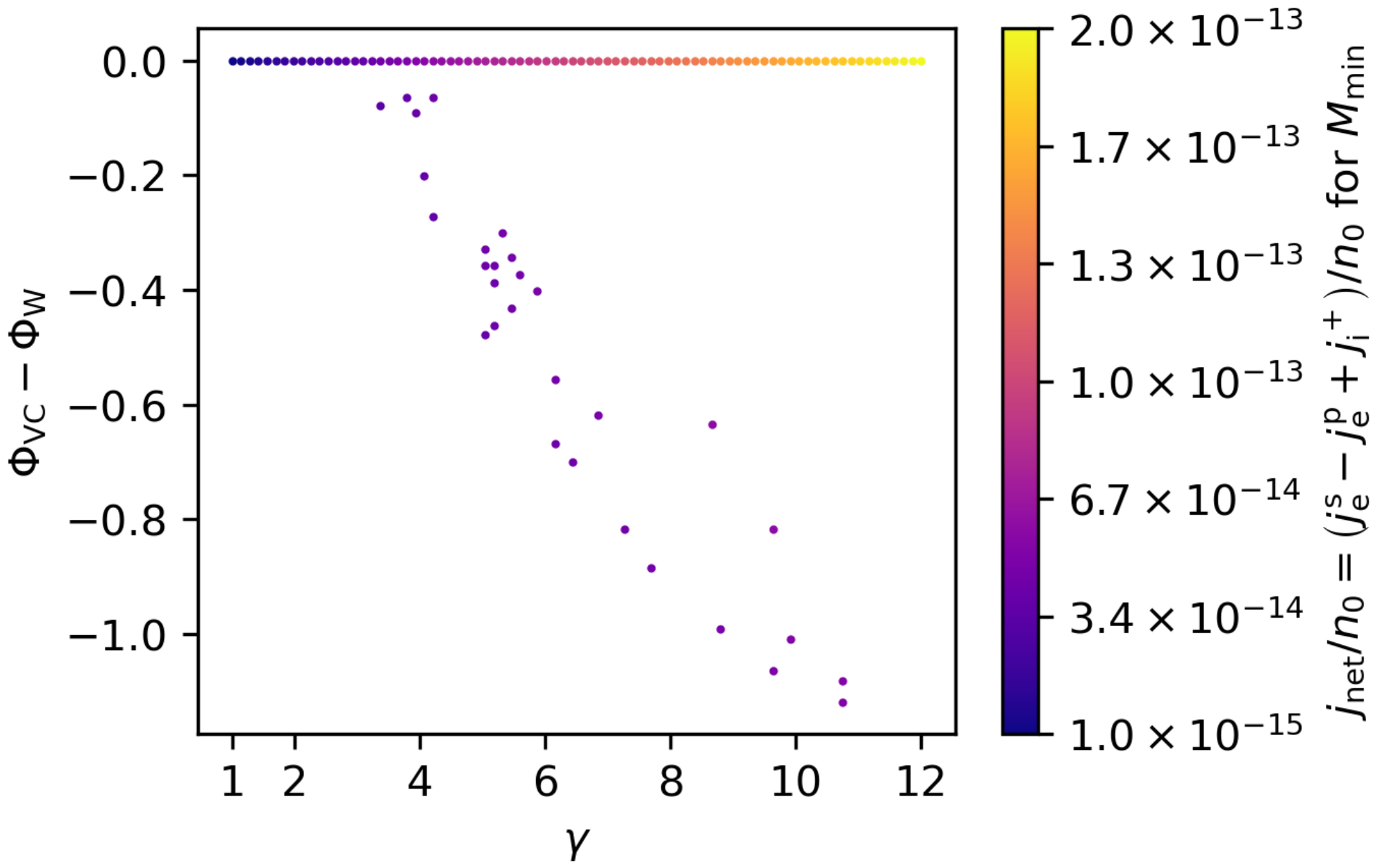}
    \caption{\(j_\text{net}/n_0\) for \(M_\text{min}\) in Fig. \ref{fig:CylHighM} as a function of current ratio \(\gamma \in \bs*{1,\,12}\) and virtual cathode potential with respect to the wall \(\Phi_\text{VC} - \Phi_\text{W}\in \bs*{0,-1.25}\), for material work function \(\phi_\text{w} = \SI{4.5}{eV}\).}
    \label{fig:CylHighJnet}
\end{figure}

\begin{figure}[H]
    \centering
    \begin{subfigure}{0.4\textwidth}
         \includegraphics[width=\textwidth]{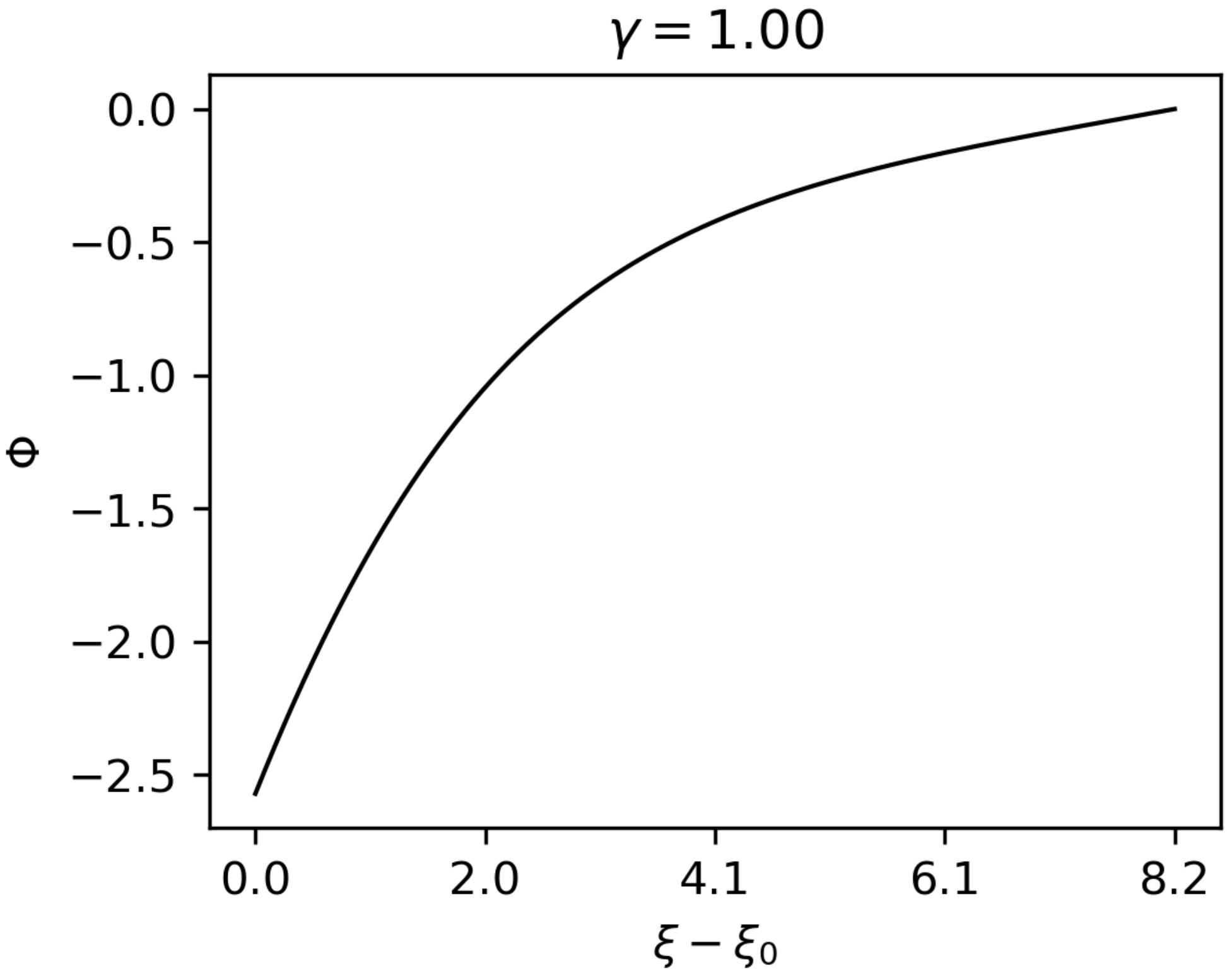}
    \end{subfigure}
    \begin{subfigure}{0.4\textwidth}
         \includegraphics[width=\textwidth]{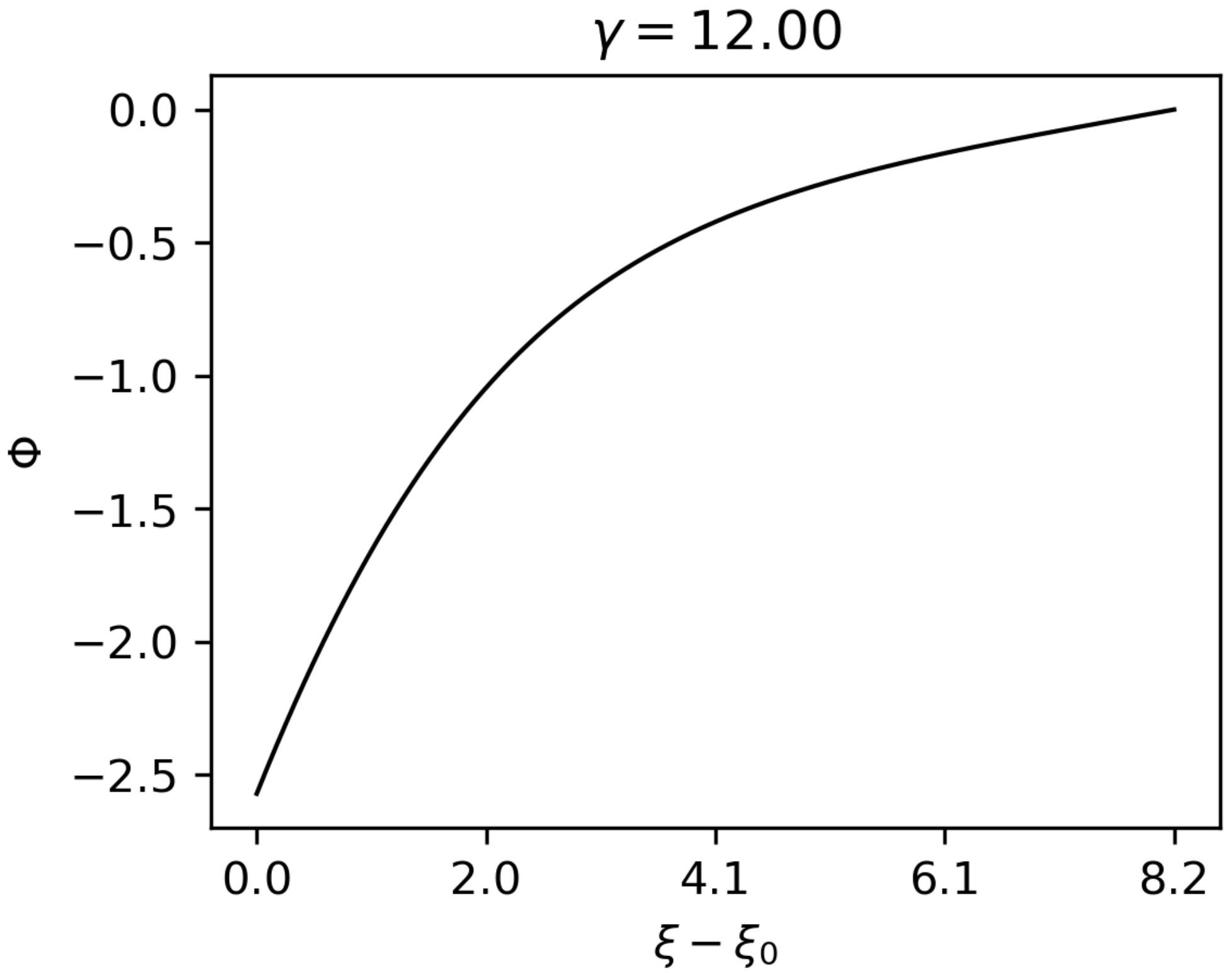}
    \end{subfigure}
    \hfill
    \begin{subfigure}{0.4\textwidth}
         \includegraphics[width=\textwidth]{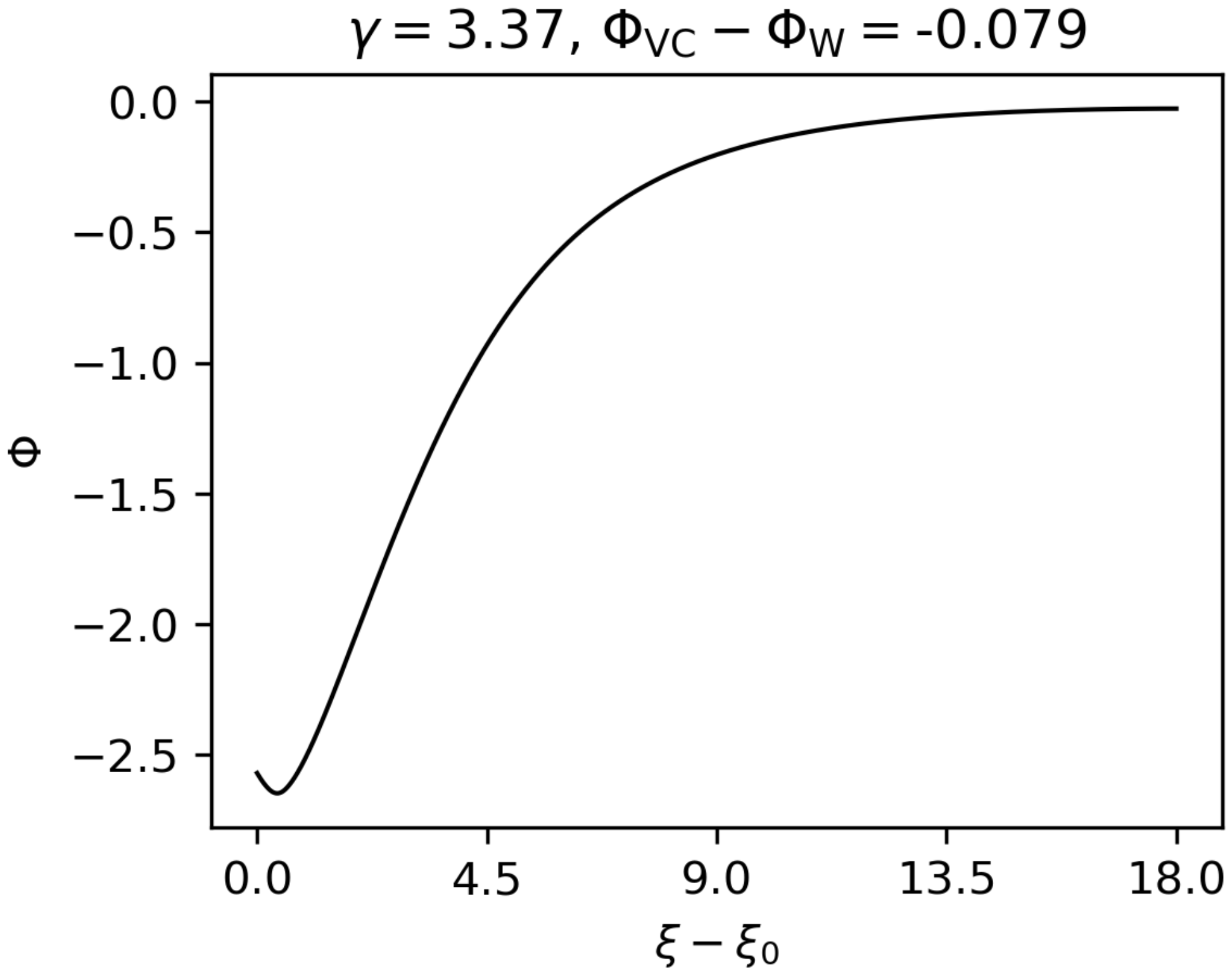}
    \end{subfigure}
    \begin{subfigure}{0.4\textwidth}
         \includegraphics[width=\textwidth]{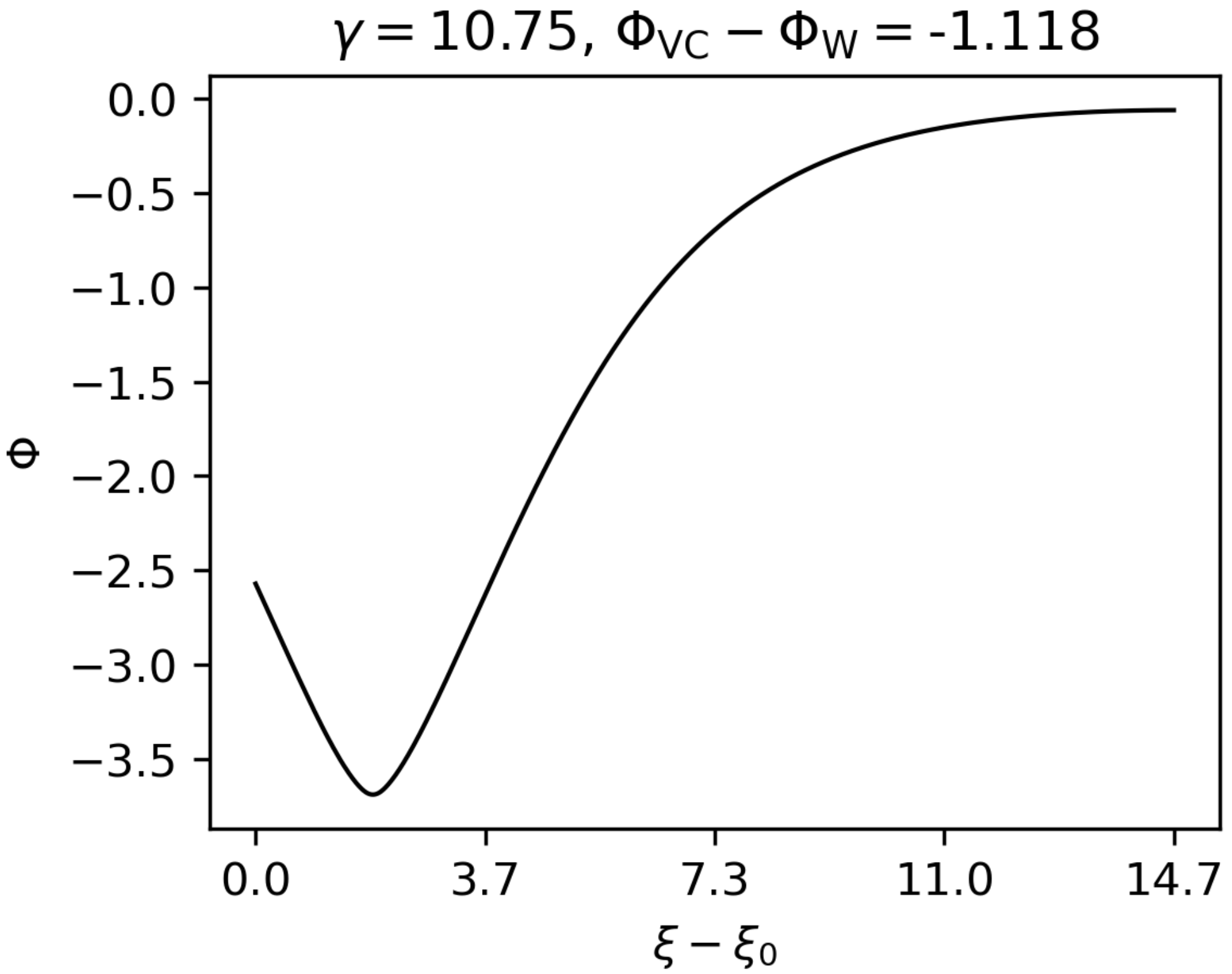}
    \end{subfigure}
\caption{Example potential sheaths at extremes of the \(\ps*{\gamma,\,\Phi_\text{VC} - \Phi_\text{W}}\) parameter space, for material work function \(\phi_\text{w} = \SI{4.5}{eV}\); (top) potential spaces without virtual cathode, (bottom) potential spaces with virtual cathode.}
\label{fig:CylHighExamplePotentials}
\end{figure}

\subsubsection{Cylindrical model spaces for low work function \(\phi_\text{w} = \SI{3.0}{eV}\)}
\begin{figure}[H]
    \centering
    \includegraphics[width=0.7\linewidth]{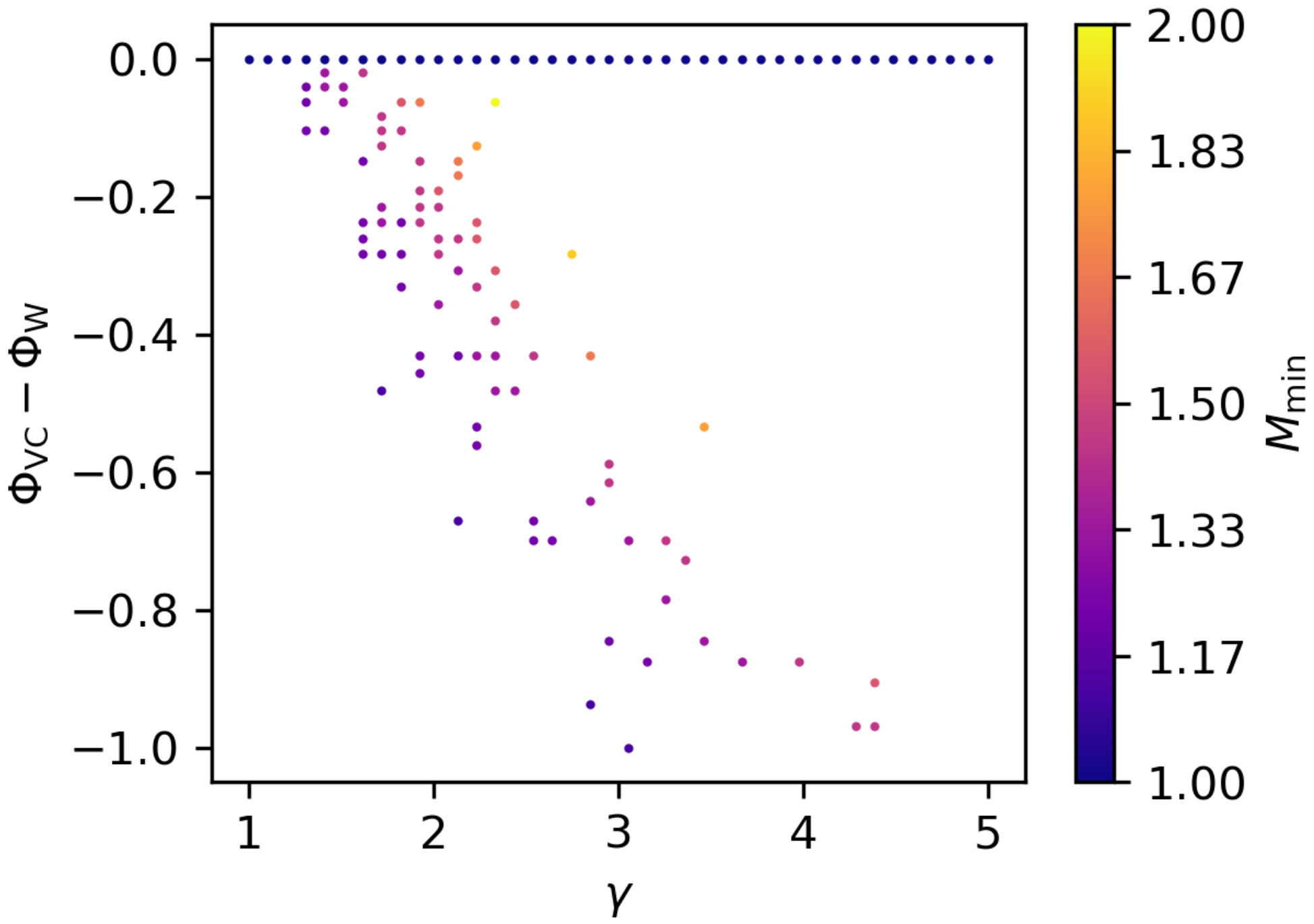}
    \caption{\(M_\text{min}\) as a function of current ratio \(\gamma \in \bs*{1,\,12}\) and virtual cathode potential with respect to the wall \(\Phi_\text{VC} - \Phi_\text{W}\in \bs*{0,-1.25}\), for material work function \(\phi_\text{w} = \SI{3.0}{eV}\).}
    \label{fig:CylLowM}
\end{figure}

\begin{figure}[H]
    \centering
    \includegraphics[width=0.7\linewidth]{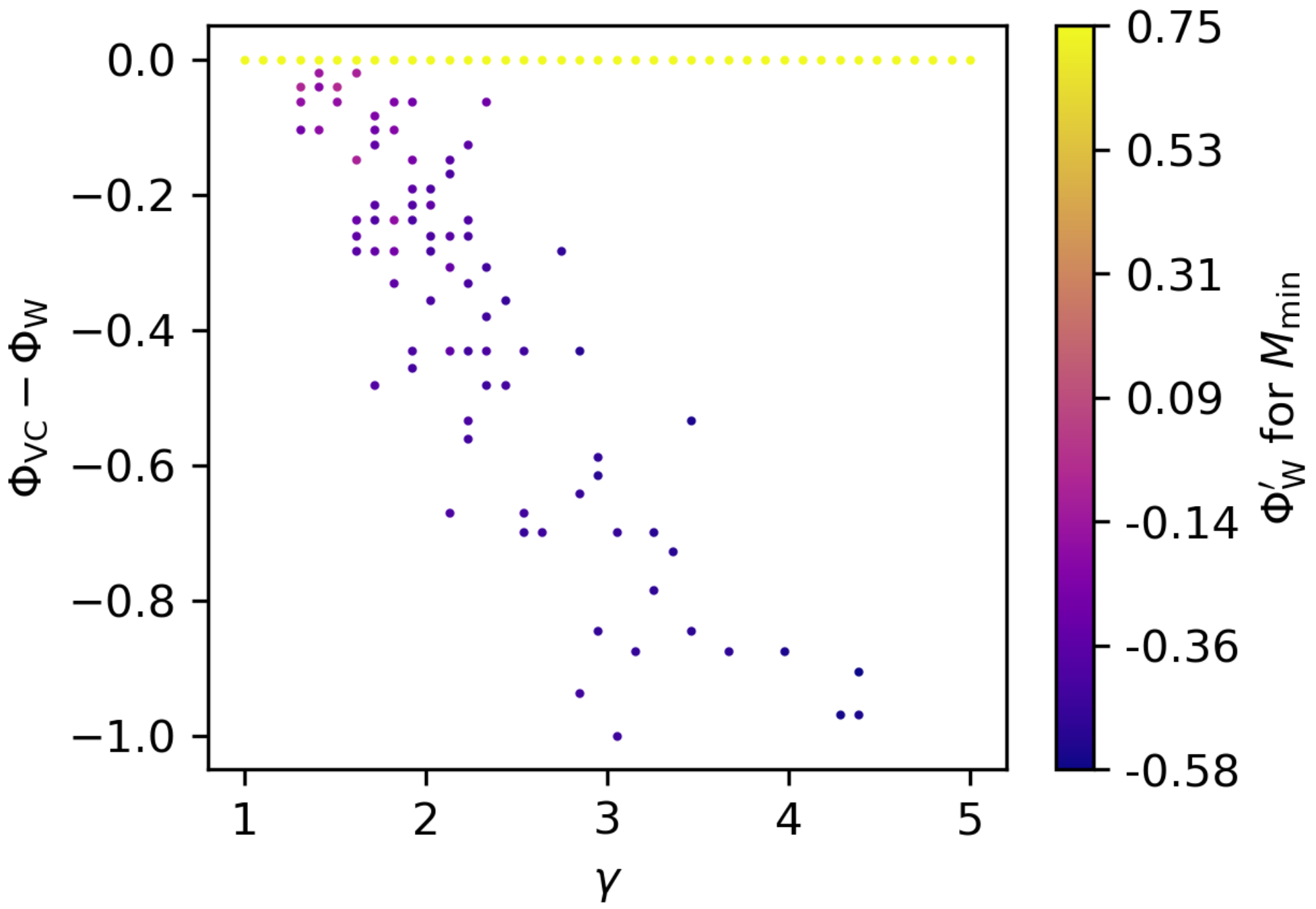}
    \caption{\(\Phi'_\text{W}\) for \(M_\text{min}\) in Fig. \ref{fig:CylLowM} as a function of current ratio \(\gamma \in \bs*{1,\,12}\) and virtual cathode potential with respect to the wall \(\Phi_\text{VC} - \Phi_\text{W}\in \bs*{0,-1.25}\), for material work function \(\phi_\text{w} = \SI{3.0}{eV}\).}
    \label{fig:CylLowPhiWPrime}
\end{figure}

\begin{figure}[H]
    \centering
    \includegraphics[width=0.7\linewidth]{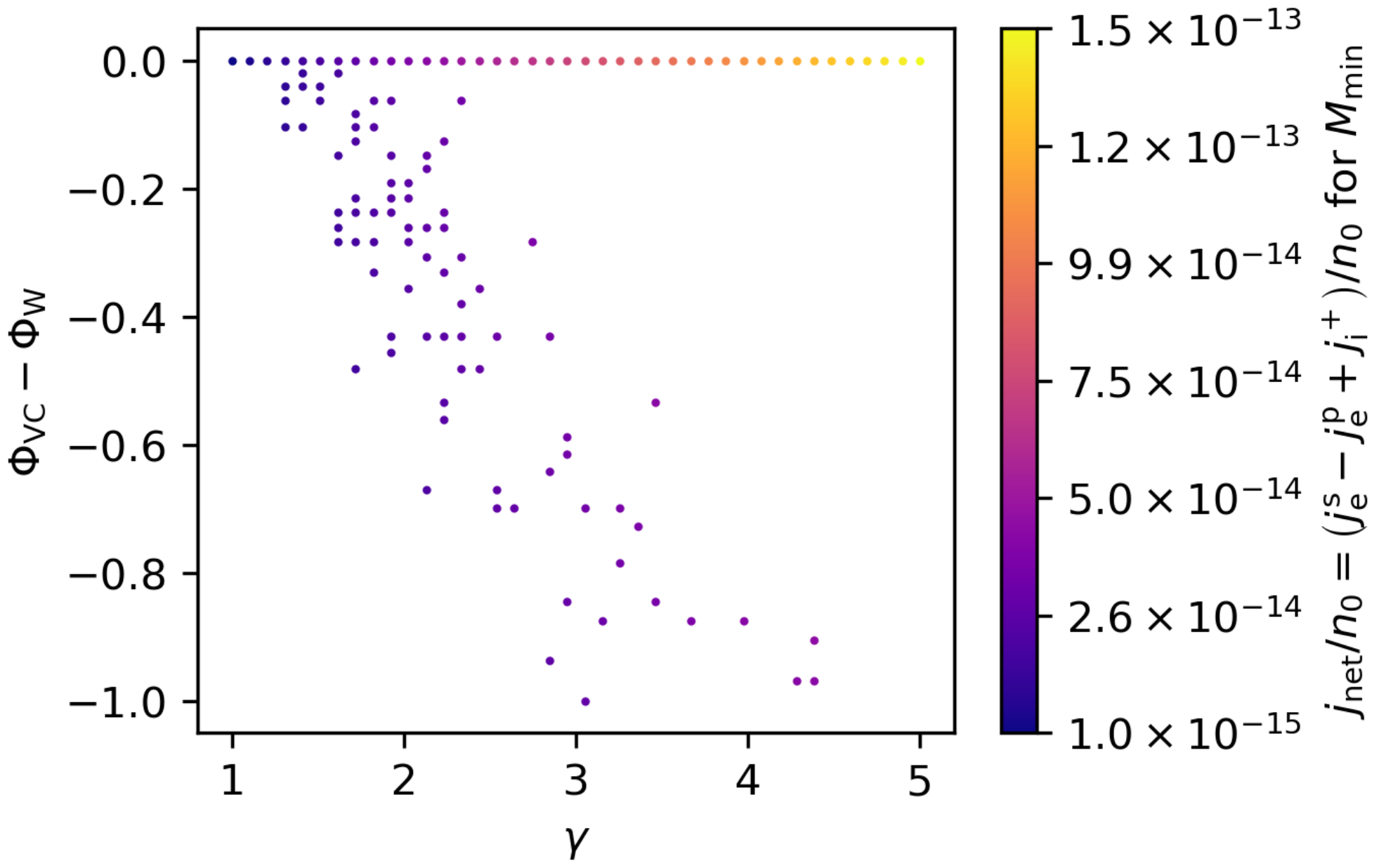}
    \caption{\(j_\text{net}/n_0\) for \(M_\text{min}\) in Fig. \ref{fig:CylLowM} as a function of current ratio \(\gamma \in \bs*{1,\,12}\) and virtual cathode potential with respect to the wall \(\Phi_\text{VC} - \Phi_\text{W}\in \bs*{0,-1.25}\), for material work function \(\phi_\text{w} = \SI{3.0}{eV}\).}
    \label{fig:CylLowJnet}
\end{figure}

\begin{figure}[H]
    \centering
    \begin{subfigure}{0.4\textwidth}
         \includegraphics[width=\textwidth]{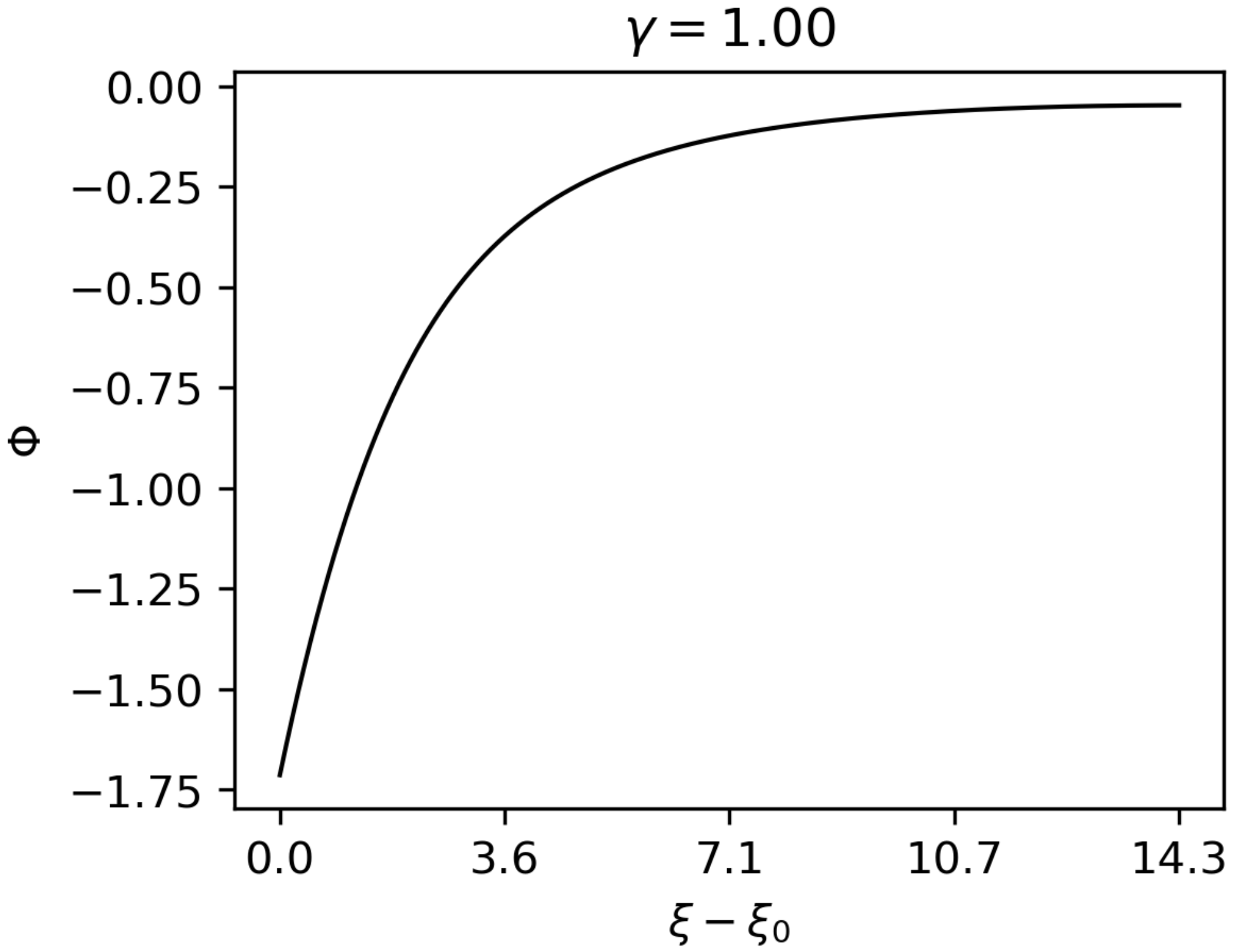}
    \end{subfigure}
    \begin{subfigure}{0.4\textwidth}
         \includegraphics[width=\textwidth]{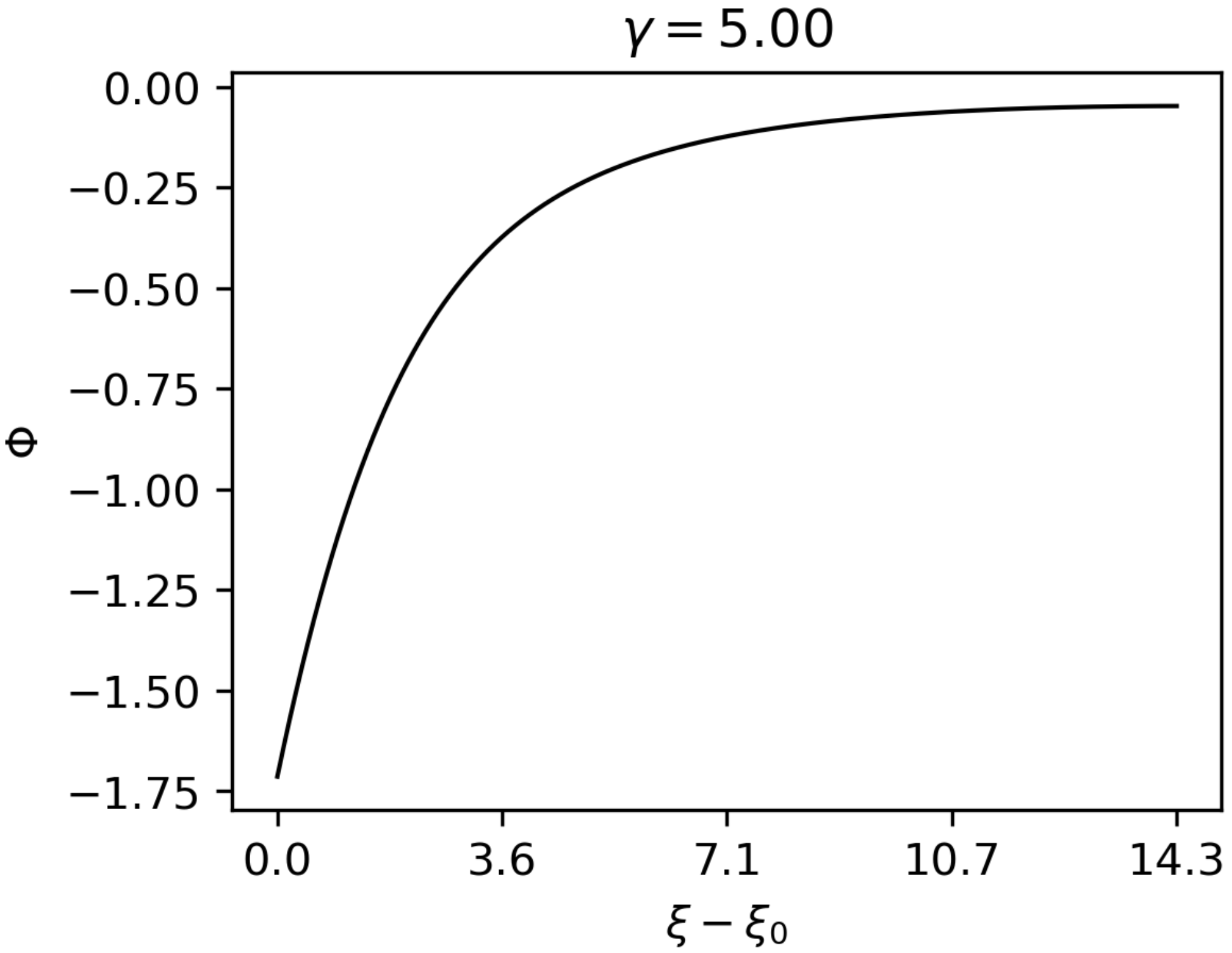}
    \end{subfigure}
    \hfill
    \begin{subfigure}{0.4\textwidth}
         \includegraphics[width=\textwidth]{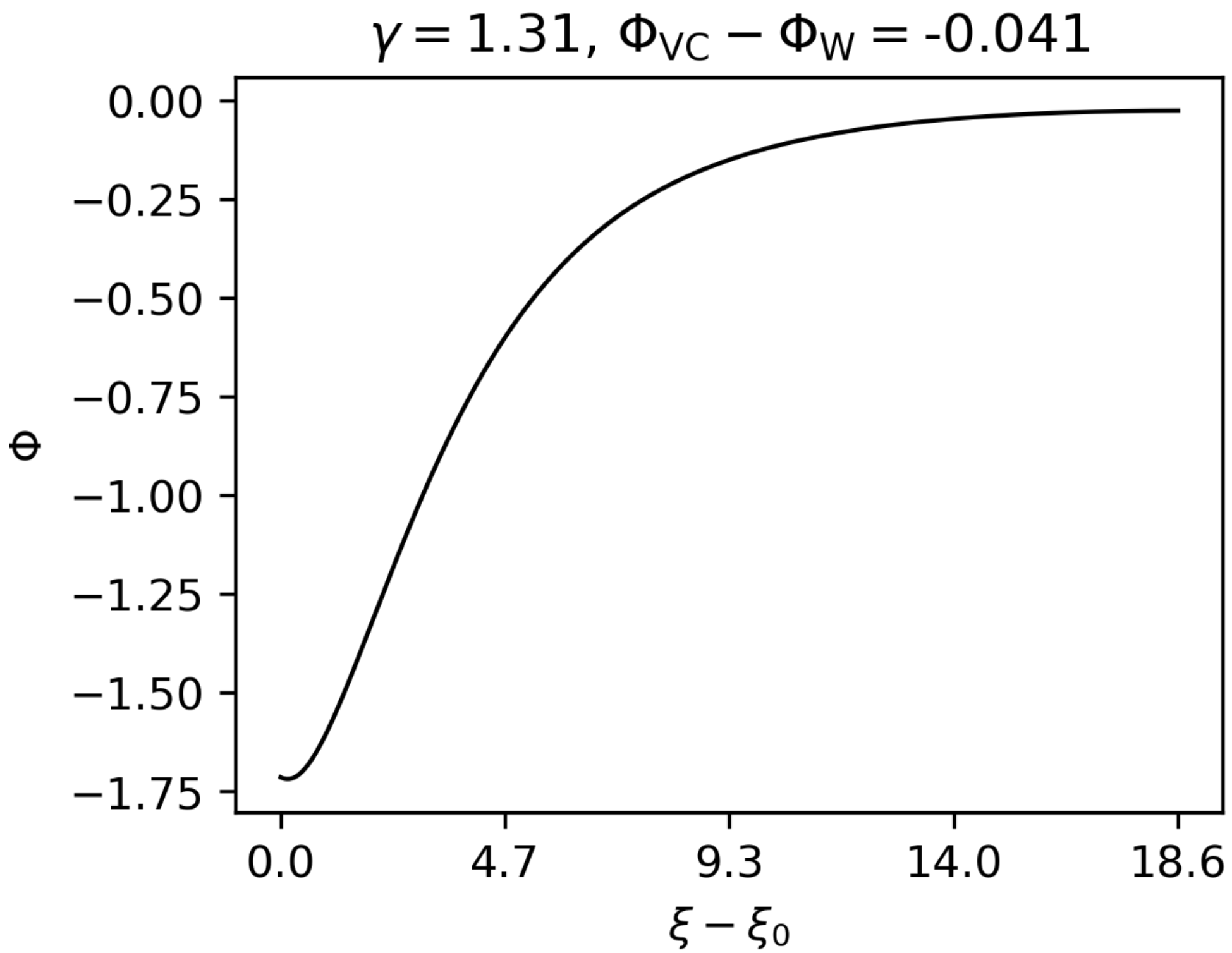}
    \end{subfigure}
    \begin{subfigure}{0.4\textwidth}
         \includegraphics[width=\textwidth]{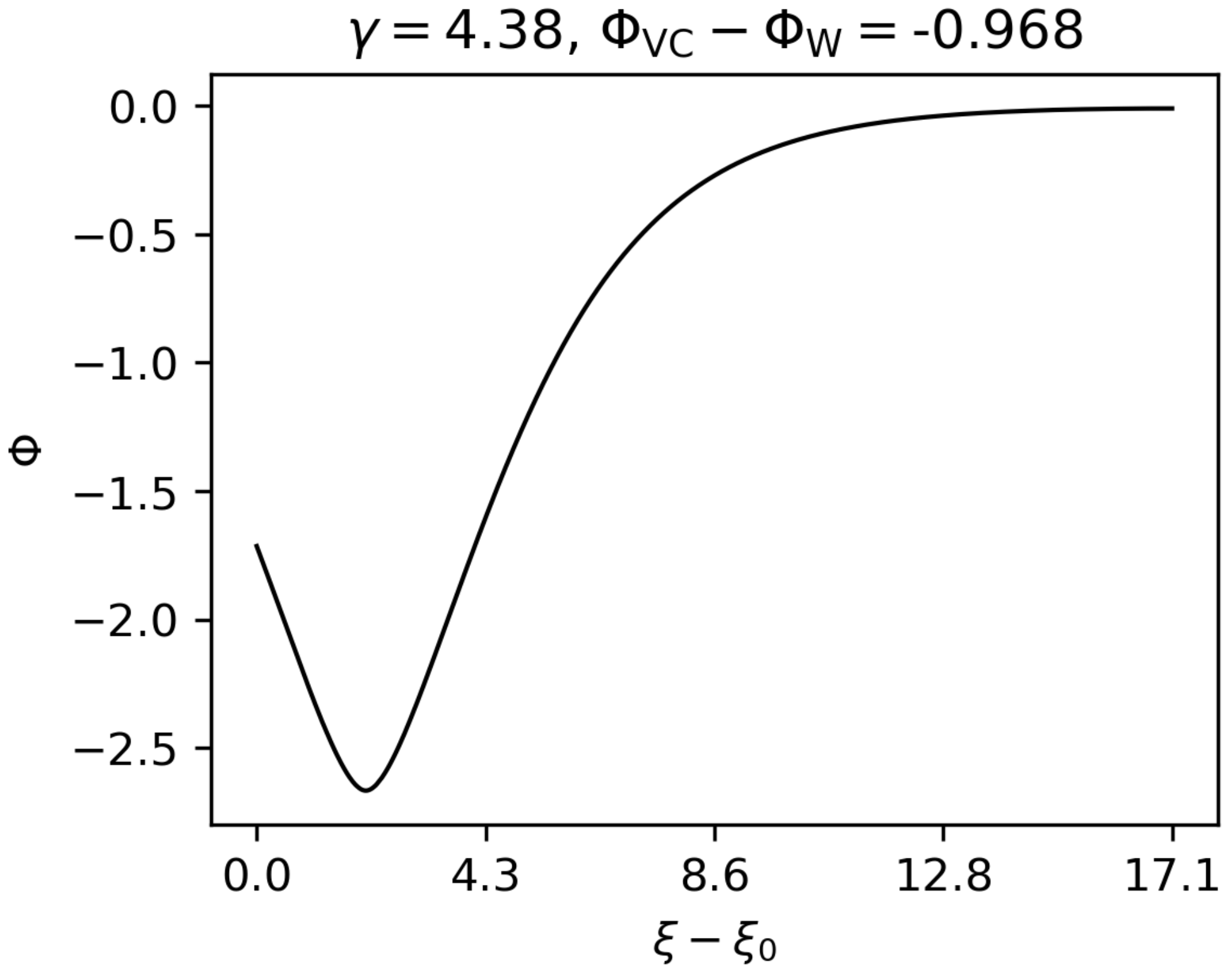}
    \end{subfigure}
\caption{Example potential sheaths at extremes of the \(\ps*{\gamma,\,\Phi_\text{VC} - \Phi_\text{W}}\) parameter space, for material work function \(\phi_\text{w} = \SI{3.0}{eV}\); (top) potential spaces without virtual cathode, (bottom) potential spaces with virtual cathode.}
\label{fig:CylLowExamplePotentials}
\end{figure}

Next, as we search for valid sheathes in the cylindrical and spherical coordinate models, we discover that the allowable range of \(\gamma\) narrows significantly, to between 1 and 12 for cylindrical coordinates in subsection \ref{Cylindrical Model Spaces} above, and to between 0.5 and 2 for spherical coordinates in subsection \ref{Spherical Model Spaces} below. For the cylindrical and spherical parameter spaces, we do not see the same obvious trend of sheathes existing along a curve that we saw in the Cartesian spaces. And while  \(M_\text{min}\) still converges to a minimum or near it for sheathes without a virtual cathode as in the Cartesian spaces, for the cylindrical and spherical parameter spaces \(M_\text{min}\) is now restricted between 1 to 2, and \(M_\text{min}\) seems to decrease with decreasing \(\Phi_\text{VC} - \Phi_\text{W}\) and increase with increasing \(\gamma\).

On the net electron current, we find the same phenomena discussed for the Cartesian spaces in subsection \ref{Cartesian Model Spaces}, in which net electron current increases with current ratio \(\gamma\) and decreases with decrease in \(\Phi_\text{VC} - \Phi_\text{W}\). While the Cartesian model cases with virtual cathode reach a maximum net current density over \(n_0\) value of \(1.9\times 10^{-13}\), the cylindrical and spherical model cases with virtual cathode only reach a maximum on the order of \(10^{-15}\) to \(10^{-14}\), but we recognize that this may be in part due to the Cartesian system supporting a larger maximum \(\gamma\) of at least 500. On the effect of low versus high work function, the cylindrical parameter spaces show that low work function limits \(\Phi_\text{VC} - \Phi_\text{W}\) to a slightly higher signed value, compared to the case of high work function. This is opposite to the effect observed in the spherical spaces, as the spherical spaces share the same behavior present for the Cartesian spaces in which lower work function leads to lower possible \(\Phi_\text{VC} - \Phi_\text{W}\). Also, for the cylindrical and spherical model spaces, the size of the \(\gamma\) range and the maximum \(\gamma\) itself are both smaller for the low work function case than in the high work function case.

\subsection{Spherical model spaces} \label{Spherical Model Spaces}

\subsubsection{Spherical model spaces for high work function \(\phi_\text{w} = \SI{4.5}{eV}\)}
\begin{figure}[H]
    \centering
    \includegraphics[width=0.7\linewidth]{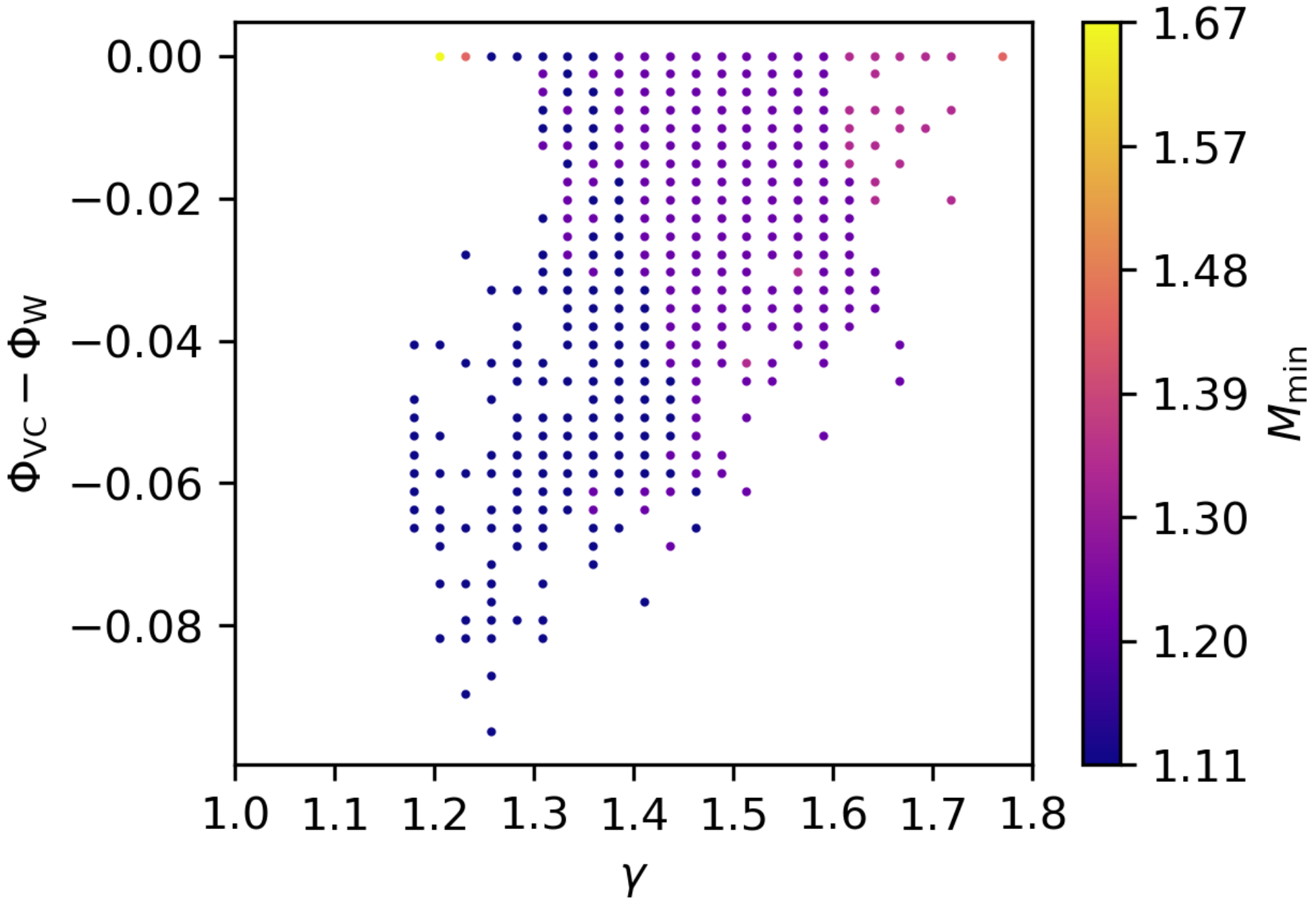}
    \caption{\(M_\text{min}\) as a function of current ratio \(\gamma \in \bs*{1,\,2}\) and virtual cathode potential with respect to the wall \(\Phi_\text{VC} - \Phi_\text{W}\in \bs*{0,-0.1}\), for material work function \(\phi_\text{w} = \SI{4.5}{eV}\).}
    \label{fig:SphHighM}
\end{figure}

\begin{figure}[H]
    \centering
    \includegraphics[width=0.675\linewidth]{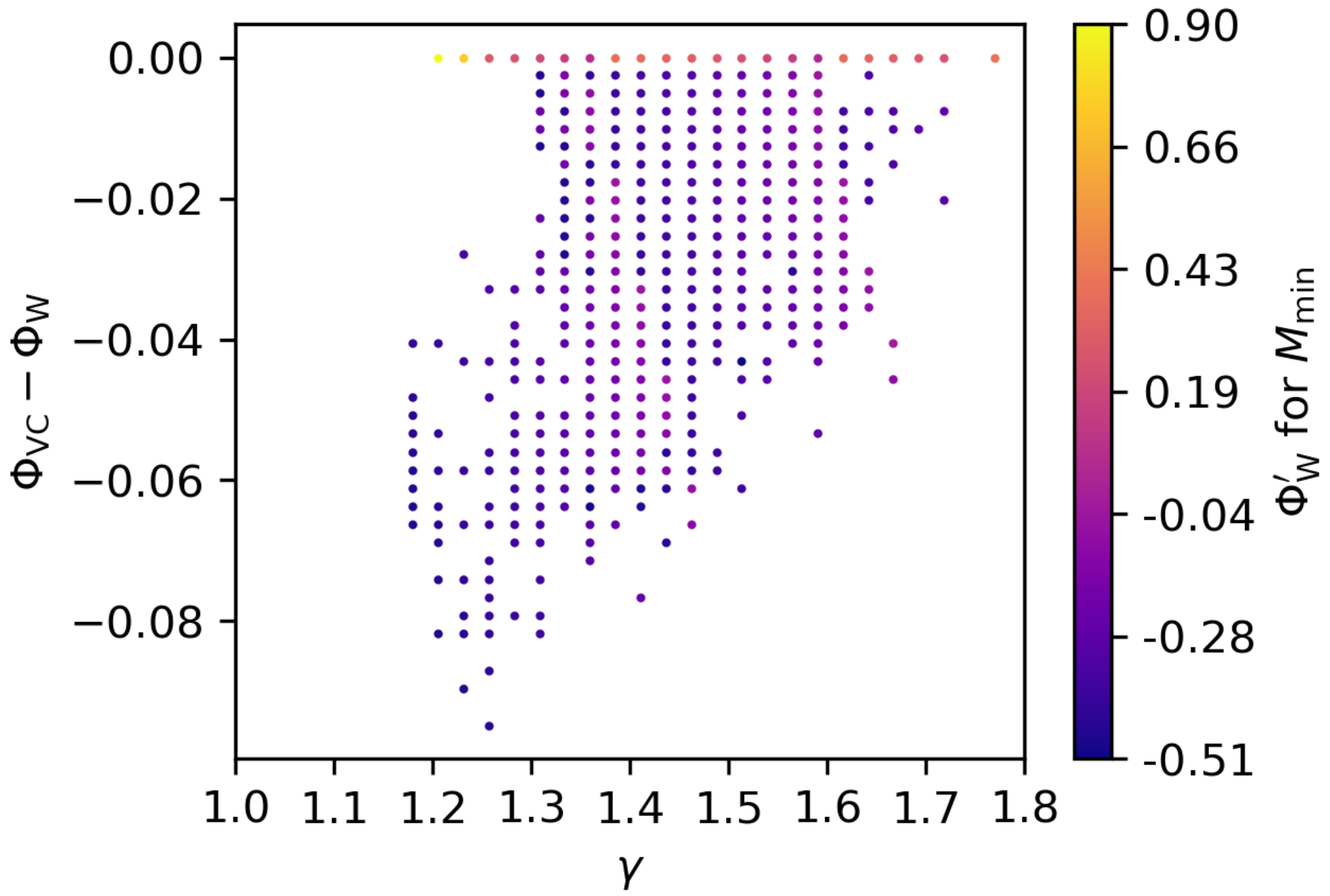}
    \caption{\(\Phi'_\text{W}\) for \(M_\text{min}\) in Fig. \ref{fig:SphHighM} as a function of current ratio \(\gamma \in \bs*{1,\,2}\) and virtual cathode potential with respect to the wall \(\Phi_\text{VC} - \Phi_\text{W}\in \bs*{0,-0.1}\), for material work function \(\phi_\text{w} = \SI{4.5}{eV}\).}
    \label{fig:SphHighPhiWPrime}
\end{figure}

\begin{figure}[H]
    \centering
    \includegraphics[width=0.675\linewidth]{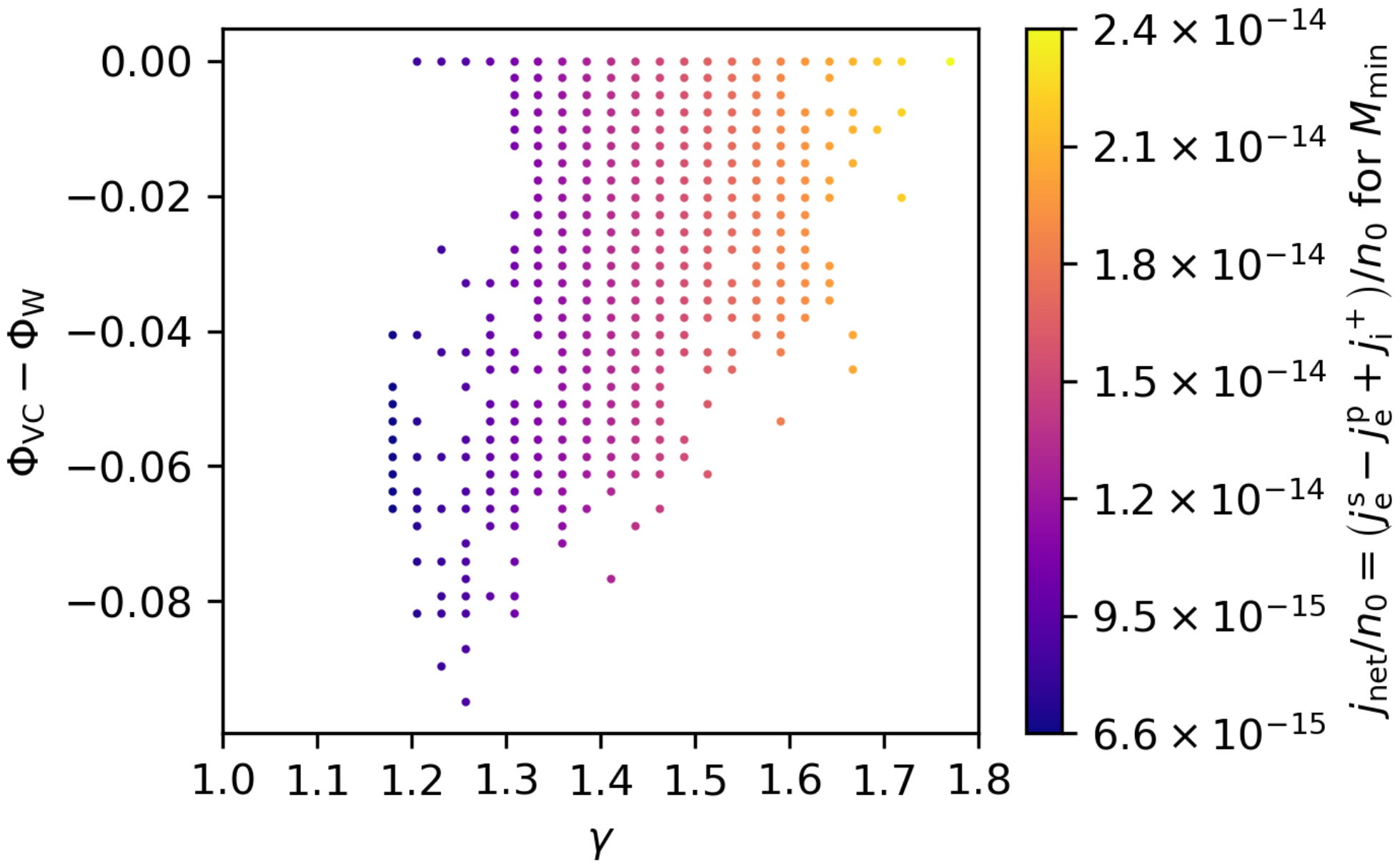}
    \caption{\(j_\text{net}/n_0\) for \(M_\text{min}\) in Fig. \ref{fig:SphHighM} as a function of current ratio \(\gamma \in \bs*{1,\,2}\) and virtual cathode potential with respect to the wall \(\Phi_\text{VC} - \Phi_\text{W}\in \bs*{0,-0.1}\), for material work function \(\phi_\text{w} = \SI{4.5}{eV}\).}
    \label{fig:SphHighJnet}
\end{figure}

\begin{figure}[H]
    \centering
    \begin{subfigure}{0.4\textwidth}
         \includegraphics[width=\textwidth]{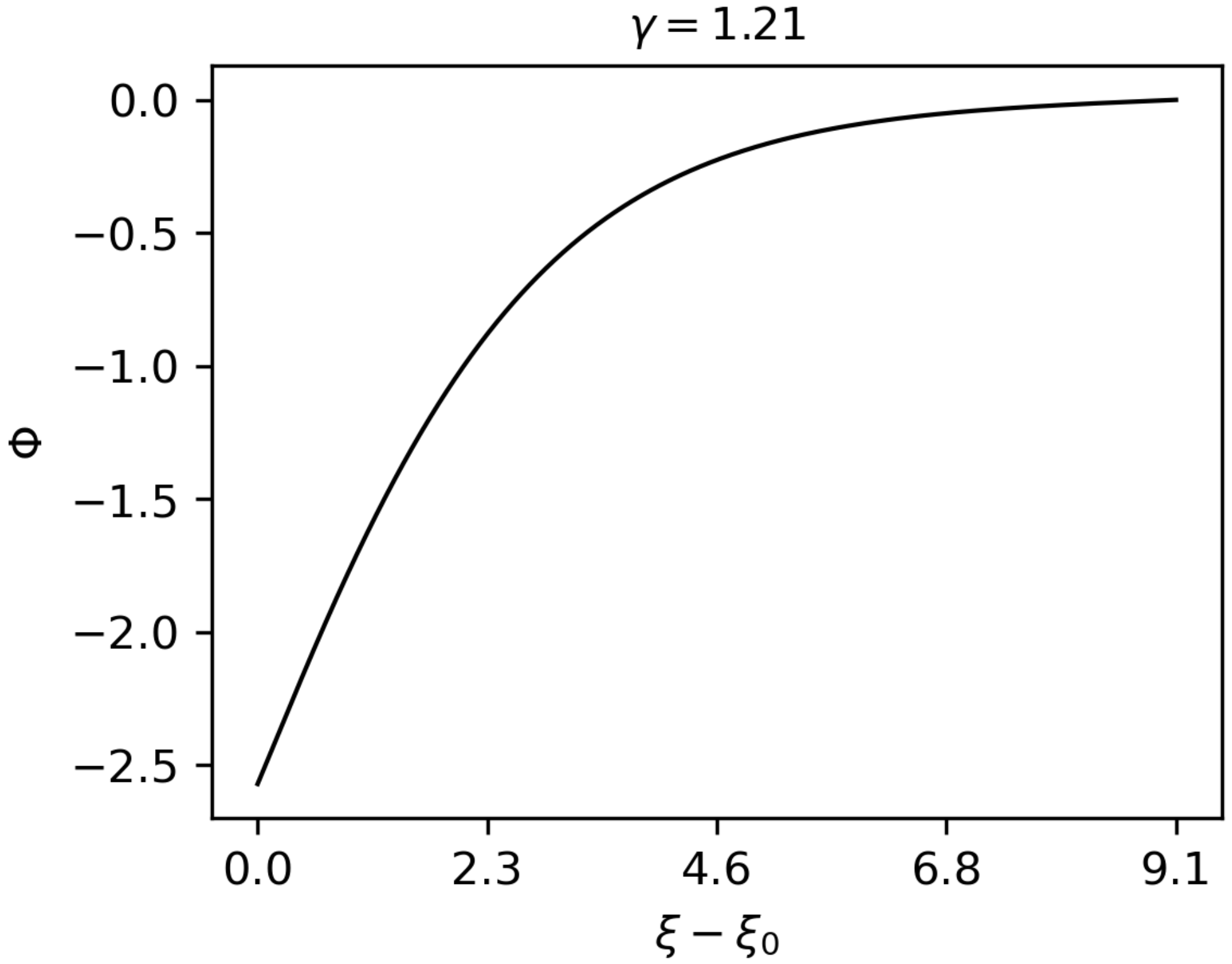}
    \end{subfigure}
    \begin{subfigure}{0.4\textwidth}
         \includegraphics[width=\textwidth]{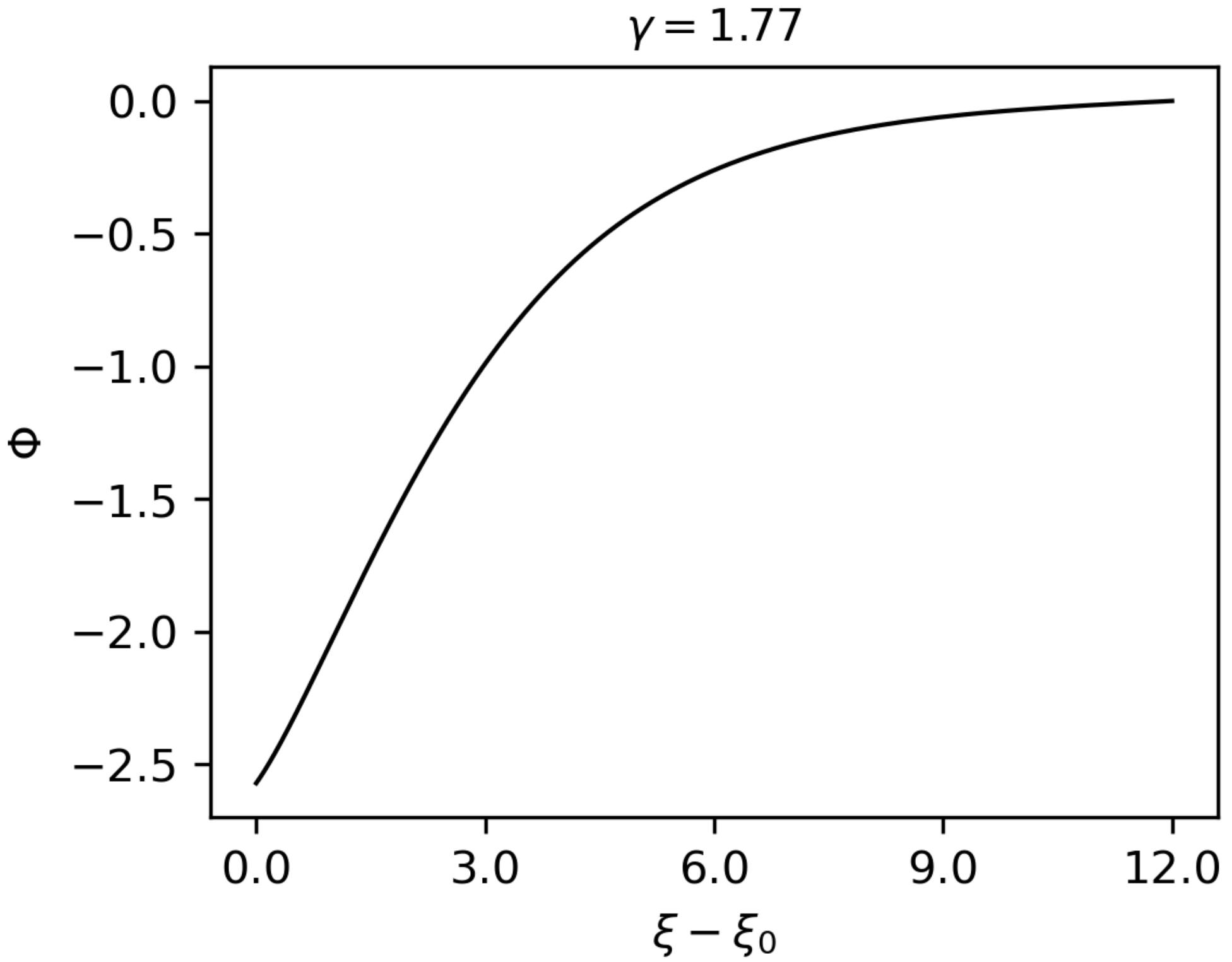}
    \end{subfigure}
    \hfill
    \begin{subfigure}{0.4\textwidth}
         \includegraphics[width=\textwidth]{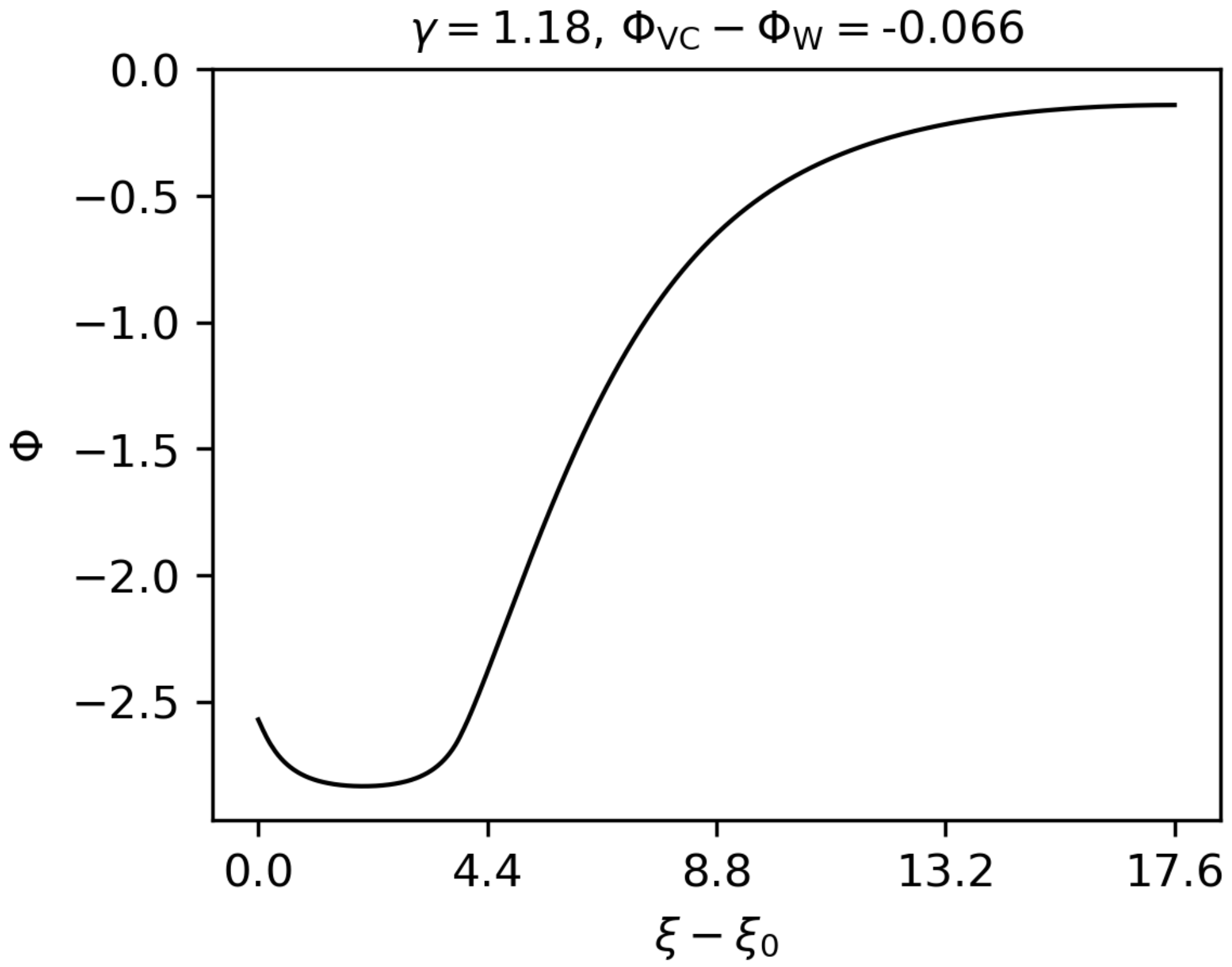}
    \end{subfigure}
    \begin{subfigure}{0.4\textwidth}
         \includegraphics[width=\textwidth]{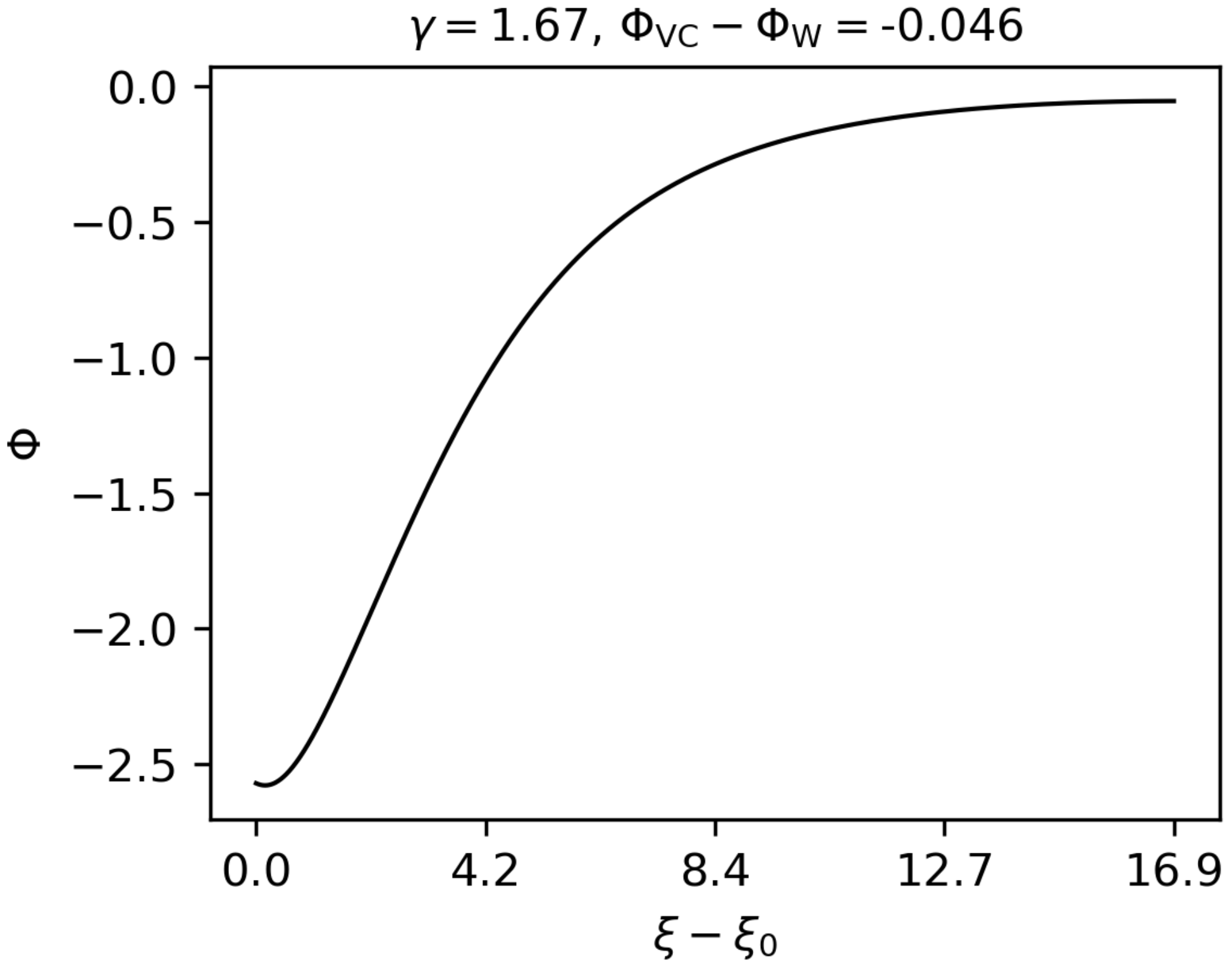}
    \end{subfigure}
\caption{Example potential sheaths at extremes of the \(\ps*{\gamma,\,\Phi_\text{VC} - \Phi_\text{W}}\) parameter space, for material work function \(\phi_\text{w} = \SI{4.5}{eV}\); (top) potential spaces without virtual cathode, (bottom) potential spaces with virtual cathode.}
\label{fig:SphHighExamplePotentials}
\end{figure}

\subsubsection{Spherical model spaces for low work function \(\phi_\text{w} = \SI{3.0}{eV}\)}
\begin{figure}[H]
    \centering
    \includegraphics[width=0.7\linewidth]{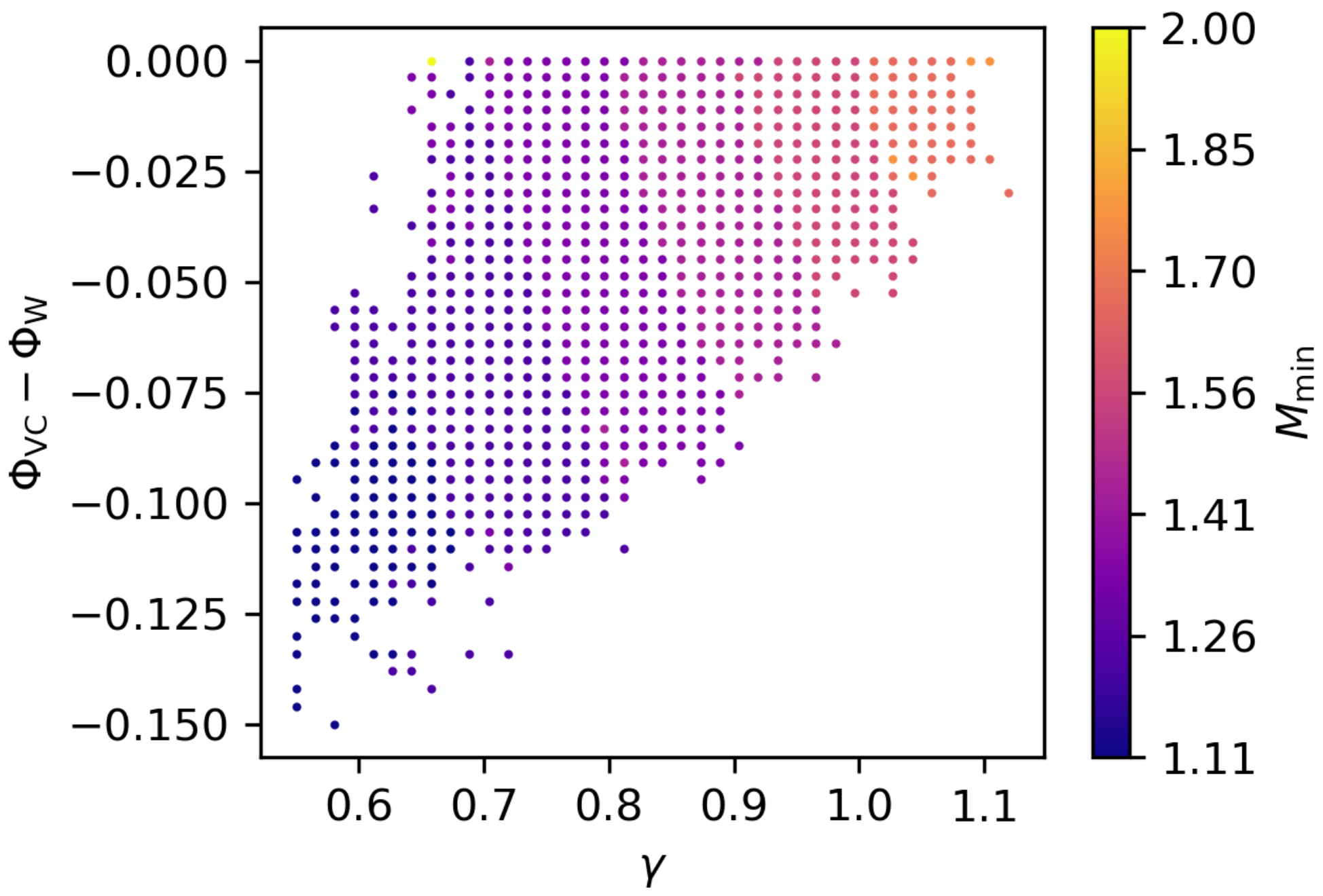}
    \caption{\(M_\text{min}\) as a function of current ratio \(\gamma \in \bs*{0.55,\,1.15}\) and virtual cathode potential with respect to the wall \(\Phi_\text{VC} - \Phi_\text{W}\in \bs*{0,-0.15}\), for material work function \(\phi_\text{w} = \SI{3.0}{eV}\).}
    \label{fig:SphLowM}
\end{figure}

\begin{figure}[H]
    \centering
    \includegraphics[width=0.7\linewidth]{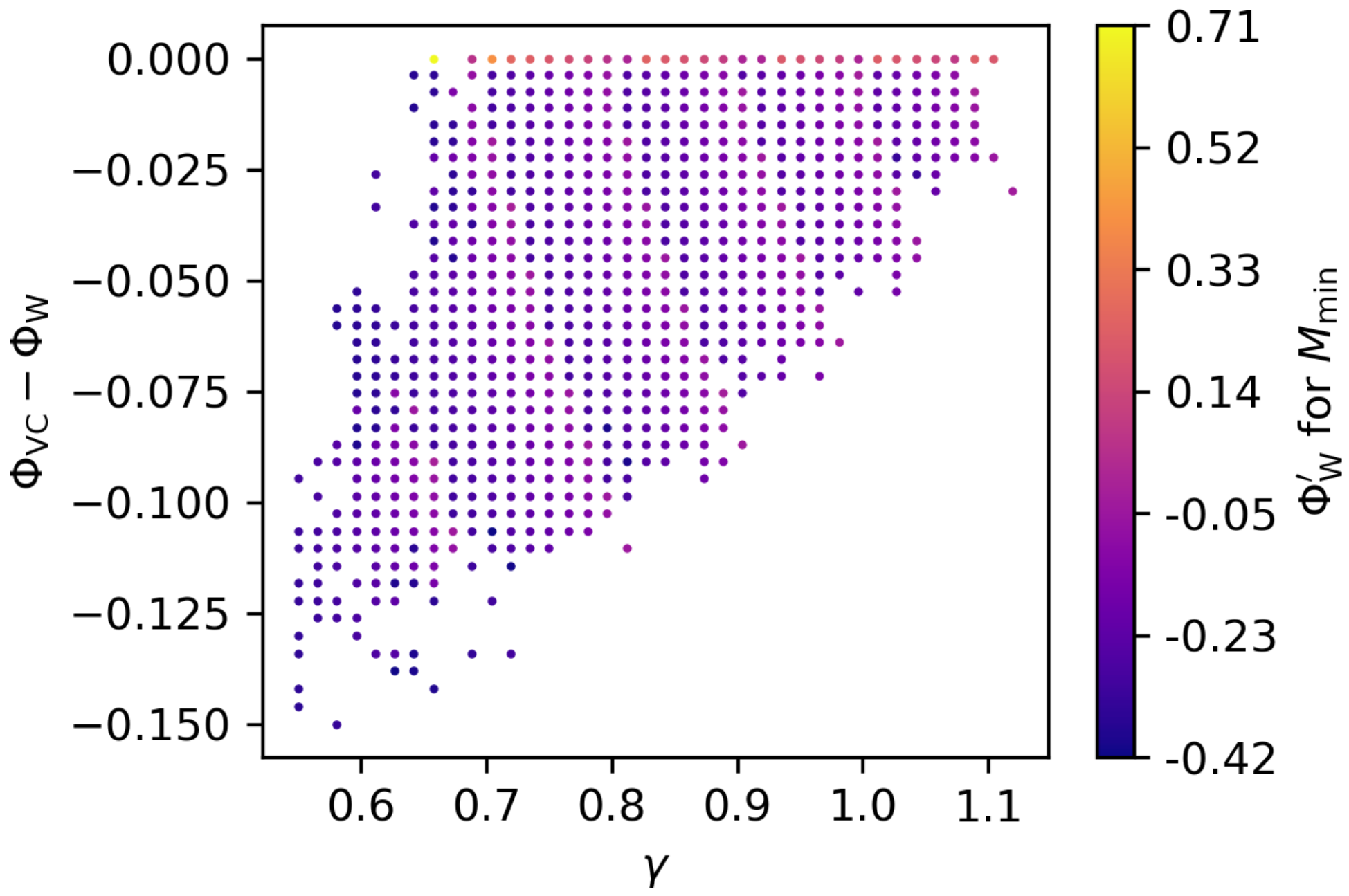}
    \caption{\(\Phi'_\text{W}\) for \(M_\text{min}\) in Fig. \ref{fig:SphLowM} as a function of current ratio \(\gamma \in \bs*{0.55,\,1.15}\) and virtual cathode potential with respect to the wall \(\Phi_\text{VC} - \Phi_\text{W}\in \bs*{0,-0.15}\), for material work function \(\phi_\text{w} = \SI{3.0}{eV}\).}
    \label{fig:SphLowPhiWPrime}
\end{figure}

\begin{figure}[H]
    \centering
    \includegraphics[width=0.7\linewidth]{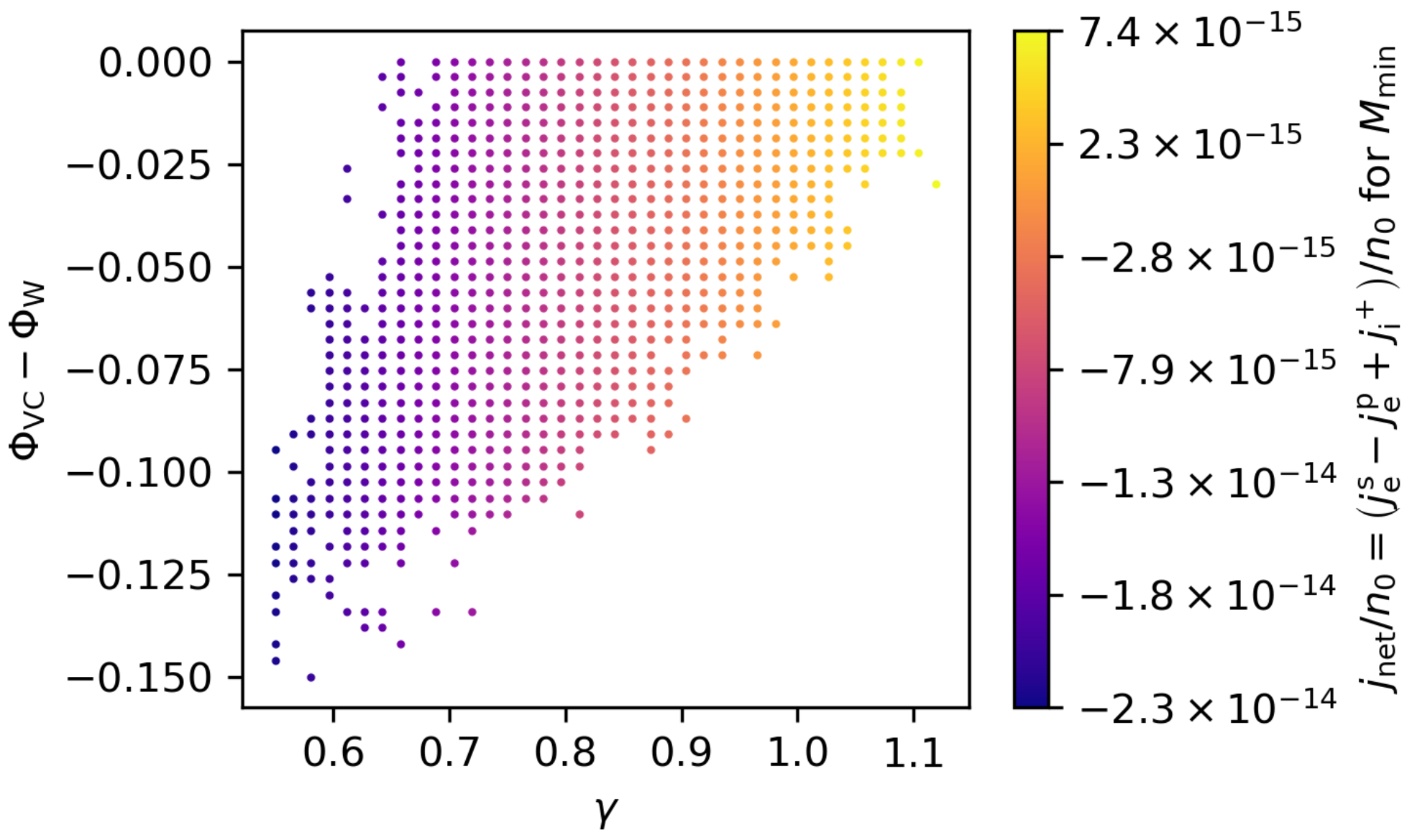}
    \caption{\(j_\text{net}/n_0\) for \(M_\text{min}\) in Fig. \ref{fig:SphLowM} as a function of current ratio \(\gamma \in \bs*{0.55,\,1.15}\) and virtual cathode potential with respect to the wall \(\Phi_\text{VC} - \Phi_\text{W}\in \bs*{0,-0.15}\), for material work function \(\phi_\text{w} = \SI{3.0}{eV}\).}
    \label{fig:SphLowJnet}
\end{figure}

\begin{figure}[H]
    \centering
    \begin{subfigure}{0.4\textwidth}
         \includegraphics[width=\textwidth]{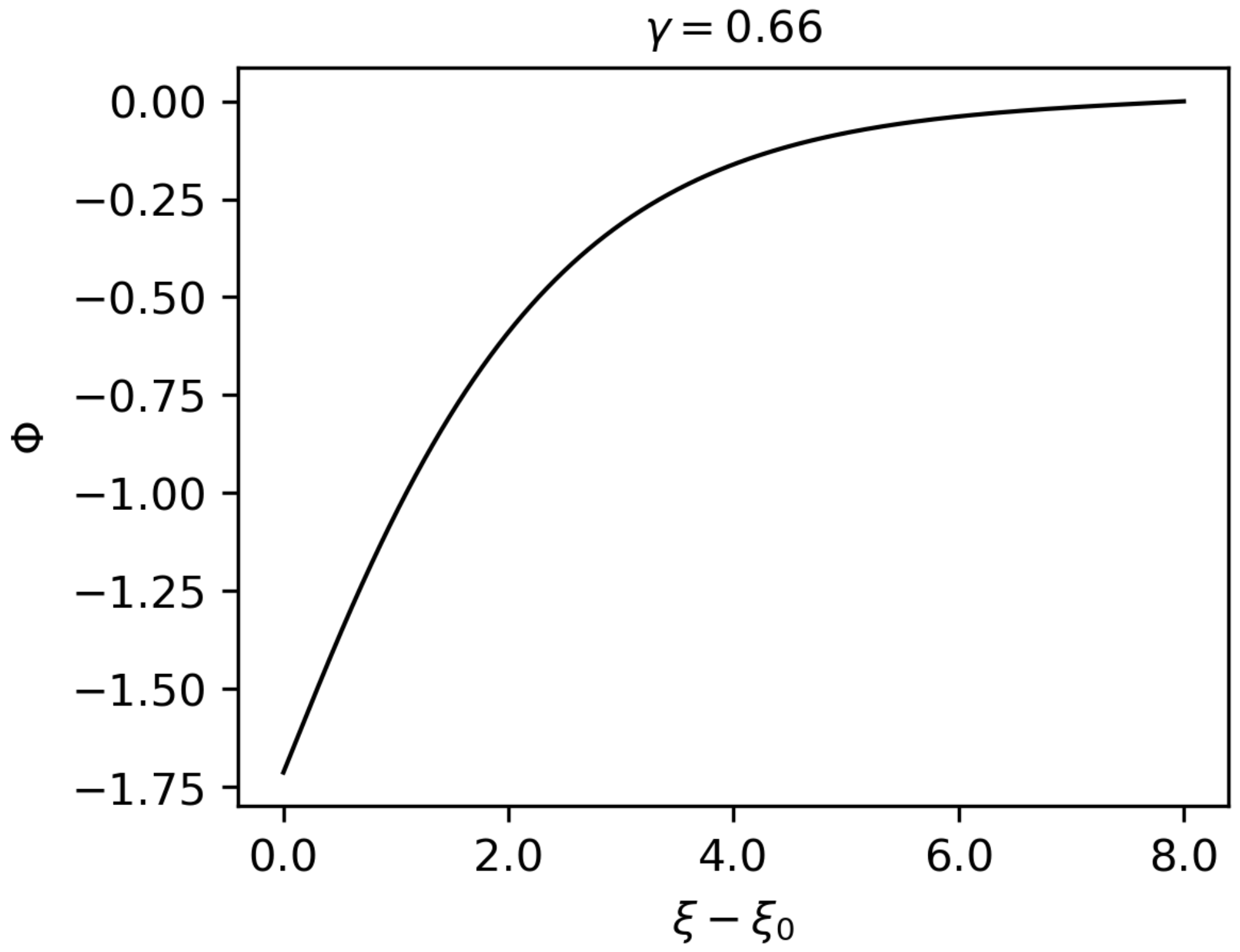}
    \end{subfigure}
    \begin{subfigure}{0.4\textwidth}
         \includegraphics[width=\textwidth]{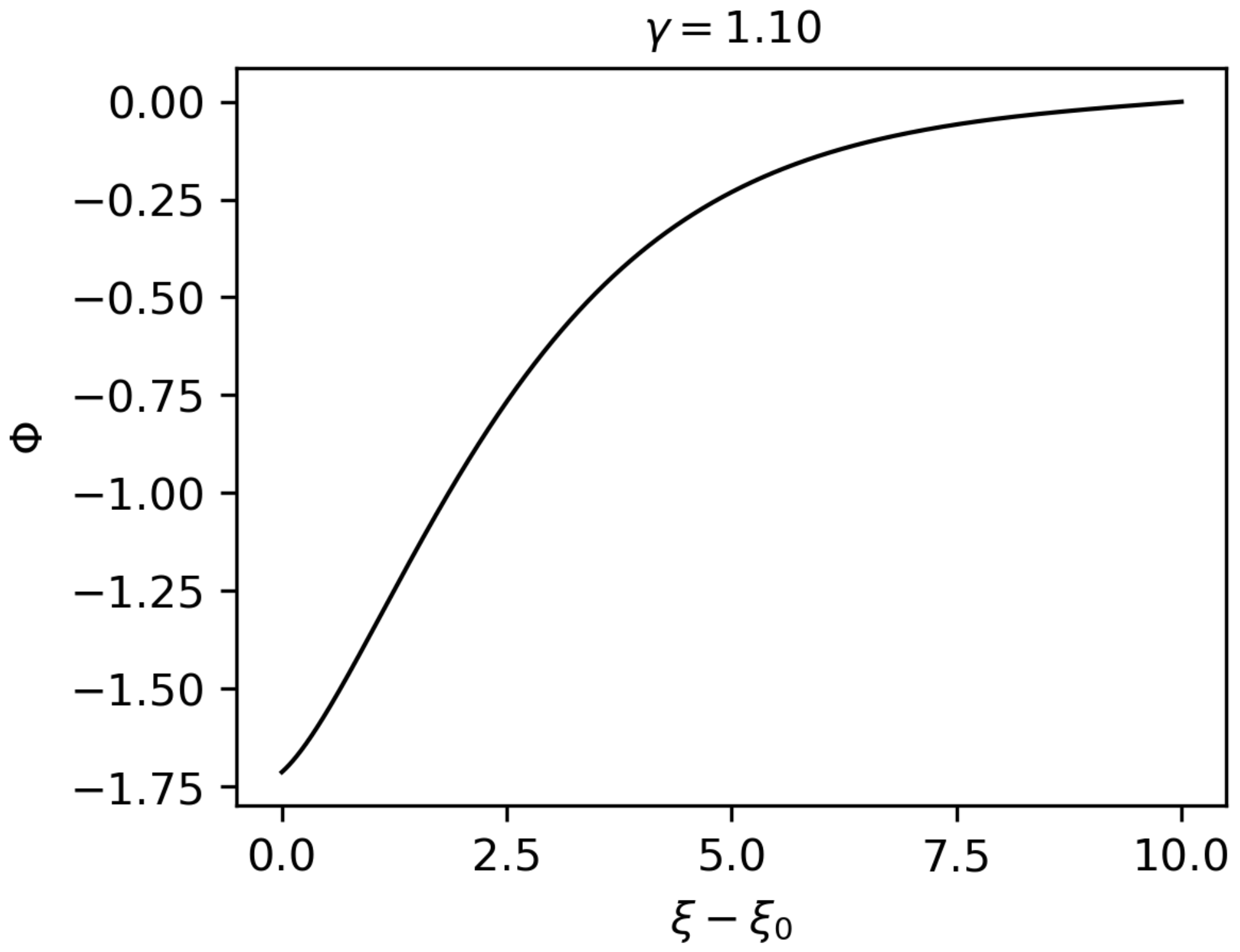}
    \end{subfigure}
    \hfill
    \begin{subfigure}{0.4\textwidth}
         \includegraphics[width=\textwidth]{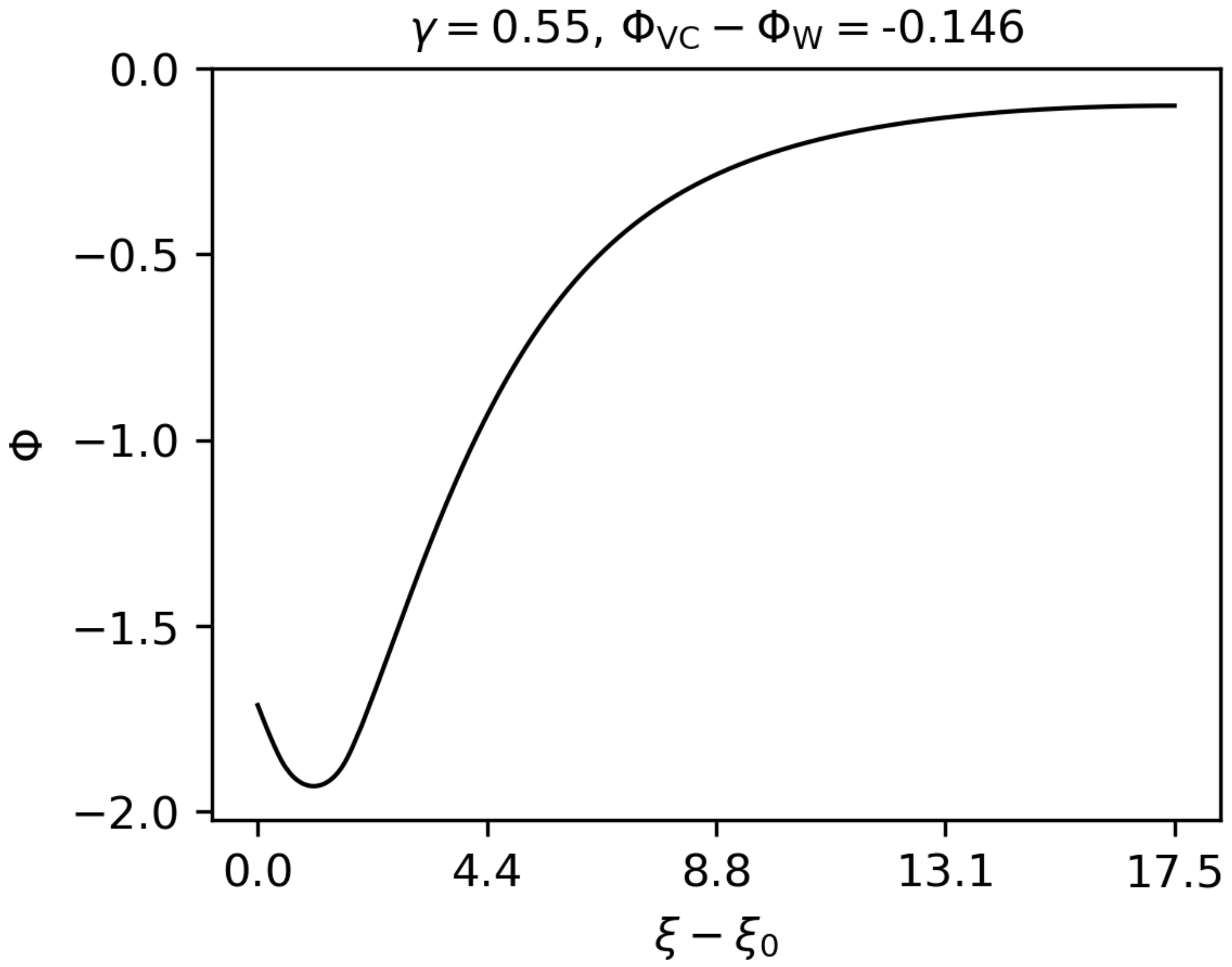}
    \end{subfigure}
    \begin{subfigure}{0.4\textwidth}
         \includegraphics[width=\textwidth]{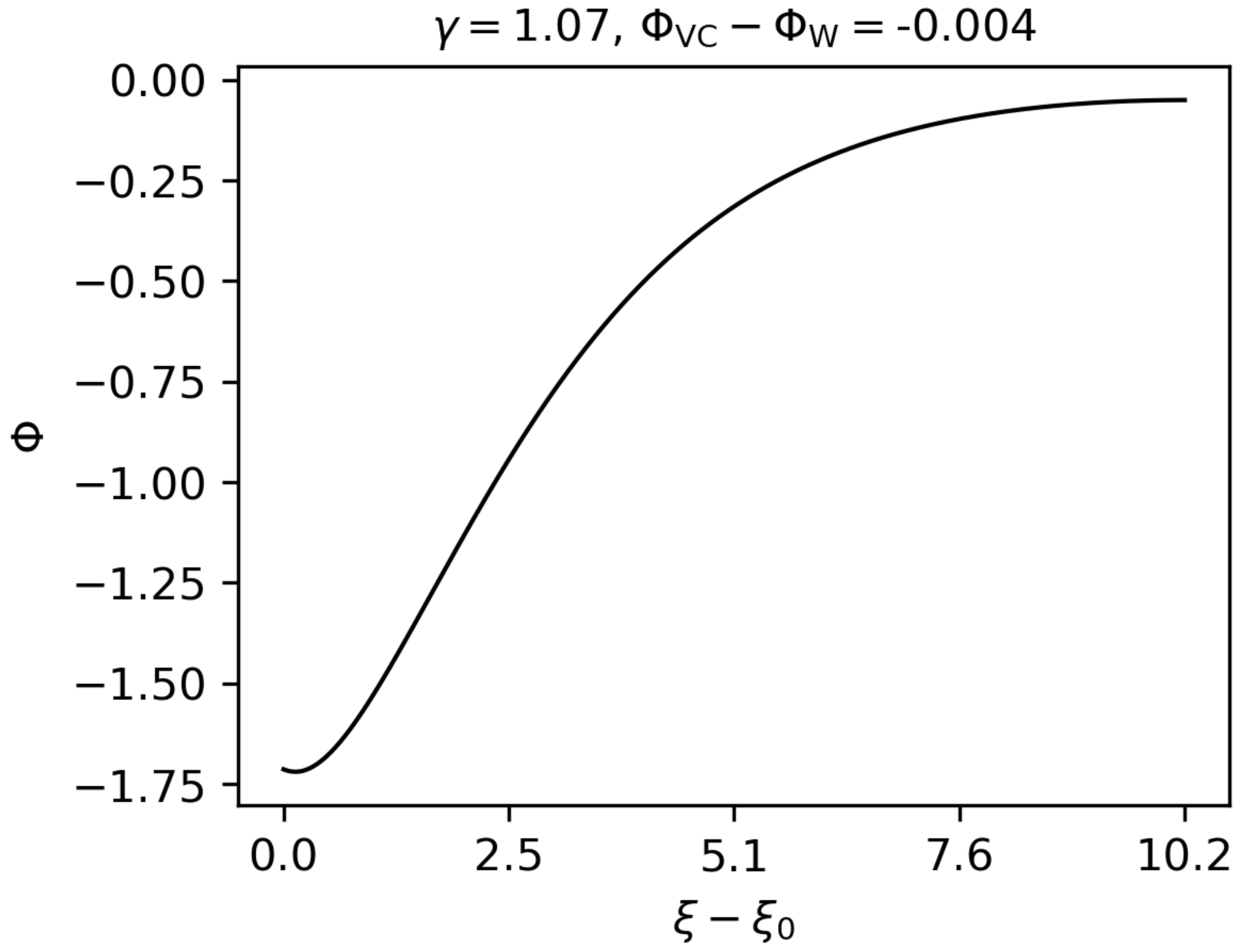}
    \end{subfigure}
\caption{Example potential sheaths at extremes of the \(\ps*{\gamma,\,\Phi_\text{VC} - \Phi_\text{W}}\) parameter space, for material work function \(\phi_\text{w} = \SI{3.0}{eV}\); (top) potential spaces without virtual cathode, (bottom) potential spaces with virtual cathode.}
\label{fig:SphLowExamplePotentials}
\end{figure}

Concerning the derivative of potential at the wall \(\Phi'_\text{W}\), we note that for all coordinate systems, it is consistently at a negative extreme for all virtual cathode magnitudes. Knowing this output parameter \(\Phi'_\text{W}\) is critical to generating a potential sheath, as it and the wall potential constitute the initial values needed to solve the initial value problem for the potential space. Also, the potential derivative at the wall is representative of the negative of electric field at the wall, which may be a useful piece of information for future studies. On the shape of the potential sheathes themselves, we find that the spherical coordinate model's virtual cathodes are particularly more rounded for large virtual cathode magnitude, implying a lower second derivative around the virtual cathode, compared to the other coordinate systems. Also, we observe that in all coordinate systems, the potential sheath does not exceed 20 Debye lengths before reaching the normalized potential \(\Phi = 0\) of the quasi-neutral plasma, and the virtual cathode is always positioned within 5 Debye lengths of the wall. 

Lastly, we find that the discrepancy between the actual \(\Phi_\text{VC} - \Phi_\text{W}\) and the input \(\Phi_\text{VC} - \Phi_\text{W}\) of a graphed potential space does not exceed \(\text{threshold}_\text{VC}\), but that it may be more apparent for \(\Phi_\text{VC} - \Phi_\text{W}\) in the cylindrical and spherical coordinates' results due to the choice of \(\text{threshold}_\text{VC} = 0.3\). However, by choosing a large enough threshold, we yield far more parameter space points in our brute force search; the benefit of this is that within the resulting larger parameter space, it is easier to glean what parameter combinations give low discrepancy in \(\Phi_\text{VC} - \Phi_\text{W}\), just by plotting the potential near output parameter combinations that already have relatively low discrepancy, than for a small parameter space that leaves few combinations to interrogate in the first place.

\section{Conclusion}
We have expanded Takamura's methodology to encompass Cartesian, cylindrical, and spherical coordinate systems, developing the integrals needed for all plasma source terms that contribute to Poisson's differential equation, and modularizing their presentation for easy use. Then, we demonstrated a generalized algorithm utilizing Runge-Kutta solution to initial value problems to create potential spaces for any of the three coordinate systems, and applied this method to reproduce Takamura's potential space results qualitatively. By a simple brute force technique, we discovered the parameters for valid potential sheathes with and without virtual cathodes, including minimum Mach number, potential derivative at the wall, and net electron current. We further show example potential sheathes at extremes of the parameter spaces that could be used to study heat spreading or the trajectory dynamics of seeded ions.

In the future, we expect to make correlations for the potential sheathes themselves, fitting them to the Mittag-Leffler class of functions that allow continuous interpolation between the Gaussian and Lorentzian functions \cite{MittagLeffler1903}. This advancement poses a mathematical optimization challenge due to the number of input parameters that yield a valid sheath, and significant number of output characteristics that quantify those valid sheathes. Finally, we would like to perform parametric studies on input parameters that we have so far held constant, i.e., producing datasets of potential sheathes and their characteristics as a function of wall and plasma temperature combinations, as well as a function of a wider range of work functions. Such investigations would inform how to engineer the material and plasma conditions of the leading edge to potentially increase the cooling effect of thermionically driven electron current.

\end{document}